\documentclass[twocolumn]{aastex61}
\pdfoutput=0 %for arXiv submission
\usepackage{amsmath,amstext}
\usepackage[T1]{fontenc}
\usepackage{apjfonts} 
\usepackage{natbib}
\usepackage{amsmath}
\usepackage{color}
\usepackage{longtable}
\usepackage{graphicx}
\usepackage{textcomp}

\usepackage{hyperref}

\shorttitle{Disk winds}
\shortauthors{Fang et al.}

\newcommand{\kms}{\,km\,s$^{-1}$}

\newcommand{\mum}{\,$\mu$m}

\newcommand{\Msun}{$M_{\odot}$}

\newcommand{\Msunyr}{$M_{\odot}$yr$^{-1}$}
\newcommand{\accunit}{$_{\odot}$ yr$^{-1}$}

\newcommand{\sect}{Sect.\,}
\newcommand{\rev}{ }

\newcommand{\OIa}{[O\,{\scriptsize I}]\,$\lambda$6300}
\newcommand{\OIb}{[O\,{\scriptsize I}]\,$\lambda$5577}
\newcommand{\SII}{[S\,{\scriptsize II}]\,$\lambda$4068}

\newcommand{\SIIb}{[S\,{\scriptsize II}]\,$\lambda$6731}
\def\arcsec{\hbox{$^{\hbox{\rlap{\hbox{\lower4pt\hbox{$\,\prime\prime$}}}\hbox{$\frown$}}}$}}

\begin{document}

\title{A New Look at T Tauri Star Forbidden Lines: MHD Driven Winds from the Inner Disk}
\author{Min Fang}
\affiliation{Department of Astronomy, University of Arizona, 933 North Cherry Avenue, Tucson, AZ 85721, USA}
\affiliation{Earths in Other Solar Systems Team, NASA Nexus for Exoplanet System Science}
\author{Ilaria Pascucci}
\affiliation{Department of Planetary Sciences, University of Arizona, 1629 East University Boulevard, Tucson, AZ 85721, USA}
\affiliation{Earths in Other Solar Systems Team, NASA Nexus for Exoplanet System Science}
\affiliation{Max Planck Institute for Astronomy (MPIA), K\"onigsthul 17, 69117, Heidelberg, Germany}
\author{Suzan Edwards}
\affiliation{Five College Astronomy Department, Smith College, Northampton, MA 01063, USA}
\author{Uma Gorti}
\affiliation{SETI Institute/NASA Ames Research Center, Mail Stop 245-3, Moffett Field, CA 94035-1000, USA}
\author{Andrea Banzatti}
\affiliation{Department of Planetary Sciences, University of Arizona, 1629 East University Boulevard, Tucson, AZ 85721, USA}
\affiliation{Earths in Other Solar Systems Team, NASA Nexus for Exoplanet System Science}
\author{Mario Flock}
\affiliation{Max Planck Institute for Astronomy (MPIA), K\"onigsthul 17, 69117, Heidelberg, Germany}
\author{Patrick Hartigan}
\affiliation{Department of Physics and Astronomy, Rice University, 6100 S. Main street, Houston, TX 77005-1892, USA}
\author{Gregory J. Herczeg}
\affiliation{Kavli Institute for Astronomy and Astrophysics, Peking University, Yiheyuan 5, Haidian Qu, 100871 Beijing, China}
\author{Andrea K. Dupree}
\affiliation{Harvard-Smithsonian Center for Astrophysics, Cambridge, MA 02138, USA}
\begin{abstract}
Magnetohydrodynamic (MHD) and photoevaporative winds are thought to play an important role in the evolution and dispersal of planet-forming disks. We report the first high-resolution ($\Delta v\sim$\,6\,\kms) analysis of \SII, \OIb, and \OIa\ lines from a sample of 48 T~Tauri stars. Following \cite{2016ApJ...831..169S}, we decompose them into three kinematic components: a high-velocity component (HVC) associated with jets, and a low-velocity narrow (LVC-NC) and broad (LVC-BC) components. We confirm previous findings that many LVCs are blueshifted by more than 1.5\,\kms\, thus most likely trace a slow disk wind. We further show that the profiles of individual components are similar in the three lines. We find that most LVC-BC and NC line ratios are explained by thermally excited gas with  temperatures between 5,000$-$10,000~K and electron densities $\sim10^{7}-10^{8}$\,cm$^{-3}$. The HVC ratios are better reproduced by shock models with a pre-shock H number density of $\sim10^{6}-10^{7}$\,cm$^{-3}$. Using these physical properties, we estimate $\dot{M}_{\rm wind}/\dot{M}_{\rm acc}$ for the LVC and $\dot{M}_{\rm jet}/\dot{M}_{\rm acc}$ for the HVC. In agreement with previous work, the mass carried out in jets is modest compared to the accretion rate. {\rev With the likely assumption that the NC wind height is larger than the BC}, the LVC-BC $\dot{M}_{\rm wind}/\dot{M}_{\rm acc}$  is found to be higher than the LVC-NC. These results suggest that most of the mass loss occurs close to the central star, within a few au, through an MHD driven wind. Depending on the wind height, MHD winds might play a major role in the evolution of the disk mass.
\end{abstract}


\section{Introduction}
Circumstellar disks form as a result of angular momentum conservation during the protostellar core collapse \citep{1977ApJ...214..488S} and play an important role both in star and planet formation. At early times, a significant fraction of the stellar mass is accreted through the disk and what is not accreted or dispersed via other mechanisms provide the raw material to build planets. As such it is important to understand how disks evolve and disperse.

 In the current paradigm, there are three main stages of disk evolution and dispersal (e.g., \citealt{2016SSRv..205..125G}, \citealt{2017RSOS....470114E} for recent reviews). For most of the disk lifetime (Stage 1), evolution is primarily set by  accretion.  Beyond a fraction of the radius where the sound speed equals the local Keplerian orbital speed ($\sim 2$~au for $10^4$~K gas, and $\gtrsim 10-100$~au for $100-1,000$~K gas for a sun-like star) gas becomes unbound and a photoevaporative thermal wind  can be established. When the disk accretion rate through this radius drops below the wind mass loss rate, photoevaporation limits the supply of gas to the inner disk, a gap is
formed (Stage 2), and the inner disk drains onto the star on the local viscous timescale - of order 100,000 years. In the last stage (Stage 3), there is no accretion onto the star and the disk is rapidly cleared from inside-out by stellar high-energy photons directly irradiating the outer disk.

Although photoevaporative winds are believed to play an important role in disk dispersal, large
scale jets/outflows, which contribute to significant mass loss at the earliest stages, are usually attributed to an origin in an MHD wind (e.g., \citealt{2014prpl.conf..451F} for a recent review). Such winds can be effective in removing disk angular momentum and in facilitating accretion of disk gas onto the star (e.g., \citealt{2007prpl.conf..277P,2007prpl.conf..261S}). 

Recently, significant theoretical effort has been devoted to understanding MHD disk winds and
their relation to accretion and disk dispersal. It has long been realized that the inclusion of non-ideal MHD effects suppresses magnetorotational instability (MRI) turbulence (e.g., \citealt{1996ApJ...457..355G}) and recent numerical simulations show that even a weak net vertical magnetic field, which
could be for example leftover from the earlier stage of cloud collapse, can drive MHD winds (e.g.,  \citealt{2009ApJ...691L..49S}, \citealt{2013A&A...552A..71F}). 
Local box simulations (e.g., \citealt{2013ApJ...769...76B}) and radially global, but vertically restricted, simulations \citep{2015ApJ...801...84G,2017A&A...600A..75B}  show that the launch of MHD winds is robust, can occur over the planet-forming region ($\sim$\,1--30\,au), and can drive accretion at the observed levels. However, the predicted wind mass loss rate, hence the accretion rate, strongly depends on the magnetic field strength and how it evolves (e.g., \citealt{2013ApJ...778L..14A}, \citealt{2017ApJ...836...46B}), both of which are poorly constrained. Therefore, identifying diagnostics that trace disk winds and deriving wind mass loss rates is important not only to understand how disks disperse but also how they evolve. 

Low-excitation optical forbidden lines, especially from the \OIa\ transition, have long been used to study jets/outflows from T~Tauri stars (e.g., \citealt{1987ApJ...321..473E,1995ApJ...452..736H}). Their line profiles typically present two distinct components: a high-velocity component (HVC), blueshifted by 30--200\,\kms\ from the stellar velocity, and a low-velocity component (LVC), typically blueshifted by $\sim$5\,\kms. Spatially resolved observations have demonstrated that HVCs are formed in extended collimated jets (e.g., \citealt{2000A&A...356L..41L,2000ApJ...537L..49B,2002ApJ...580..336W}),  most likely linked to MHD winds (e.g., \citealt{2006A&A...453..785F}).

Employing much higher spectral resolution than in previous studies ($\Delta v \sim$\,4\,\kms), \cite{2013ApJ...772...60R} found that the  [O\,{\scriptsize I}]~LVC itself can be described by the combination of a narrow component (NC) and a broad component (BC). More recently,  \cite{2016ApJ...831..169S}  confirmed this finding on a much larger sample of 33 T~Tauri stars  observed with Keck/HIRES ($\Delta v \sim $\,7\,\kms). Using the line profiles and inclinations from resolved disk images, they inferred that most of the LVC-BC arises within $\sim 0.5$\,au while most of the LVC-NC arises outside this radius. By combining measured velocities with line widths and disk inclinations, they also found that
the LVC-BC tend to be narrower and more blueshifted for closer to face-on disks, as shown in some disk wind models (e.g., \citealt{2008MNRAS.391L..64A}). Finally, since the emitting region is within the photoevaporating radius even for 10,000\,K gas, they could conclude that the LVC-BC traces an MHD disk wind. The origin of the LVC-NC remains unclear as the inferred radial extent is consistent with a thermally driven photoevaporative wind but no trend between blueshifts and disk inclinations was seen \citep{2016ApJ...831..169S}. 

\begin{figure*}
\begin{center}
   \includegraphics[width=2\columnwidth]{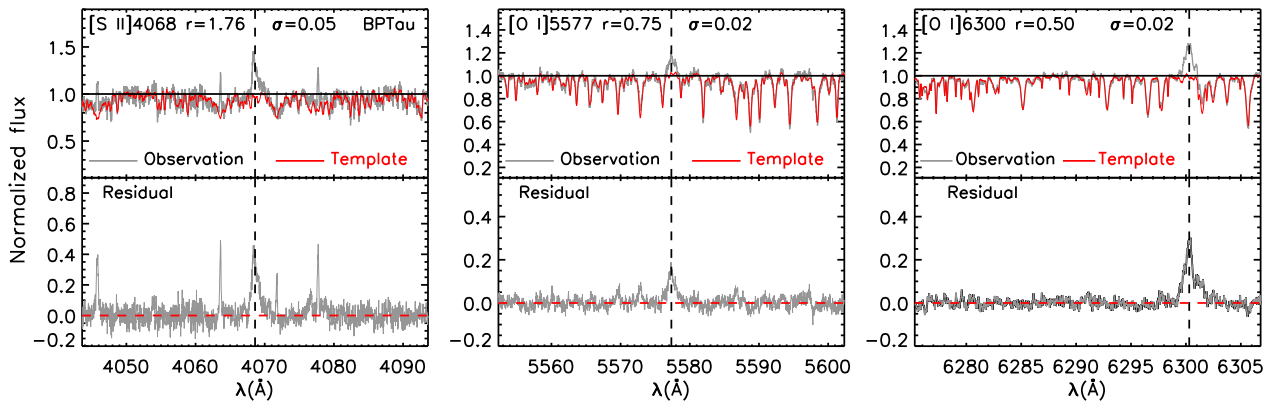}
  \caption{Top: Example of the removal of telluric (only for \OIa) and photospheric absorption features near  \SII, \OIb, and \OIa\ lines for BP~Tau. In each panel, the normalized target spectrum is shown in gray. The telluric and photospheric absorption features have not been removed from the spectrum.  The best-fit photospheric template is shown in red, and used to subtract the photospheric absorption features (including the telluric absorption features for \OIa) from the target spectrum. Bottom: the final corrected line profiles of \SII, \OIb, and \OIa\ of BP~Tau.}\label{Fig:sky_phot_example}
\end{center}
\end{figure*}

 An important step in computing wind mass loss rates is to constrain the properties of the emitting gas associated with the wind. Ratios of lines tracing the same kinematic component can be employed for this task (e.g., \citealt{2010LNP...793..213D}). However, so far the only lines analyzed at similarly high spectral resolution to distinguish BC and NC are the \OIa\ and \OIb.  Assuming that the lines are thermally excited their ratios constrain the range of temperature-electron densities for the emitting gas and imply higher electron density in the region traced by the LVC-BC for gas at similar temperatures \citep{2016ApJ...831..169S}. However, FUV photodissociation of OH molecules can produce a similar range of
[O\,{\scriptsize I}] ratios in much cooler ($\sim$1,000\,K) gas (e.g., \citealt{2000JChPh.11310073H} and \citealt{2011ApJ...735...90G}  for an application to disks). 
In this scenario of non-thermal
emission, line ratios do not constrain the temperature and density of the gas. In order to estimate wind mass loss rates, it is therefore necessary to first establish that the \OIa\ and \OIb\ emission is thermal.
As first pointed out
in \cite{2014A&A...569A...5N}, the \SII\ has a critical density very similar to the \OIa\ line and is likely to be thermally excited, 
hence the [S\,{\scriptsize II}] and [O\,{\scriptsize I}] line profiles, as well as their line ratio, can be used to distinguish between thermal vs non-thermal excitation.

In this work, we expand upon previous high-resolution studies by analyzing the
\SII, \OIb, and \OIa\  from a sample of 48 T~Tauri stars  with the main goal of determining whether the [O\,{\scriptsize I}] emission is thermal or non-thermal. In a parallel work, we focus on the kinematic behavior of individual [O\,{\scriptsize I}] components to clarify the link between jets and winds (Banzatti~et~al. 2018).
First, we describe our sample, observations, and data reduction ($\S$\ref{sect2}). Then, we present our analysis ($\S$\ref{sect3}), which includes line decomposition in HVC LVC-BC, and LVC-NC following \cite{2016ApJ...831..169S}. In $\S$\ref{sect4} we show that the  \SII , \OIb , and \OIa\ line profiles are very similar within each kinematic component strongly suggesting that the [O\,{\scriptsize I}] emission is thermally produced, and not the result of photodisocciation of OH
in a cool gas. We use line ratios to constrain the properties of the emitting gas and compute wind mass loss rates in $\S$\ref{sect5}. We also calculate mass accretion rates from several permitted lines and compare accretion vs mass loss rates ($\S$\ref{sect5}). We discuss the implications of our results in $\S$\ref{Sect:discussion} and summarize our findings in $\S$7.

\section{Sample and Data Reduction}\label{sect2}

 \begin{table*}
\scriptsize
\renewcommand{\tabcolsep}{0.02cm}
\caption{A list of the sources in this work, as well as their stellar and disk properties and the used photspheric templates}\label{Table:sample}
\begin{center}
\begin{tabular}{lcccccccccccccccccccccccccc}
  \hline
ID &Name &Region &Dist  &Spt      &Log~$L_{\star}$  &$A_{\rm V}$ &$R_{\star}$  &$M_{\star}$ &Disk &$i$  &Ref &  RV(Helio) &Correction    &\multicolumn{2}{c}{Photospheric template} \\
       \cline{15-16} 
&   &  &(pc)  &Type &($L_{\odot}$) &(mag)    &($R_{\odot}$)   &($M_{\odot}$) & & (deg) &   &(\kms)&(\kms)    &[S\,{\tiny II}]\,$\lambda$4068   &[O\,{\tiny I}]\,$\lambda$5577, [O\,{\tiny I}]\,$\lambda$6300 \\
% \multicolumn{14}{c}{Taurus (1--2\,Myr)}\\
\hline
1&DP~Tau  &Taurus &130  &M0.8 & $-1.24$&0.80 &0.56 &0.55 &Full & &1 &16.55$\pm$0.89 &$-0.64$    &no correction  & TWA~13  \\
2&CX~Tau  &Taurus&126.8&M2.5 &$-0.61$ &0.25  &1.36 &0.35&TD  &61 &1, 9 &19.15$\pm$1.31 &$-0.40$    &\nodata        &TWA~8A  \\
3&FP~Tau  &Taurus&127.2&M2.6 &$-0.81$ &0.60  &1.10 &0.35&TD  & 66&1, 9 &16.90$\pm$2.11 &$-0.75$ & TWA~8A  & TWA~8A     \\
4&FN~Tau  &Taurus&129.7&M3.5 &$-0.29$ &1.15  &2.22&0.24&Full  & &1 &16.29$\pm$1.23&$-0.09$ &TWA~8A   & TWA~8A    \\
5&V409~Tau&Taurus&130.2&M0.6 &$-0.19$ &1.00  &1.87&0.47&Full  & &1 &17.78$\pm$0.73&$-0.49$ &TWA~13  &TWA~13      \\
6&BP~Tau  &Taurus&127.7&M0.5 &$-0.40$ &0.45  &1.45&0.54&Full  &39 &1, 10 &16.76$\pm$0.54 &$-0.43$ &TWA~13  &V819~Tau    \\
7&DK~Tau~A  &Taurus&127.1&K8.5 &$-0.35$ &0.70  &1.41&0.66&Full &26 &1, 11&16.60$\pm$0.42 &$-0.71$ &TWA~13   &V819~Tau      \\
8&HN~Tau~A&Taurus&133.3&K3   &$-0.81$ &1.15  &0.63&0.69&Full &75 &1, 9 &18.91$\pm$1.96 &$-0.46$ &no correction  &2MASS~J15584772-1757595\\
9&UX~TauA&Taurus&137.5 &K0   &$0.18$  &0.65  &1.74&1.40&TD$^{a}$ &39  &1, 11 &18.80$\pm$0.64 &$-0.44$ &EPIC~212021375$^{b}$  &EPIC~203476597         \\
10&GK~TauA&Taurus&128.1&K6.5 &$-0.11$ &1.50  &1.78&0.67&Full &71  &1, 9 &18.52$\pm$0.64 &$-0.49$  &HD~201092$^{b}$ & V819~Tau \\ 
11&GI~Tau &Taurus&129.2&M0.4 &$-0.32$&2.05&1.58&0.53&Full & &1&19.02$\pm$0.55 &$-0.40$&TWA~13   & V819~Tau           \\ 
12&DM~Tau &Taurus&143.5&M3.0 &$-0.87$ &0.10&1.05&0.31 &TD$^{a}$ &34  &1, 11 &19.64$\pm$1.50 &$-1.36$&TWA~8A  &TWA~8A     \\
13&LKCa~15&Taurus&156.9&K5.5 &$-0.11$&0.30 &1.69&0.76&TD$^{a}$ &51 &1, 11 &18.71$\pm$0.46 &$-0.55$&HD36003$^{b}$&V819~Tau      \\
14&DS~Tau &Taurus&157.2&M0.4 &$-0.62$ &0.25  &1.12&0.62&Full  &71&1, 9 &16.51$\pm$0.97 &$-0.28$&TWA~13    &TWA~13       \\
%\hline
% \multicolumn{14}{c}{Lupus~1 and Lupus~3 (1--3\,Myr)}\\
%\hline
15&SZ~65   &Lupus &153.4 &K6 & $-0.03$  &0.80 &1.90 &0.68&TD  &61&1, 19 &$-$2.85$\pm$0.90 &$-0.60$&HD~36003$^{b}$ & V819~Tau      \\ 
16&SZ~68A  &Lupus&152.1 &K2  &$0.75$ &1.00  &3.57&1.27&Full &33 &1, 19 &$-$1.41$\pm$1.28 &$-0.74$&EPIC~212021375$^{b}$ & 2MASS~J15584772-1757595     \\
17&SZ~73    &Lupus&154.8 &K8.5 &$-0.74$ &2.75 &0.90&0.75&Full &50 &1, 12 &$-$3.56$\pm$1.59 &$-1.10$ &HD~201092$^{b}$      &V819~Tau    \\
18&HM~Lup   &Lupus&154.1 &M2.9 & $-0.78$&0.60  &1.16&0.32&Full &53 &1, 12 &$-$1.62$\pm$1.82 &$-0.90$&TWA~8A   &TWA~8A     \\
19&GW~Lup   &Lupus&153.9 &M2.3& $-0.61$&0.55 &1.34 &0.37&Full &40 &1, 12 &$-$1.17$\pm$1.12 &$-0.77$&TWA~8A   &TWA~8A     \\
20&GQ~Lup   &Lupus&150.1 &K5.0 &$-0.04$&1.60  &1.80&0.78&Full &60 &1, 13 &$-$2.13$\pm$0.29 &$-1.00$&HD~36003$^{b}$  & V819~Tau            \\
21&SZ~76    &Lupus&157.5 &M3.2 & $-0.69$&0.90 &1.33 &0.28&TD$^{a}$  &  &1&$-$1.14$\pm$2.16 &$-1.01$&TWA~8A     &TWA~8A   \\
22&RU~Lup   &Lupus&157.2 &K7.0 &$0.16$ &0.00   &2.48 &0.55&Full &3 &2, 19 &$-$0.8$\pm$2 &$-0.98$          &no correction & V819~Tau           \\
23&IM~Lup   &Lupus&156.4 &K6.0 & $0.01$&0.40 &1.98&0.67&TD &48 &1, 19&$-$0.64$\pm$0.58&$-1.18$  &HD~36003$^{b}$ & V819~Tau              \\
24&RY~Lup   &Lupus&156.6 &K2   & $0.26$ &0.40   &2.02&1.27&Full &68 &3, 22 &$-$0.43$\pm$1.09 &$-0.90$&EPIC~212021375$^{b}$ & EPIC 203476597     \\
25&Sz~102  &Lupus&160 &K2  & $-2.02$ &0.70  &0.15&\nodata&Full  &73 &3, 20 &12$\pm$2 &  $-1.82$        &no correction         &no correction     \\
%25&Sz~102$^{c}$  &336.1 &K2  & $-1.38$ &0.70  &0.31&\nodata&Full  &73 &3, 20 &12$\pm$2           &no correction         &no correction     \\
26&Sz~111   &Lupus&156.9 &M1.2 & $-0.69$&0.85 &1.11&0.50&TD$^{a}$   &53 &1, 22 &$-$0.16$\pm$0.74 &$-1.32$ &HD~209290$^{b}$ & V819~Tau           \\
27&Sz~98    &Lupus&154.3 &M0.4 & $-0.49$&1.25 &1.30&0.58&Full  &47 &1, 19&$-$0.32$\pm$0.70 &$-0.86$ &TWA~13        &TWA~13      \\
28&EX~Lup    &Lupus &157.0 &M0  &$-0.12$ &1.1  &2.04  &0.44 &Full &38  &3, 28  &2$\pm$1 &$-0.88$  &no correction   &V819~Tau     \\
%\hline
% \multicolumn{14}{c}{$\rho$~Oph (1--3\,Myr)}\\
%\hline
29&AS~205A  &$\rho$~Oph&125.9  &K5 &$0.89$ &1.75 &5.26 &0.68&Full &25 &1, 4, 14   &$-$5.31$\pm$0.55 &$-1.78$ &HD~36003$^{b}$  & V819~Tau        \\
30&DoAr~21  &$\rho$~Oph&132.6  &G1  &$1.32$ &7.10 &4.50 &2.79&TD$^{a}$ &  &1   &$-$1.63$\pm$2.44 &$-1.46$        &\nodata         &EPIC~203476597     \\
31&DoAr~24E &$\rho$~Oph&135.7  &K0 &$0.20$ & 4.32   &1.76 &1.41&Full  &20&5, 30&$-$5.77$\pm$0.60 &$-1.66$          &\nodata          &EPIC~203476597        \\
32&DoAr~44  &$\rho$~Oph&144.3  &K2  &$-0.03$  &1.7  &1.45 &1.22  &TD$^{a}$ &16 &6, 11 &$-$4.50$\pm$0.53 &$-1.50$           &EPIC~212021375$^{b}$ & 2MASS~J15584772-1757595       \\
33&EM*~SR~21A  &$\rho$~Oph&136.8  &F7  & $0.98$ &6.2 &2.62 &1.79 &TD$^{a}$ &18  &1, 11, 15, 16 &$-$5.66$\pm$3.65  &$-1.41$          &HIP~42106$^{b}$    &HIP~42106$^{b}$     \\
34&V853~Oph &$\rho$~Oph&138  &M2.5 & $-0.30$&0.14 &1.95 &0.32&Full&54  &5, 11  &$-$5.8$\pm$1.10 &$-1.96$           &TWA~8A   &TWA~8A     \\
%33&V853~Oph$^{c}$ &76.6  &M2.5 & $-0.81$&0.14 &1.08 &0.36&Full&54  &5, 11  &$-$5.8$\pm$1.10            &TWA~8A   &TWA~8A     \\
35&RNO~90   &$\rho$~Oph&115.7  &G8  &$0.42$&4.3  &2.01 &1.68 &Full  &37&17, 30&$-$9.11$\pm$1.09 &$-1.70$            &KW27$^{b}$   &  EPIC~203476597   \\
36&V2508~Oph &$\rho$~Oph&122.7  &K7 &$-0.13$&1.7  &1.78 &0.63&Full  &41&17, 11 &$-$7.62$\pm$0.79 &$-1.94$            &HD~201092$^{b}$      &V819~Tau   \\
37&V1121~Oph &$\rho$~Oph&119.7  &K4 &$-0.06$&1.08 &1.62   &0.96& Full &31&5, 11&$-$6.27$\pm$0.32 &$-1.78$  &HD~36003$^{b}$ & TWA~9A          \\

%\hline
% \multicolumn{14}{c}{Corona Australis (2\,Myr)}\\
%\hline
38&RX~J1842.9$-$3532 &Corona Australis&152.2   &K3  & $-0.23$ &0.6 &1.24 &1.07 &TD$^{a}$ &54 &1, 18 &$-$0.92$\pm$0.53 &$-1.31$     &HD~36003$^{b}$ & 2MASS~J15584772-1757595    \\   
39&RX~J1852.3$-$3700 &Corona Australis&144.1   &K4 &$-0.40$ &0.25 &1.10 &0.96 &TD$^{a}$&16 &1, 18  &0.64$\pm$0.53 &$-1.16$       &HD~36003$^{b}$  & 2MASS~J15584772-1757595    \\
40&VV~CrA            &Corona Australis&146.6   &K7 &$0.32$&3.95  &2.99&0.53&Full&49&1, 21  & $-$1.0$\pm$1  &$-1.39$     &no correction  & V819~Tau             \\
41&SCrA~A+B            &Corona Australis &150.3   &K6  &$0.24$ & 0.5 &2.58 &0.61&Full&10 &8, 30&1.41$\pm$2.42 &$-1.00$ &no correction & 2MASS~J15584772-1757595      \\
%\hline
% \multicolumn{14}{c}{TWA (8-9 Myr)}\\
%\hline
42&TW~Hya   &TWA&59.8 &M0.5 &$-0.63$ &0.00  &1.11&0.61&TD$^{a}$ &7 &1, 11 &13.55$\pm$0.34 &$-1.24$ &TWA~13  &  V819~Tau    \\
43&TWA~3A  &TWA&36.4 &M4.1     &$-1.19$ &0.05  &0.85&0.16&TD&&1  &11.84$\pm$2.52 &$-0.83$ &TWA~7   &TWA~7        \\
%\hline
% \multicolumn{14}{c}{Other regions}\\
%\hline
%43&UScoCTIO~33    &147.5  &M4.5  &$-1.58$ &0.40 &0.59 & &  &  &1   &  &  &     \\
%46&LkH$\alpha$~348&1333.3 &  &  &   & & &  &  &   &  &  &     \\
%44&V1057~Cyg$^{c}$ &Cygnus     &897.7 &$\sim$G0 &2.47 &4.63   &15.62      &5.80 &Full     &   &23, 17 &$-14.3\pm1$  &KW27$^{b}$  &	EPIC 203476597     \\
%45&V1515~Cyg$^{c}$ &Cygnus  &980.8 &$\sim$G3 &2.01 &2.93  &10.23 &4.55 &Full  &  &24, 25, 17   &$-11.13\pm0.53$  &EPIC~212021375$^{b}$  &2MASS~J15584772-1757595     \\
44&V1057~Cyg$^{c}$ &Cygnus     &897.7 &$\sim$G0$?$ &2.47 &4.63   &      & &Full     &   &23, 17 &$-14.3\pm1$ &$-1.17$ &no correction  &	EPIC 203476597     \\
45&V1515~Cyg$^{c}$ &Cygnus  &980.8 &$\sim$G3$?$ &2.01 &2.93  & & &Full  &  &24, 25, 17   &$-11.13\pm0.53$ &$-1.40$ &no correction &2MASS~J15584772-1757595     \\
46&HD~143006$^{a}$ &Upper~Sco    &165.5 &G3  &0.50 &0.45  &1.81   &1.52 &TD$^{a}$ &$\sim$28  &1, 17, 29  &$-0.19\pm0.21$ &$-1.30$  &KW27$^{b}$   &KW541$^{b}$     \\
%46&EX~Lup         &157.0 &M0  &$-0.12$ &1.1  &2.04  & &Full &  &3  &2$\pm$1   &  &  &     \\
47&DI~Cep     &Cepheus   &430.0 &G8&0.85  &0.25  &3.27 &2.28 &Full  &  &26, 17 &$-7.00\pm0.56$ &$-2.36$&KW27$^{b}$  &EPIC~203476597     \\
48&As~353A    &LDN 673   &404.0 &K5&0.51  &0.00  &3.50 &0.60 &Full &   &27, 17 &$-7.87\pm0.60$ &$-1.10$&no correction &HBC~427  \\
\hline

\end{tabular}
\end{center}
\scriptsize  Note: $^{a}$ listed as the TDs in \cite{2016A&A...592A.126V}; $^{b}$ main sequence templates; $^{c}$ The stellar masses and radii for the two FU~Ori are unknown since their spectral types are very uncertain. References: {1. \citet{2014ApJ...786...97H}, 2. \citet{2014A&A...561A...2A}, 3. \citet{2017A&A...600A..20A}, 4. \citet{2012ApJ...758...31L}, 5. \citet{2003A&A...404..913S}, 6. \citet{2014A&A...568A..18M}, 7. \cite{2010ApJ...724..835W}, 8. \cite{2000ApJ...539..815J}, 9 \cite{2017ApJ...844..158S}, 10. \cite{2011A&A...529A.105G}, 11. \cite{2017ApJ...845...44T}, 12. \cite{2016ApJ...828...46A}, 13. \cite{2017ApJ...835...17M}, 14. \cite{2010ApJ...723.1241A},   15. \cite{2015MNRAS.450.3559N}, 16. \cite{2016A&A...592A.126V}, 17. This work, 18. \cite{2010AJ....140..887H}, 19. \cite{2017A&A...606A..88T}, 20. \cite{2016A&A...596A..88L}, 21. \cite{2016MNRAS.458.2476S}, 22. \cite{2018arXiv180106154V}, 23. \cite{2003ApJ...595..384H}, 24. \cite{1977ApJ...217..693H}, 25. \cite{1988cels.book.....H}, 26. \cite{1979ApJS...41..743C}, 27. \cite{2006ApJ...647..432R}, 28. \cite{2018arXiv180410340H}, 29. \cite{2017A&A...599A..85L}, 30. \cite{2011ApJ...733...84P}} %{\scriptsize \tablerefs{1. \citet{2014ApJ...786...97H}, 2. \citet{2014A&A...561A...2A}, 3. \citet{2017A&A...600A..20A}, 4. \citet{2012ApJ...758...31L}, 5. \citet{2003A&A...404..913S}, 6. \citet{2014A&A...568A..18M}, 7. \cite{2010ApJ...724..835W}, 8. \cite{2000ApJ...539..815J}, 9 \cite{2017ApJ...844..158S}, 10. \cite{2011A&A...529A.105G}, 11. \cite{2017ApJ...845...44T}, 12. \cite{2016ApJ...828...46A}, 13. \cite{2017ApJ...835...17M}, 14. \cite{2010ApJ...723.1241A},   15. \cite{2015MNRAS.450.3559N}, 16. \cite{2016A&A...592A.126V}, 17. This work, 18. \cite{2010AJ....140..887H}, 19. \cite{2017A&A...606A..88T}, 20. \cite{2016A&A...596A..88L}, 21. \cite{2016MNRAS.458.2476S}, 22. \cite{2018arXiv180106154V}, 23. \cite{2003ApJ...595..384H}, 24. \cite{1977ApJ...217..693H}, 25. \cite{1988cels.book.....H}, 26. \cite{1979ApJS...41..743C}, 27. \cite{2006ApJ...647..432R}, 28. \cite{2018arXiv180410340H}, 29. \cite{2017A&A...599A..85L}, 30. \cite{2011ApJ...733...84P}}}
\end{table*}
\normalsize
%15. \cite{2007MNRAS.378..369N}, 16. \cite{2015ApJ...801...31R},  17. \cite{2015AJ....150...32R}, 18. \cite{2015ApJ...814...14P}, 19. \cite{2017A&A...600A..20A}, 20. \cite{2015MNRAS.450.3559N}, 21. This work, 22. \cite{2010AJ....140..887H}, 23. \cite{2013A&A...551A..34S}, 24. \cite{2013A&A...555A..67R}, 25. \cite{2016A&A...592A.126V}
%\let\thefootnote\relax{$^{a}$ The reference for the available spectral types of individual sources 1
%$^{a}$:  LS:  low-mass disks with small holes, ML: massive disks with large holes, MS: massive disks with small holes,  LL:  low-mass disks with large holes, defined in \cite{2016A&A...592A.126V}. 
%\clearpage

\setcounter{table}{2}

\subsection{Sample}

Our sample comprises 48 young stars with disks, see Table~\ref{Table:sample}. Most of them belong to the following five star-forming regions and associations: Taurus, Lupus~I, Lupus~III, $\rho$~Oph, and Corona Australis. 
The TW~Hya and Upper~Sco associations have an average age of 5--10\,Myr \citep{2013A&A...549A..15F,2016ApJ...833...95D,2017ApJ...842..123F} while all other regions are younger, with ages 1--3\,Myr \citep{2009ApJ...703..399L,2017arXiv170301251F,1999ApJ...525..440L,2011ApJ...736..137S}. 
We retrieve {\it Gaia} Data Release 2 parallactic distances for 47 of them from the  geometric-distance table provided by \citet{2018arXiv180410121B}. No parallax is reported for DP~Tau, hence  we adopt the median distance of nearby Taurus members which is $\sim$130~pc. With these new distances, Sz~102 is two times farther away than other Lupus members while V853~Oph is $\sim$50~pc closer than the $\rho$~Oph star-forming region. However, we note that both sources have  high Astrometric Excess Noise in Gaia~DR2 (2.881 and 5.944 for Sz~102 and V853~Oph, respectively), indicating that their astrometric solution is unreliable \citep{2018arXiv180409366L}. Therefore, for these two sources we take the mean distance to Lupus~III and $\rho$~Oph.  Using stellar members collected by \cite{2017A&A...600A..20A} for Lupus and \cite{2015A&A...579A..66M} for $\rho$~Oph, we find 160~pc for Sz~102 and 138~pc for V853~Oph.

In addition to source distance and association, 
Table~\ref{Table:sample} also lists stellar spectral type, visual extinction ($A_{\rm V}$), luminosity, radius and mass  ($L_{\star}$, $R_{\star}$, $M_{\star}$), disk type and inclination ($i$), and corresponding references. Note that stellar luminosities from the literature have been scaled to the new distances in our table. Stellar radii are calculated from the Stefan-Boltzman equation where the effective temperature is derived from the source spectral type according to the spectral type-effective temperature relation in \cite{2014ApJ...786...97H}. Stellar masses are calculated from the luminosity and effective temperature using the non-magnetic pre-main sequence evolutionary tracks of \citet{2016A&A...593A..99F}. Our sample covers a large range in spectral type (F7 to M4), hence in stellar mass, from $\sim$2.8 to $\sim$0.2\,\Msun . Note that the two sources in Cygnus (V1057~Cyg and V1515~Cyg) are well known FU~Ori objects, young stars showing strong episodic accretion bursts (e.g., \citealt{2014prpl.conf..387A}).

On the disk type, we distinguish "Full disks" from "Transitional disks" (TDs). TDs are known to have reduced near- and mid-infrared excess emission with respect to the median of T Tauri stars pointing to a dust depleted inner region \cite[e.g.,][]{2014prpl.conf..497E}. Hence, we classify our sources by comparing the source spectral energy distribution (SED) to the  median SEDs of CTTs of similar spectral type, see Appendix~\ref{Appen:SED} for details. With our approach we identify 28 full disks and 15 TDs. 
The 5 disks we could not classify are around G and F stars. As these stars are rare and their disks disperse faster \cite[e.g.,][]{2009ApJ...695.1210K}, we simply lack a median SED for this spectral type range. For SR~21A and HD~143006  we adopt the literature classification of TDs \citep{2015MNRAS.450.3559N,2016A&A...592A.126V}. As there is no hint of reduced infrared emission in the SED of the other three G-type stars (V1057~Cyg, V1515~Cyg, and DI~Cep) we classify their disks as full.
Recently, \cite{2016A&A...592A.126V} used color criteria and SED modeling to identify a large sample of TD candidates in nearby star-forming regions. Of the 14 sources in common 13 are classified in the same way. Only FN~Tau is classified as full disk here but listed as  TD in \cite{2016A&A...592A.126V}. Note, however, that \cite{2016A&A...592A.126V} estimate an uncertainty on the cavity size that is as large as the cavity itself, hence FN~Tau might actually have a full disk as we adopt here.

\subsection{Observations and data reduction}
Our targets were observed during two nights, January~23 and May~23 2008, through the program C199Hb (PI: Greg~J.~Herczeg). The Keck/HIRES spectrograph \citep{1994SPIE.2198..362V} was used with the C1 decker  and a 0\farcs861$\times$7$''$ slit resulting in a nominal resolution of 48,000. The sources were observed with the kv380 filter, wavelength coverage 3900--8500\,\AA, during the night of January~23 and  with the wg335 filter, wavelength coverage 3480--6310\,\AA,  during the night of May~23. In addition to the science targets, three telluric standards at least were observed each night. These standards are used to correct for telluric features (\sect~\ref{Sect:corrected_profiles}) near the \OIa\, line as well as for flux calibration (Appendix~\ref{Sect:flux_calibration}).

The raw data were reduced using the Mauna Kea Echelle Extraction (MAKEE) pipeline written by Tom~Barlow\footnote{http://www2.keck.hawaii.edu/inst/common/makeewww/index.html}. MAKEE is designed to run non-interactively using a set of default parameters, and carry out bias-subtraction, flat-fielding, and spectral extraction,  including sky subtraction, and wavelength calibration with ThAr calibration lamps. The wavelength calibration is performed in air by setting the keyword "novac" in the pipeline inputs. Heliocentric correction is also applied to the extracted spectra.

\subsection{Stellar Radial velocity}\label{Sect:RV}
The stellar radial velocity (RV) is derived by cross-correlating each star optical spectrum with the synthetic spectrum of a star that has the same effective temperature.
The grid of model spectra is from  \cite{2013A&A...553A...6H}  for a solar abundance and surface gravity log\,$g$=4.0. The model spectrum is first degraded to match the spectral resolution of HIRES, rotationally broadened to match the source photospheric features, and 'veiled' by adding a constant flux value to account for the fact that photospheric lines from T Tauri stars are less deep than those of main-sequence stars of the same spectral type.
The cross-correlation is carried out separately for each echelle order that does not have strong emission lines or is highly veiled. Hence, the total number of available orders varies from source to source but it is at least 5,  except for RU~Lup, EX~Lup, and V1057~Cyg. 
For RU~Lup, we only detect photospheric features near the \OIa\ line and use them to estimate the stellar radial velocity. For EX~Lup, we detect photospheric features within 5733--5745\,\AA\ and 5936--5948\,\AA\ and the cross-correlation is performed within these wavelength ranges. The stellar radial velocity of V1057~Cyg is obtained by cross-correlating the observed spectrum with the model spectrum within 5556--5580\,\AA\ and 5645--5673\,\AA.
Table~\ref{Table:sample} provides the stellar RV and associated uncertainty computed as the mean and standard deviation of the RVs obtained from individual orders. 

\begin{figure*}
\begin{center}
  \includegraphics[width=1.8\columnwidth]{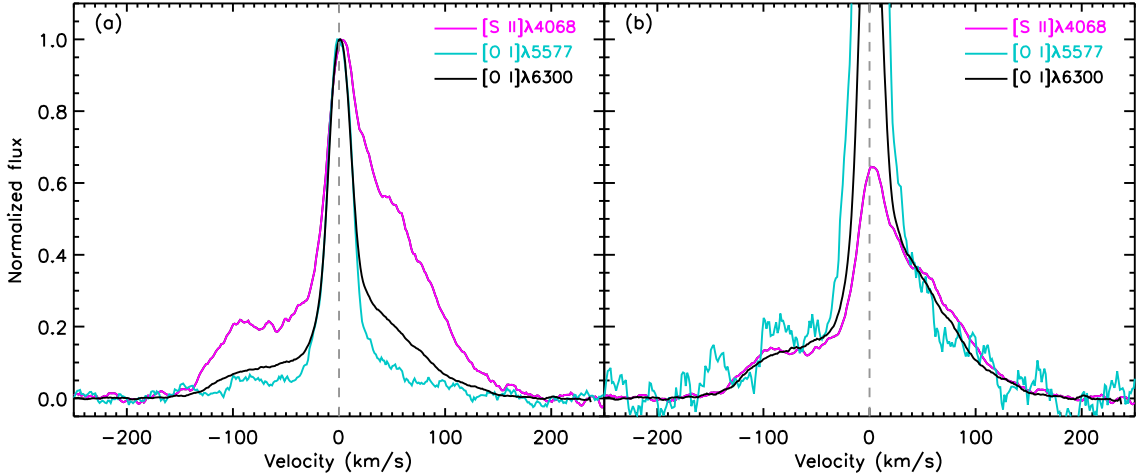}
  \caption{Comparison of  \SII\ (magenta), \OIb\ (cyan), and \OIa\ (black) line profiles for DP~Tau. The line profiles are scaled to show the similar velocity structure in the narrow core and broad wings among all three lines. \label{Fig:line_profile_example}}
\end{center}
\end{figure*}

Sz~102 and VV~CrA have highly veiled spectra with strong emission lines, hence no HIRES order can be used to derive their RVs. For these two sources, we download archival ESO Phase 3 X-Shooter spectra, which are wavelength and flux calibrated and have a resolution of $\sim$17,400 between 5,500--10,000\AA. For VV~CrA, we derive its RV by fitting its Li~I~$\lambda$6708 absorption line with a Gaussian function as in \cite{2015ApJ...814...14P}.  For Sz~102, we cannot  clearly identify the Li~I~$\lambda$6708 absorption line, hence we derive its RV by cross-correlating the observed spectrum with the model spectrum within 7485--7530\AA, 8020--8112\AA, and 8762--8845\AA\ where we can identify  photospheric absorption lines. The RVs of these two sources are also listed in Table~\ref{Table:sample}.  

Note that synthetic spectra are only used for the computation of stellar RVs. As discussed in the next section, photospheric subtraction is carried out mostly with the observed spectra of weak-line T Tauri  stars (WTTS). We do not have 
WTTS templates for 18 sources in the \SII\ setting and for one source, SR~21A (F7), around the \OIa\ and the \OIb\ settings, hence we used the main sequence (MS) templates in these instances, see Table~\ref{Table:sample}.  

Arc calibration frames for the absolute wavelength solution were taken only at the beginning of each night. Therefore, systematic shifts on the wavelength solution are possible when moving the telescope. We assess these possible shifts by cross-correlating the telluric lines with the model atmospheric transmission curve for the Keck Observatory which is calculated with TAPAS (a web-based service of atmospheric transmission
computation for astronomy, see \citealt{2014A&A...564A..46B}). We find that the wavelength calibration tends to be redshift by $\sim$0.5--2\,\kms. In Table~\ref{Table:sample}, we list the RV correction for each source. Except for Sz~102 and VV~CrA, whose RVs are estimated from X-Shooter spectra, the RVs of all other sources need to be corrected by adding the values in the "Correction" column to their RVs. However, the corrections only affect the heliocentric radial velocity derived from the HIRES spectra. They do not change the relative shifts between the stellar photospheric absorption features and the forbidden lines, which are the focus of this paper. We list these corrections in case one wants to compare the RVs calculated here with those obtained via other methods. 

\section{Line Profiles and Classification}\label{sect3}

\subsection{Corrected line profiles and detection rates}\label{Sect:corrected_profiles}
To produce corrected line profiles, we first remove any telluric absorption and then subtract the  photospheric features near the lines. Telluric removal is only necessary for the \OIa\ line and is achieved with the procedure summarized in the HIRES Data Reduction manual\footnote{see the description in https: \\
//www2.keck.hawaii.edu/inst/common/makeewww/Atmosphere/index.html} and using the telluric standards acquired in the two nights.

After telluric removal, we subtract the stellar photosphere following \citet{1989ApJS...70..899H}. In short, we select a photospheric standard with a spectral type similar to the target. We then broaden (through the rotation velocity $Vsini$), veil (parameter $r_{\lambda}$), and shift in velocity its spectrum to best match the target's photospheric lines. Veiling  is defined as the ratio of the excess ($F_{\rm excess}$) to the photospheric flux ($F_{\rm phot}$): $r_{\lambda}=F_{\rm excess}/F_{\rm phot}$ and it is used to mimic the filling in of the photospheric lines due to the excess emission from the accretion shocks.  The best-fit photospheric spectrum is the one that minimizes the $\chi^{2}$, defined as  $\chi^{2}=\sum (F_{\rm target}-F_{\rm Standard})^{2}$. Corrected line profiles are produced by subtracting the best-fit photospheric spectra from the target spectra.
We do not apply any photospheric subtraction on extremely veiled spectra that have no photosheric features, e.g. several of the spectra in the \SII\ setting.
Figure~\ref{Fig:sky_phot_example} shows an example of the technique described here, including telluric removal near the \OIa\,\AA{} line (rightmost panel). Table~\ref{Table:sample} lists the photospheric standards used for individual lines. 

\begin{figure*}
\begin{center}
\includegraphics[width=2\columnwidth]{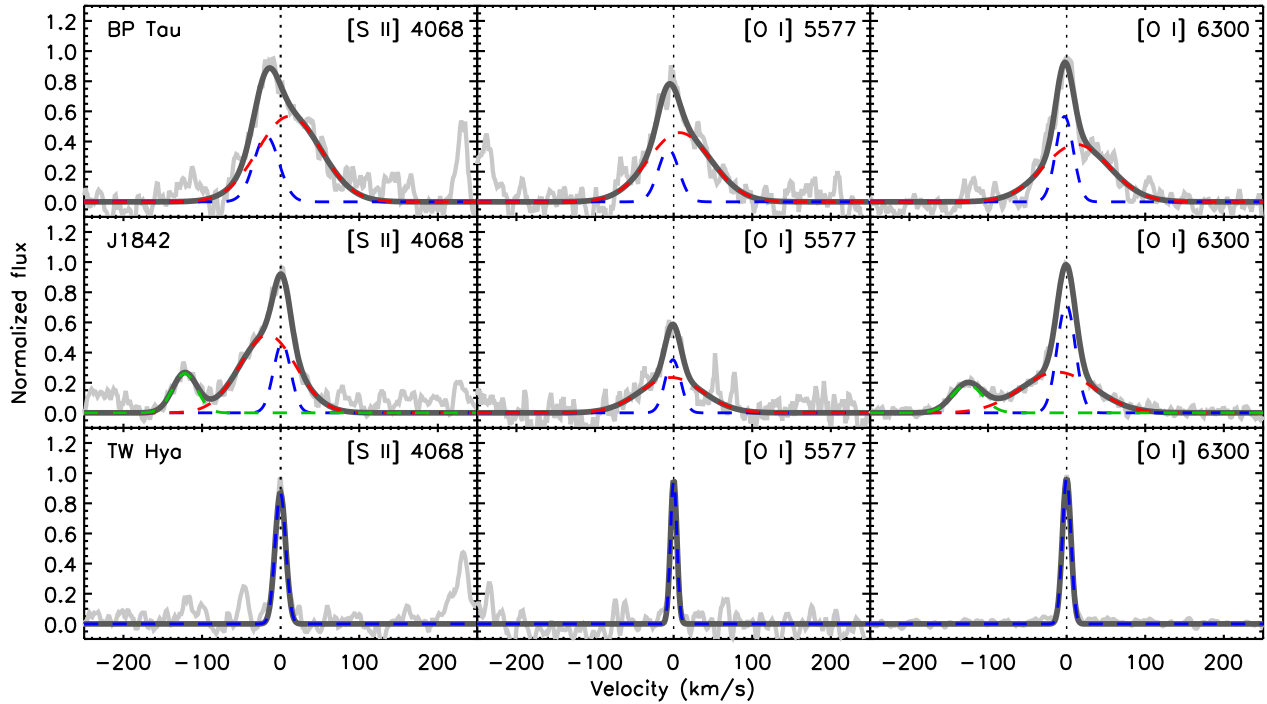}
\caption{Examples of \SII\ ,  \OIa\ , and \OIb\ line profile fits. A green dashed line is used for the HVC, a red dashed line for the LVC-BC, while  a blue dashed line for the LVC-NC. The dark solid line is the sum of the  individual components. Line profile fits for all sources are shown in Fig.~\ref{Fig:line2}.}\label{Fig:example_line2}
\end{center}
\end{figure*}

Overall, we detect the \OIa, the \OIb, and the \SII\  lines  from 45, 26, and 22 sources, respectively\footnote{CX~Tau, DoAr~21, and DoAr~24E are so faint at short wavelengths that their \SII\  spectra cannot be extracted with the MAKEE pipeline. SR~21A shows only marginal detections in the \OIa\ line, See Fig.~\ref{Fig:line5}. Of the three sources (Sz~68A, DoAr~21, and DoAr~24E) without \OIa\ detection, one is a TD and two are full disks; all have spectral types early-K to G. A detail discussion of these sources is presented in Appendix~\ref{Appen:detail}.}. Hence, the detection rate is 92\%, 54\%, and 49\% in these lines\footnote{For the \SII\ line, the three sources without extracted spectra have been excluded when calculating the detection rates.}. 
A total of 18 sources have detections in all three lines.

\subsection{Line decomposition}\label{Sect:line_decomposition}
As recently shown in \cite{2016ApJ...831..169S}, oxygen forbidden lines can be well reproduced by the superimposition of a few Gaussian profiles, presumably tracing different kinematic components. Figure~\ref{Fig:line_profile_example} compares the \SII, \OIb, and \OIa\ emission lines from DP~Tau, one of the sources with the highest S/N spectra. Note how a narrow component centered at the stellar velocity is present in all lines (panel a). Broader higher velocity emission is also apparent in the three tracers (Panel b).

Motivated by the similarity of the profiles in the high S/N spectrum of DP~Tau and following \cite{2016ApJ...831..169S}, we use a combination of Gaussians  to decompose  the \SII, \OIb, and \OIa\, lines. 
 To find the minimum number of Gaussians that describes the observed profile, we use an iterative approach based on the IDL procedure {\it mpfitfun}. 
We start by fitting each profile with one Gaussian and compute the residual spectrum by subtracting the best fit from the original profile.  If the root mean square (rms) of the residual is higher than that of the original spectrum next to the line of interest (by more than $\sim$2$\sigma$), we add another Gaussian and re-fit the original profile simultaneously with two Gaussians. We re-compute the residual spectrum and add an extra Gaussian until the rms of the residual is within 2$\sigma$ of the rms of the original spectrum. 

\begin{figure*}
\begin{center}
\includegraphics[width=1\columnwidth]{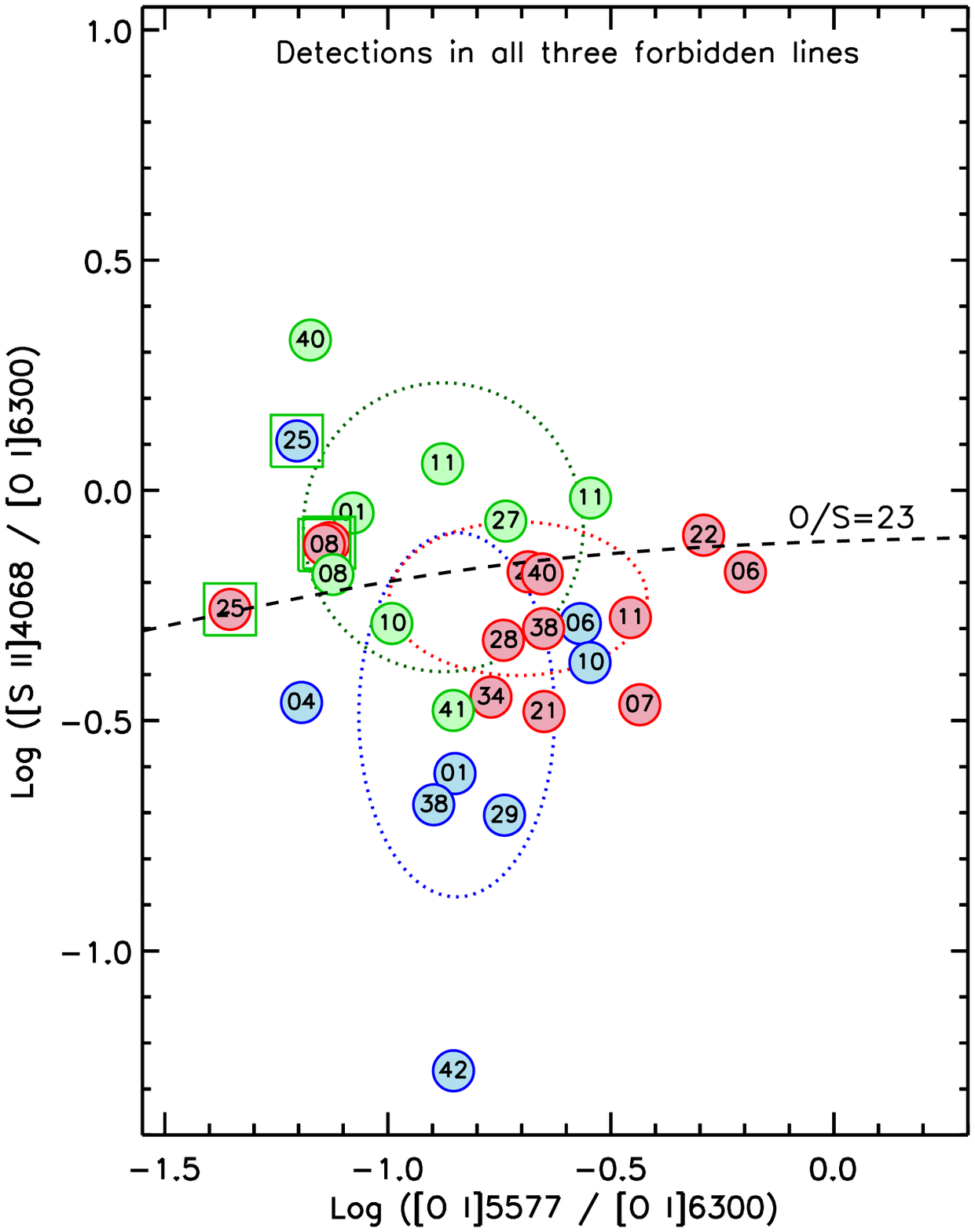}
\includegraphics[width=1\columnwidth]{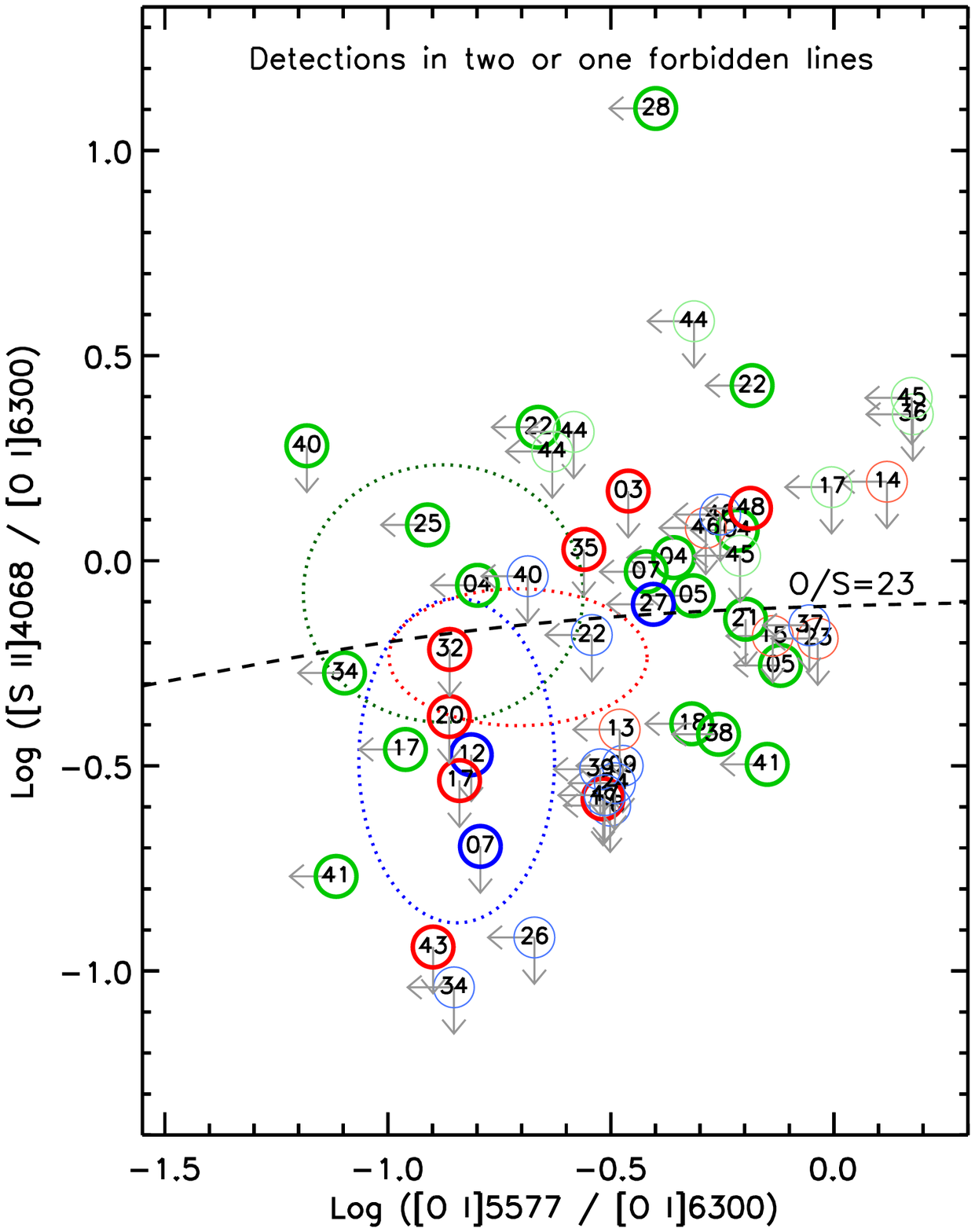}
\caption{Line ratios with colors following the preliminary classification: green for HVC, red for LVC-BC, and blue for LVC-NC. The number inside the circle gives the source ID as listed in Table~\ref{Table:sample}. Left panel: only sources that are detected in all three forbidden lines. Dotted-line ellipses encircle the mean $\pm$ standard deviation of the HVCs (green), LVC-BCs (red), and LVC-NCs (blue) from the preliminary classification.  Green squares mark those LVC-BCs that are re-classified as HVCs. Right panel: like the left panel but for sources that are detected in two forbidden lines (thick-line circles) and only in the \OIa\ line (thin-line circles). In both panels the black dashed line shows predicted ratios for 10$^{4}$~K gas, assuming an Oxygen to Sulfur abundance ratio of 23, where collisions with electrons (densities $10^{5.8-8.5}$\,cm$^{-3}$) excite the upper levels of the transitions, see  \sect~\ref{Sect:LVC_lineratio} for more details. 
} \label{Fig:twolineratio_com}
\end{center}
\end{figure*}

Table~\ref{Tab:para_LVC1} lists  the parameters resulting from the decomposition of each individual line while Figure~\ref{Fig:example_line2} shows examples of the best fit Gaussian profiles for three sources.
All other fits are shown in Appendix~\ref{Appendix:lineprofile}, Figure~\ref{Fig:line2}. Seven sources (HN~TauA, RU~Lup, EX~Lup, As~205A, VV~CrA, SCrA~A+B, and As~353A) have \SII\ lines highly contaminated by Fe~{\scriptsize I} emission. In addition, the \OIb\ line from RU~Lup is also affected by Fe~{\scriptsize I} emission. We decontaminate these line profiles before line decomposition as discussed in detail in Appendix~\ref{Appen:line_decontamination}. Note that the 
\SII\ profiles from EX~Lup, As~205A, VV~CrA, and As~353A are not well recovered  at  velocities more blueshifited than $\sim$200\,\kms. Hence, the most blueshifted HVC \SII\ component for each of these sources is marked as  unreliable in Table~\ref{Tab:para_LVC1} and will not be used in the following analysis. 

Table~\ref{Tab:para_LVC1} also lists upper limits for undetected lines.
For sources with detections in the \OIa\ line, the upper limits on the other lines are computed assuming that they share the \OIa\ profile and have peaks of 3$\times rms$, where the $rms$ is calculated from the standard deviation of the continuum near the line. For sources with non-detections in all the lines, the upper limits are calculated assuming a Gaussian profile with the FWHM of an unresolved line ($\sim$6.3\,\kms), and a peak of 3$\times$$rms$.

\subsection{Preliminary classification}\label{Sect:Pre_class}
Different approaches have been used in the literature to separate the LVC from the HVC (e.g., \citealt{1995ApJ...452..736H,2014A&A...569A...5N}). As our spectra have a similar resolution to those in \cite{2016ApJ...831..169S},  we use their criteria for a preliminary classification, see last but one column of Table~\ref{Tab:para_LVC1}. In short, a component is called LVC (HVC) if the absolute value of the Gaussian velocity centroid is smaller (larger) than 30\,\kms. Within the LVC, an LVC-BC has a $FWHM>$40\kms\ while an LVC-NC is narrower. As a consequence even lines that can be fitted with one Gaussian fall in one of these categories (see also \citealt{2016ApJ...831..169S} and \citealt{2018arXiv180310287M}, but Banzatti~et~al. 2018 for an independent treatment of single components).

Based on this preliminary classification, we find that HVCs and LVCs are  often present in the three line profiles but with different contributions: in the \SII\ line the HVC tends to be more prominent relative to the LVC than in the \OIa\ and \OIb\ lines, the LVC dominates the \OIb\, profile, while both an HVC and an LVC are seen in the \OIa\, profile. Two extreme cases are worth discussing. The \OIa\ and \SII\ profiles of FN~Tau are the most complex in our sample with one LVC and three HVCs, while its  \OIb\, line only presents one LVC. For Sz~73, the \SII\ and \OIb\ lines only have one component, classified as HVC and LVC respectively, while its \OIa\ shows both one LVC and one HVC.

 The above classification cannot be taken too rigidly, as there is some overlap in line widths between BC and NC (Banzatti~et~al. 2018) and  HVCs could show small blueshifts in highly inclined systems and be wrongly classified as LVCs when adopting the sharp velocity boundary of 30\,\kms\ utilized here. In the next Section we will use line ratios to test and further refine our classification.

\subsection{Refined classification}\label{sect:refiments}
Ratios of lines that trace the same kinematic component can be used to constrain the properties of the emitting gas.
\cite{2016ApJ...831..169S} have already shown that the \OIa{} and \OIb{} lines have very similar LVC profiles (e.g., their Figure~15 and best fit parameters in Table~5). Section~\ref{com_SII_OI6300} will illustrate that, within a specific kinematic component, the similarity extends to the  \SII{} line.  Hence, here we use the ratios of these three forbidden lines to test if different kinematic components separate out and, if so, further refine our preliminary classification.

\begin{figure*}
\begin{center}
  \includegraphics[width=2\columnwidth]{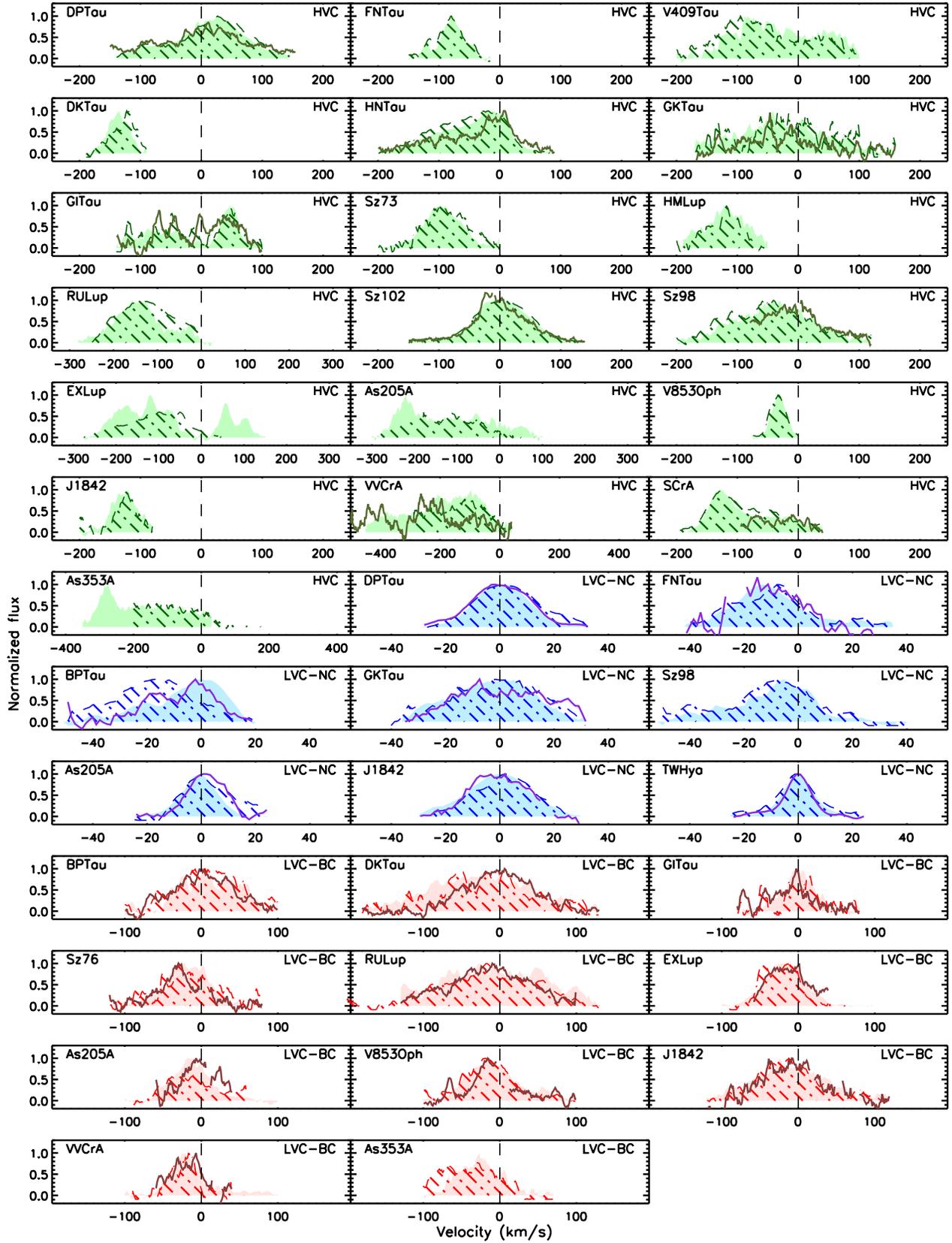}
  \caption{Comparison of the HVC (green), LVC-NC (blue), and LVC-BC (red) profiles for the \SII\ (dash-dotted line shaded area), \OIb\ (solid line), and \OIa\ (color filled area). Most individual kinematic components have similar profiles in the three different forbidden lines investigated here.}\label{Fig:threelines_HVC}
\end{center}
\end{figure*}

Figure~\ref{Fig:twolineratio_com} shows line ratios (left panel) or upper limits (right panel) for all 45 sources that are detected in the \OIa , the brightest of our forbidden lines. HVCs (green circles) have a combination of higher \SII/\OIa\ (hereafter SII40/OI63) and lower \OIb/\OIa\  (hereafter OI55/63) ratios than LVCs. This is best illustrated by the mean and standard deviation of different component line ratios which are shown as dotted-line ellipses in both panels. Note that these values are computed from our preliminary classification and only from sources with at least two detected forbidden lines which are used to calculate line ratios.

The left panel of Fig.~\ref{Fig:twolineratio_com} reveals 4 possible LVCs with line ratios more compatible with HVCs (circles surrounded by green squares:  one LVC-BC from DP~Tau (ID~1) and HN~TauA (ID~8), and one LVC-NC and one LVC-BC from Sz~102 (ID~25). Interestingly,  HN~TauA and Sz~102 have close to edge-on disks ($i > 70^{o} $). Thus, it is very likely that, due to projection effects, their HVCs have centroids $<$30\,\kms\ and have been erroneously categorized as LVCs by our preliminary classification scheme. Although the disk inclination of DP~Tau is not known, the star is about an order of magnitude underluminous compared to other Taurus members of the same spectral type \citep{2014ApJ...786...97H}, pointing to  obscuration from a highly inclined disk (see discussions on such underluminous young stars in \citealt{2009A&A...504..461F,2013A&A...549A..15F}). In addition, the \OIa{} and \SII\  LVC-BC centroids are close to our 30\,\kms\ HVC/LVC boundary.   
Thus, we re-classify all these four LVCs as HVCs, see last column of  Table~\ref{Tab:para_LVC1}.

Fig.~\ref{Fig:twolineratio_com} also shows the expected SII40/OI63 and OI55/63 ratios for gas
at 10$^{4}$~K where collisions with electrons (densities $10^{5.8-8.5}$\,cm$^{-3}$) excite the upper level of the transitions (e.g., \citealt{2014A&A...569A...5N} but also \sect~\ref{Sect:LVC_lineratio}). Most HVCs have SII40/OI63 ratios higher than the model predicted ones, while most LVCs have lower ratios. We also note that one LVC-BC from As~353A (ID~48, right panel) shows higher SII40/OI63 ratios than the model predicted ones, similar to HVCs. This, combined with possible Fe~{\scriptsize I} contamination in the \SII\ transition (see Appendix~\ref{Appen:line_decontamination}), leads us to mark the LVC-BC as suspicious, last column of Table~\ref{Tab:para_LVC1}.

Following this refined classification,  we find that 51\% (23/45) of the sources with \OIa\ detection have an HVC while 84\% (38/45) show an LVC.  Furthermore, 31\% (14/45) of the sample presents an LVC-BC while 33\% (15/45) has an LVC-NC. Finally, only 20\% (9/45) of our sources have both LVC-BC and LVC-NC components.

 \begin{table}
%\scriptsize
\renewcommand{\tabcolsep}{0.2cm}
\caption{Mean line ratios and standard deviations after HVC reclassification.}\label{Table:meanlineratio}
\begin{center}
\begin{tabular}{cccc}
\hline
 Line ratios &LVC-NC  &LVC-BC &HVC \\
\hline
Log~OI55/63 &$-$0.81$\pm$0.19 &$-$0.62$\pm$0.20 &$-$0.97$\pm$0.31  \\ 
Log~SII40/OI63&$-$0.56$\pm$0.35 &$-$0.25$\pm$0.18 &$-$0.08$\pm$0.36 \\  
 \hline
\end{tabular}
\end{center}
\end{table}

\begin{table}
\scriptsize
\renewcommand{\tabcolsep}{0.2cm}
\caption{Results from the Gehan's generalized Wilcoxon test}\label{Table:problineratio}
\begin{center}
\begin{tabular}{cc|cc}
\hline
% & LVC-NC     &LVC-BC &HVC \\
%\hline
  \multicolumn{2}{c|}{Log~OI55/63} & \multicolumn{2}{c}{Log~SII40/OI63}\\
\hline
Pairs &Probability &Paris &Probability \\ 
\hline
 LVC-NC/LVC-BC &p=$6\times10^{-3}$            &LVC-NC/LVC-BC&p=$1\times10^{-2}$ \\  
 LVC-NC/HVC    &p=$1\times10^{-1}$ &LVC-NC/HVC   &p<$5\times10^{-5}$    \\  
 LVC-BC/HVC    &p$<5\times10^{-5}$&LVC-BC/HVC   &p$=4\times10^{-4}$       \\  
 \hline
\end{tabular}
\end{center}
\end{table}

\section{Thermal [O~I] emission and associated gas properties}\label{sect4} 

\subsection{The similarity of [S~II] and [O~I] profiles of individual kinematic components}\label{com_SII_OI6300}
 Figure~\ref{Fig:threelines_HVC} shows a comparison of HVC, LVC-NC and LVC-BC profiles for the sources that are detected both in  \SII\ and \OIa. Whenever detected, \OIb\ lines are also superimposed. To test how similar are the profiles of individual kinematic components, we remove from the observed profile the other components using our best fit parameters reported in Table~\ref{Tab:para_LVC1}.

 Regardless of kinematic component (HVC, LVC-BC or NC), most profiles are similar in the [O~{\scriptsize I}] and [S~{\scriptsize II}] lines. Only BP~Tau has an LVC-NC that is clearly more blueshifted in the \SII\ than in the [O~{\scriptsize I}] lines, although the line widths are actually similar. This similarity is strengthened when comparing the centroids and FWHMs of individual components (see Fig.~\ref{Fig:com_SII_OI6300} and \ref{Fig:com_OI2}  in Appendix~\ref{Appen:com_SII_OI6300}).
 Thus, we conclude that, within a kinematic component, the three lines trace a similar region.  This means that their line ratios can be used to constrain the properties of the emitting gas, in particular temperature and density.

\begin{figure*}
\begin{center}
\includegraphics[width=1.\columnwidth]{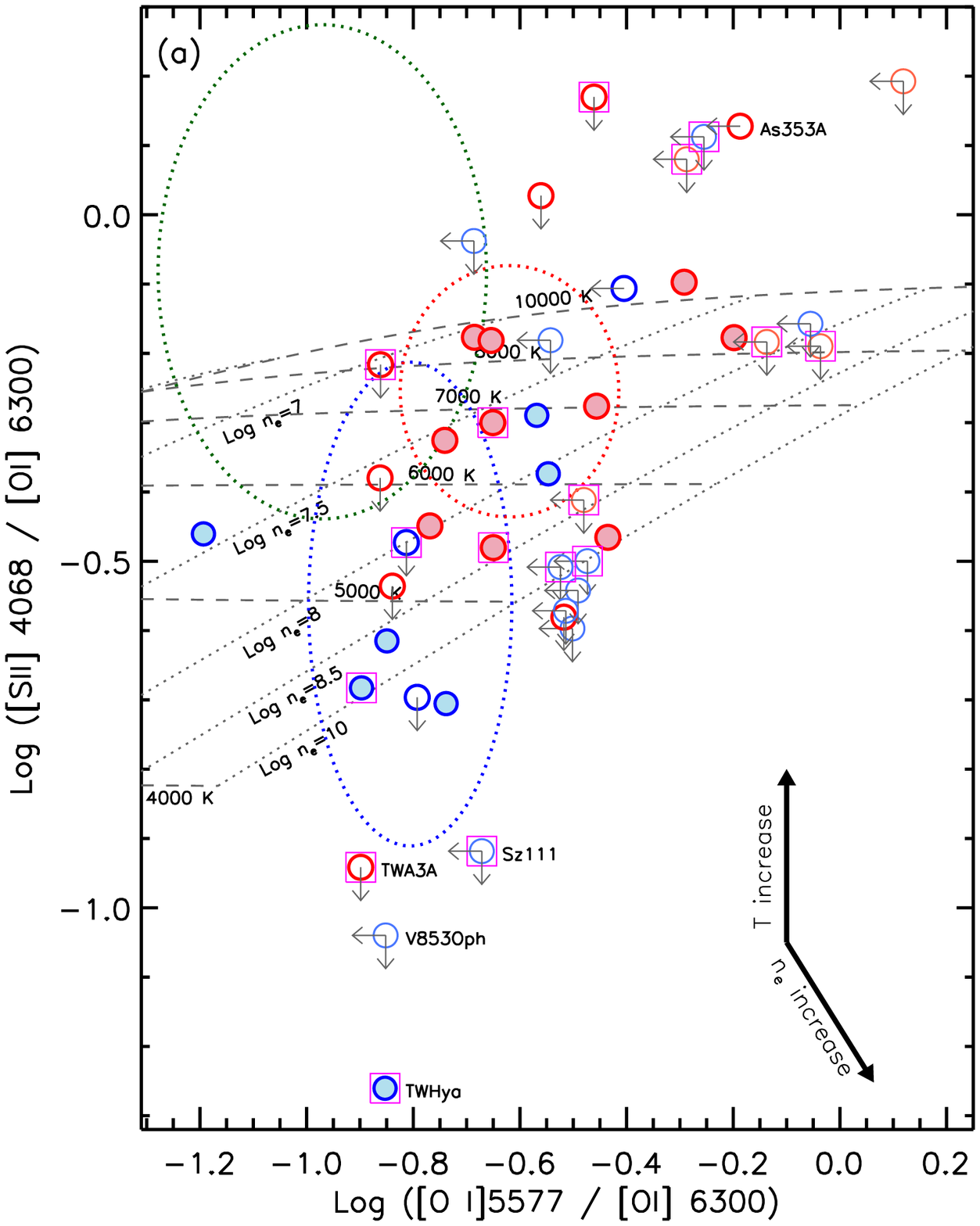}
\includegraphics[width=1.\columnwidth]{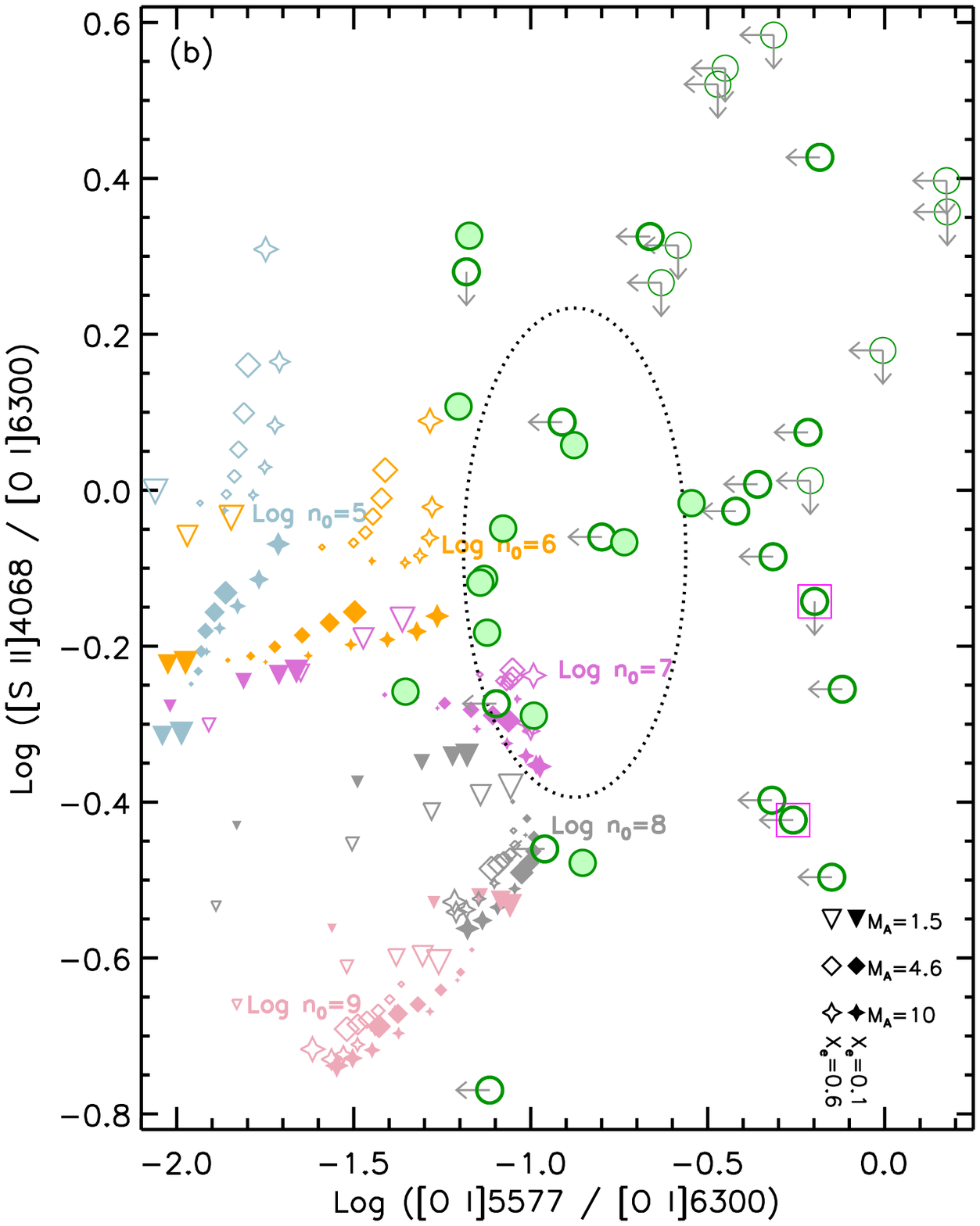}
\caption{SII40/OI63 vs OI55/63 line ratios for LVC-NC (blue) and LVC-BC (red) in Panel~(a) and HVC (green) in Panel~(b). In each panel, the filled circles mark sources with detections in all three forbidden lines, thick-line open ones for those with detections in two forbidden lines, and thin-line ones for sources that have only an \OIa\ detection. Magenta squares mark TDs. Dotted-line ellipses encircle the mean $\pm$ standard deviation of the HVCs (green), LVC-BCs (red), and LVC-NCs (blue) after the refined classification (see \sect~\ref{sect:refiments}). In Panel~(a), observed line ratios are compared with those predicted by thermally excited gas. Dashed grey lines locate gas at temperatures from 4,000\,K (bottom) to 10,000\,K (top) while  dotted lines at different electron densities from $Log~n_{\rm e}$~(cm$^{-3}$)=6.5 (left) to $Log~n_{\rm e}$~(cm$^{-3}$)=10 (right). In panel~(b) HVC line ratios are compared with those predicted by shock models \citep{2015ApJ...811...12H}. Open symbols are for $X_{e}$=0.6 while filled ones for $X_{e}$=0.1, triangles are for  $M_{\rm A}$=1.5, diamonds are for  $M_{\rm A}$=4.6 while stars are for a higher Mach number of 10. 
Different colors indicate different preshock number densities of nucleons with $Log~n_{0}({\rm cm^{-3}})$ from 5 to 9. Finally, the  size of the symbols scales with the shock velocity ($V_{\rm s}$) which ranges from 30 to 80\,\kms\ in these models. }\label{Fig:twolineratio_com2}
\end{center}
\end{figure*}

\subsection{Gas properties constrained by line ratios}\label{Result:Line_ratios}
 The difference  in the HVC and LVC forbidden line ratios was already noted in \sect~\ref{sect:refiments} and used to improve upon our preliminary kinematic classification. Here, we search for and quantify any difference in line ratios between LVC-NC, LVC-BC, and HVC. Table~\ref{Table:meanlineratio} lists the mean and standard deviation of Log~OI55/63 and Log~SII40/OI63 for three types of components when considering only detections. LVC-BCs have the largest mean Log~OI55/63 ratio while HVCs have the lowest one. HVCs have the highest mean Log~SII40/OI63 ratio while LVC-NCs the lowest one.

 Next, we include upper limits and use 
the Gehan's generalized Wilcoxon test in the ASURV code \citep{1985ApJ...293..192F} to quantify any statistically significant difference between pairs of kinematic components. The test finds a low probability that the HVC Log~OI55/63 and Log~SII40/OI63 are the same as the LVC-BC. HVC and LVC-NC also have statistically different Log~SII40/OI63, but indistinguishable Log~OI55/63. Finally, the LVC-BC and NC differ both in  Log~OI55/63 ratios and Log~SII40/OI63, see Table~\ref{Table:problineratio}. These findings suggest that the combination of these three forbidden lines is sufficient to separate the kinematic components discussed in this paper.

\subsubsection{Thermally excited gas explains most LVC line ratios}\label{Sect:LVC_lineratio}
Figure~\ref{Fig:twolineratio_com2}~(a) shows the relation between the SII40/OI63 and the OI55/63 ratios for the  LVC components.  In the same figure we also plot predicted ratios for homogeneous and isothermal gas that is thermally excited by collisions with electrons (grey lines). When computing these ratios we take the solar sulfur abundance, $\alpha({\rm S})=1.4\times10^{-5}$  \citep{2005ASPC..336...25A}, and a depleted oxygen abundance as in the interstellar medium, $\alpha({\rm O})=3.2\times10^{-4}$ \citep{1996ARA&A..34..279S}. Following \cite{2014A&A...569A...5N}, we also assume that  all oxygen is neutral while all sulfur is singly ionized. Gas at higher temperature and electron density is predicted to have higher SII40/OI63  and OI55/63 line ratios.

Most of the observed LVC ratios correspond to thermally excited gas with temperatures between 5,000 and 10,000 K and electron densities$\sim10^{7}-10^{8}$\,cm$^{-3}$. In As~353A, its LVC-BC have SII40/OI63 ratio higher than expected from this simple model, with a value seen in some of the HVC. As discussed in \sect~\ref{sect:refiments}, the \SII\ line in this very high accretion rate star is blended with Fe~{\scriptsize I} emission and classified as suspicious in Table~\ref{Tab:para_LVC1}, after our attempt at deblending. Turning to the LVC NC, three of the transition disk sources, TW~Hya, TWA~3A, and Sz~111, have much lower SII40/OI63 ratios than those expected for gas at $\sim$5,000\,K.  The Gehan's generalized Wilcoxon test finds a low probability (p= $6\times10^{-2}$) that the SII40/OI63 ratios of LVC-BCs of full disks and TDs are drawn from the same parent population while their OI53/63 ratios are indistinguishable (p=0.3 for both the LVC NC and BC). This can be explained if TDs have a decreased S\,{\scriptsize II} over O\,{\scriptsize I} abundance ratio, as could happen if sulfur is sequestered in large grains in the disk midplane.

\subsubsection{Shock-heated gas explains most HVC line ratios}\label{Sect:shock_HVC}
Jets generate shock waves in the surrounding medium where part of their kinetic energy is converted into thermal motion, thereby producing bright forbidden line emission (e.g., \citealt{2015ApJ...814...52G}). To further investigate the origin of the HVCs detected in our sample we compare their [S\,{\scriptsize II}] and [O\,{\scriptsize I}] line ratios to those expected from the shock models of \citet{2015ApJ...811...12H}, see Fig.~\ref{Fig:twolineratio_com2}~(b). The model line ratios were calculated for a range of input parameters: $Log~n_{0}({\rm cm^{-3}})=2-11$ with an increment of unity in log scale, $X_{e}$=0.1 and 0.6, $V_{\rm s}$=30--80\kms, $M_{\rm A}$=1.5, 4.6 and 10. Here, $n_{0}$ is the preshock number density of nucleons, $X_{e}$ is the preshock ionization fraction,  $V_{\rm S}$ is the shock velocity, and $M_{\rm A}$ is the Alfv\'{e}nic Mach number. 
We have also scaled the ratios to the atomic abundances used in Sect.~4.2.1, $\alpha({\rm O})=3.2\times10^{-4}$ and $\alpha({\rm S})=1.4\times10^{-5}$, from the original solar abundances in \cite{2015ApJ...811...12H}. 

Shock models with a pre-shock number density of $Log~n_{0}({\rm cm^{-3}})\sim6-7$ can reproduce most of the observed SII40/OI63 and OI55/63 HVC line ratios.  As forbidden line emission occurs in the post-shock cooling zone, the specific transition critical density and excitation temperature dictates how close to the front the emitting material is located. For instance the \OIa{} line probes gas slightly closer to the front than the \SIIb (see e.g., Figure~3 in \citealt{2015ApJ...811...12H}), which has a similar excitation temperature but a critical density about hundred times lower. The \SII\ line investigated in this work has similar temperature {\it and} density to the \OIa , hence should also trace denser/hotter gas closer to the front than the \SIIb .

\section{Mass outflow to mass accretion rate}\label{sect5}

\subsection{Mass accretion rates}\label{sect:Macc}
After flux-calibrating our spectra (see Appendix~\ref{Sect:flux_calibration}), we estimate the accretion luminosity using 12 emission lines covered by the HIRES spectra, including 
five Balmer lines (H$\zeta$, H$\delta$, H$\gamma$, H$\beta$, and H$\alpha$), four He~{\scriptsize I} line (4026, 4471, 5876, and 6678\AA), the He~{\scriptsize II} line at 4686\AA, and the Ca~{\scriptsize II} lines at 3934 and 3968\,\AA. These lines are chosen because  their luminosities are known to correlate tightly with accretion luminosities (e.g., \citealt{2008ApJ...681..594H,2009A&A...504..461F,2012A&A...548A..56R,2014A&A...561A...2A,2017A&A...600A..20A}). After computing line luminosities we convert them into accretion luminosities via the empirical relations listed in Appendix~\ref{Appen:line_acc}.
Table~\ref{Table:acc_line} lists the accretion luminosities derived from the transitions discussed above as well as    the 3$\sigma$-clipping  mean of the accretion luminosities for each source. Mean accretion luminosities are converted into mass accretion rates using the following relation:
\begin{equation}
\dot{M}_{\rm acc}=\frac{L_{{\rm acc}}R_{\star}}{{\rm G}M_{\star} (1-\frac{R_{\star}}{R_{\rm in}})},
\end{equation}
\noindent where $R_{\rm in}$ denotes the truncation radius of the disk, which is taken to be 5\,$R_{\star}$ \citep{1998ApJ...492..323G}, G is the gravitational constant, $M_{\star}$ is the stellar mass, and $R_{\star}$ is the stellar radius. Mass accretion rates are also listed in Table~\ref{Table:acc_line}. Our sample covers a large range in accretion luminosities and mass accretion rates,  from Log($L_{\rm acc}/L_{\odot}$)=$-$3.48 to 0.72 and Log$(\dot{M}_{acc}/M$\accunit)=$-$10.15 to $-$6.12, estimated from the accretion-related emission lines. 

\begin{figure}
\begin{center}
\includegraphics[width=1\columnwidth]{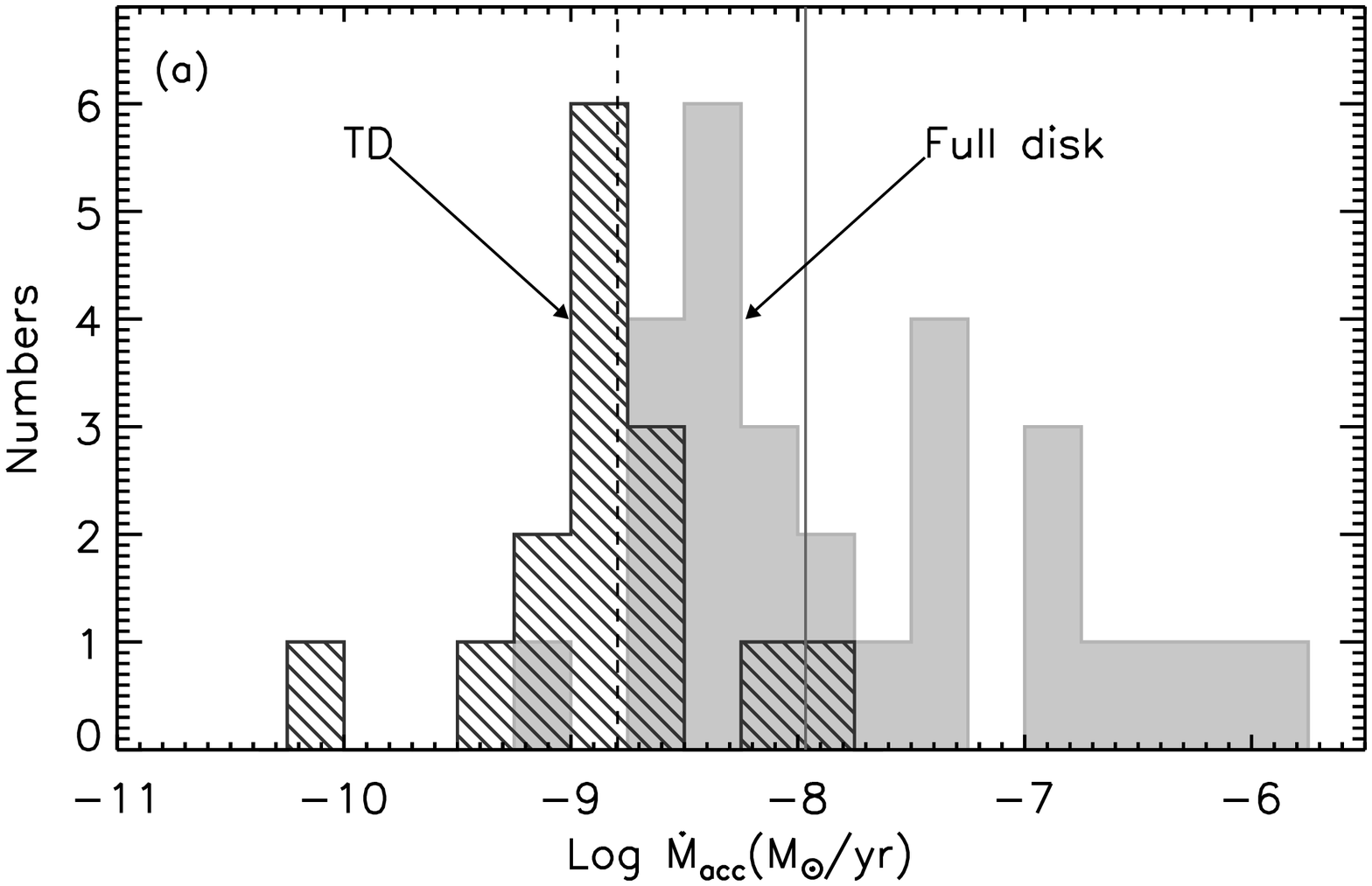}
\includegraphics[width=1\columnwidth]{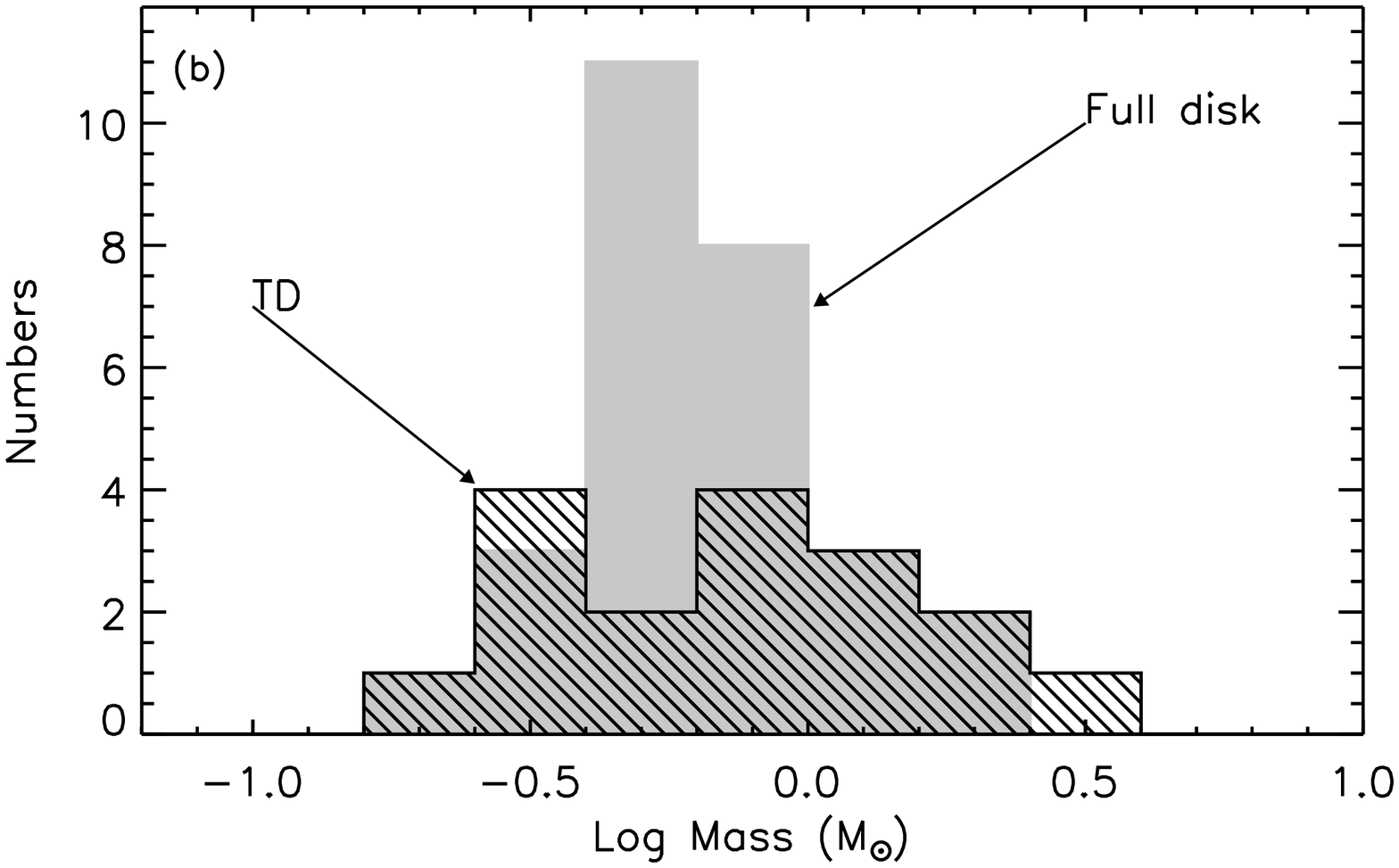}
\caption{Panel (a): Distribution of accretion rates for full disks (grey-color filled histogram) and TDs (hatch-filled histogram). The solid line and the dashed line mark the median accretion rates of full disks and TDs, respectively. Panel (b): Distribution of stellar masses for full disks (grey-color filled histogram) and TDs (hatch-filled histogram).}\label{Fig:acc_dis}
\end{center}
\end{figure}

We do not detect any accretion-related emission lines from DoAr~21 ($\lambda$=4420--6300\,\AA), DoAr~24E ($\lambda$=4420--6300\,\AA),  SR~21A ($\lambda$=3650--6300\,\AA), and V1057~Cyg ($\lambda$=3650--6300\,\AA), hence we do not list their accretion rates in Table~\ref{Table:acc_line}. Note that V1057~Cyg and V1515~Cyg are  well known FU~Ori objects. For this class of objects accretion luminosities are best derived from the emission of their self-luminous disks (e.g., \citealt{1998AJ....115.2491K}) and we use the more recent estimates from \cite{2006ApJ...648.1099G} in our paper.
Interestingly, the spectra of DoAr~21 and DoAr~24E do not show any \OIa{} emission while that of SR~21A only a marginal detection, see Appendix~\ref{Appen:detail} for more details. To clarify which sources are truly accreting, we also compare the derived accretion luminosities with typical chromospheric values (see \citealt{2017A&A...605A..86M}). Only TWA~3A has an accretion luminosity below the chromospheric emission, suggesting that chromospheric activity dominates the line emission.

\begin{figure*}
\begin{center}
\includegraphics[width=\columnwidth]{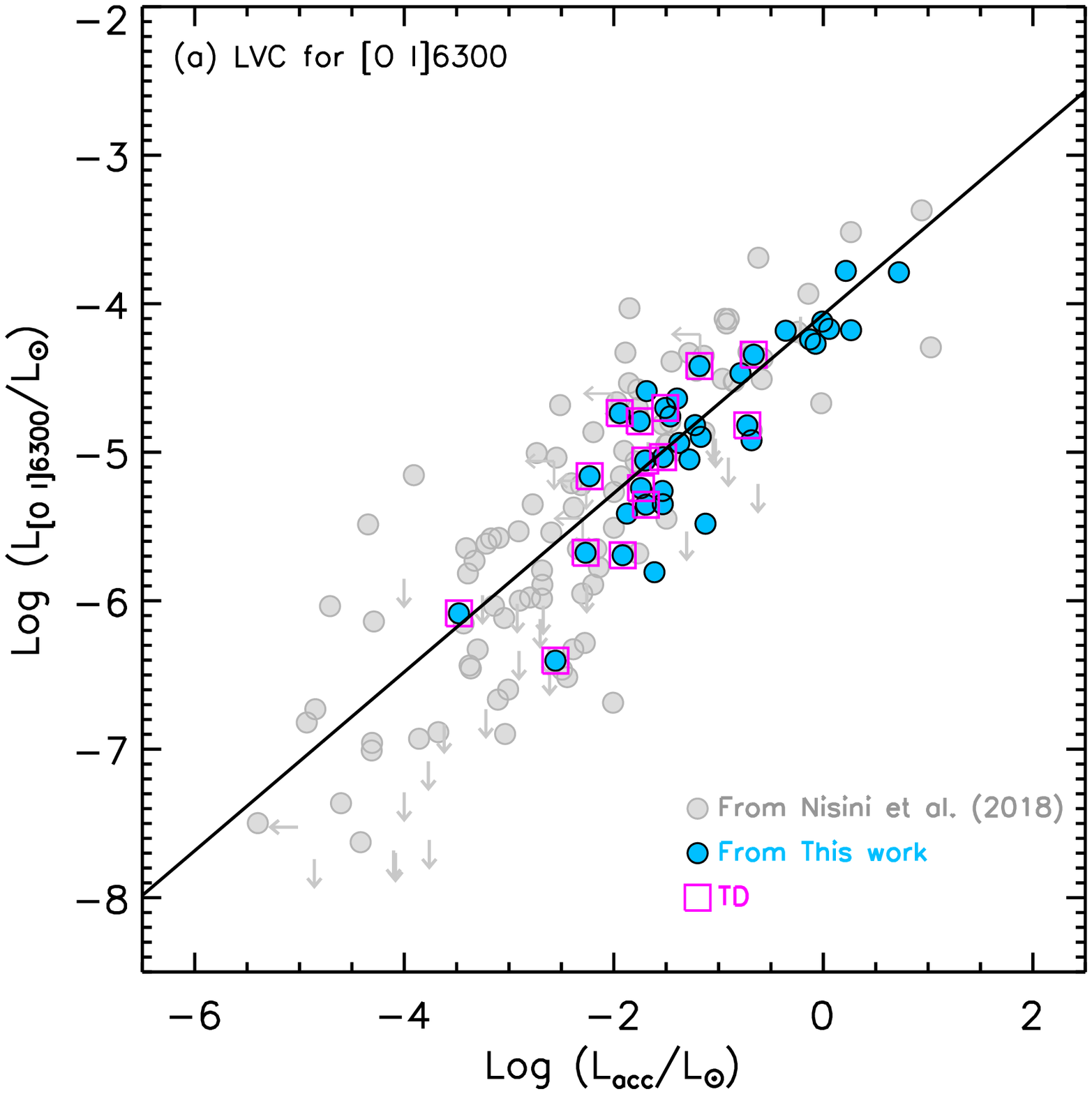}
\includegraphics[width=\columnwidth]{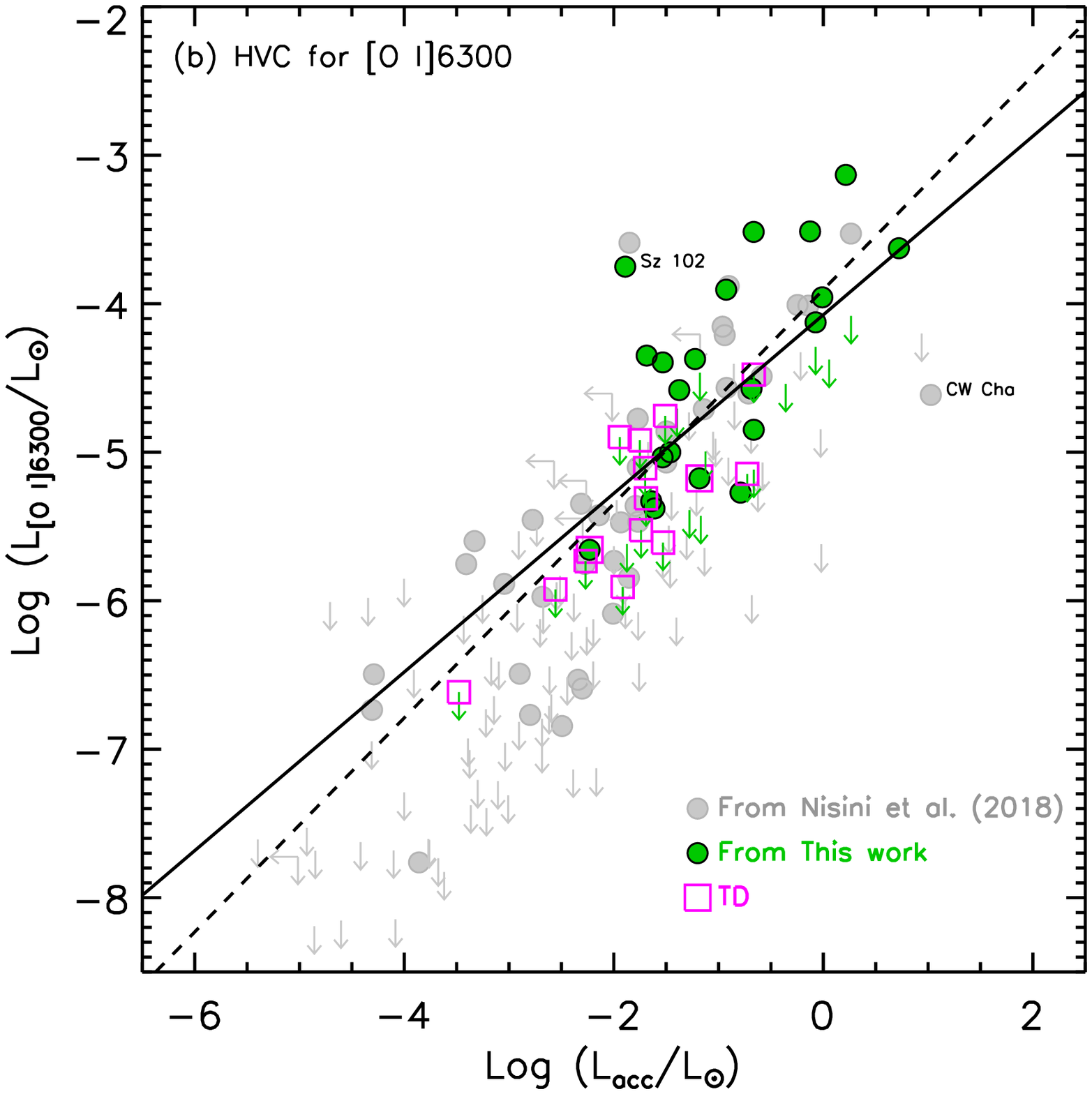}
\caption{Line luminosities for the \OIa\ LVC (Panel a) and for the HVC (Panel b) vs accretion luminosities. Magenta squares mark TDs, color-filled circles are sources  from this work, while gray circles are additional sources from \cite{2018A&A...609A..87N}. Solid lines are the best linear fit to all LVC detections present in Panel (a). The additional dashed line in Panel (b) gives the best fit linear relation to the detected HVCs. The upper limits in Panel~ (b) for our sample are calculated assuming a Gaussian profile with the FWHM of 70\kms, a median of the HVC FWHM, and a peak of 3$\times rms$}\label{Fig:line_acc}
\end{center}
\end{figure*}

Figure~\ref{Fig:acc_dis}(a) compares the distribution of accretion rates for full disks (gray filled histogram) and TDs (hatch-filled histogram). The K-S test returns a low probability (P$\sim1.8\times10^{-5}$) for the full disks and TDs to be drawn from the same parent population. The accretion rates of TDs are systematically lower  than those of full disks with a median value of 1.6$\times10^{-9}$\,\Msunyr, about seven time lower than that of the full disks in our sample. A similar result is reported in \cite{2015MNRAS.450.3559N} when comparing full and transition disks in Ophiuchus and Taurus. Fig.~\ref{Fig:acc_dis}(b) compares the distribution of stellar masses. The K-S test gives a 55\% probability that the two samples are drawn from the same stellar mass distribution, hence the difference in accretion rates most likely relates to their different evolutionary stage.

\subsection{Correlation between [O~{\scriptsize I}] and accretion luminosity}\label{Result:Line_acc_HVC}
Previous work has shown that the \OIa\  line luminosity correlates with stellar properties like stellar and accretion luminosity but not with X-ray luminosity (e.g., \citealt{2013ApJ...772...60R} and \citealt{2018A&A...609A..87N}). Here, we repeat the analysis by combining our sample with that of \cite{2018A&A...609A..87N}, as they cover a large range in stellar properties, and by implementing the new {\it Gaia} distances   \citep{2018arXiv180410121B} for all samples. Our goal is to clarify which correlations are present and provide the most up-to-date relations with accurate distances for individual sources.

\begin{figure}
\begin{center}
\includegraphics[width=\columnwidth]{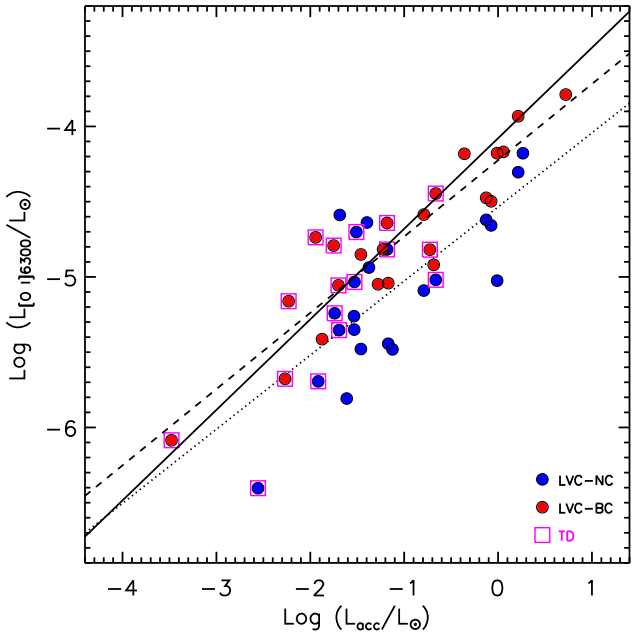}
\caption{Line luminosities for \OIa\ LVC-NC (blue) and LVC-BC (red) vs accretion luminosities.  
Magenta squares mark TDs.  The solid line is the linear fit to the total LVC (see Fig.~\ref{Fig:line_acc}, Panel (a)), while the 
dashed line and dotted line are the best fits to the LVC-BC and LVC-NC, respectively.}\label{Fig:line_acc_HVC}
\end{center}
\end{figure}

Figure~\ref{Fig:line_acc} (a, b) shows the relationship between the LVC (NC+BC) and HVC \OIa\  line luminosity versus the accretion  luminosity. 
The Spearman's rank correlation coefficients for the LVCs and HVCs  are 0.79 and 0.55, respectively,  with a probability lower than $5\times10^{-5}$ that the data are uncorrelated.  We perform outlier-resistant two-variable linear regressions including only the detections\footnote{Upper limits depend on the assumed line profile which, especially for the HVC, vary substantially from source to source. In the case of line ratios, Sect.~4, we could include upper limits because we required the detection of at least one line which then set the profile for the non-detections. } and obtain  the following relationships:
\begin{equation}
\label{Equ2:LVC_acc}
Log~L_{\rm OI63, LVC}=(0.60\pm0.03)Log~L_{\rm acc} -(4.07\pm0.06)
\end{equation}
\begin{equation}
\label{Equ3:HVC_acc}
Log~L_{\rm OI63, HVC}=(0.72\pm0.07)Log~L_{\rm acc} -(3.90\pm0.11)
\end{equation}

\begin{figure*}
 \begin{center}
 \includegraphics[width=2\columnwidth]{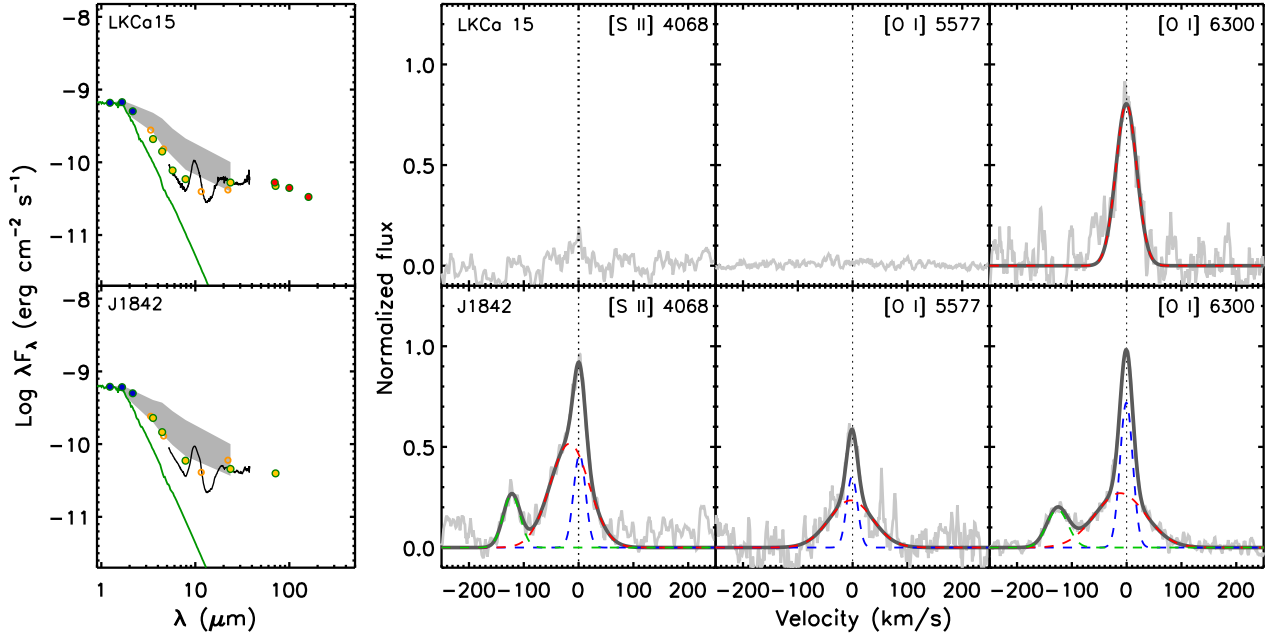}
 \caption{Left panels: SEDs of two TDs, LkCa~15 and J1842, around stars of similar spectral type and accretion rate.  {\rev In each panel, the grey region show the upper and lower quartiles of the Lynds~1641 classical T~Tauri median SED with the same spectral type. The median SED has been reddened  with the extinction of each source, and then normalized to the J-band flux.} Right panels: \SII , \OIb , and \OIa{} line profiles from the same sources. While these systems have very similar SEDs, the forbidden line profiles suggest that the inner gaseous disk of LkCa~15 is in a more evolved stage than that of J1842.}\label{Fig:SED_LR3}
 \end{center}
 \end{figure*}

The slope for the $L_{\rm OI63, LVC} $ vs. $ L_{\rm acc}$ in this work is similar to those (0.52$\pm$0.07, 0.59$\pm$0.04) in  \cite{2013ApJ...772...60R} and \cite{2018A&A...609A..87N}, but much flatter than the one (0.81$\pm$0.09) in \cite{2014A&A...569A...5N}, which could be due to their smaller sample size. Our $L_{\rm OI63, HVC} $ vs. $ L_{\rm acc}$ slope is also similar to the one (0.75$\pm$0.08) in \cite{2018A&A...609A..87N}, but slightly flatter than the one ($\sim$0.9) in \cite{2013ApJ...772...60R}. The $L_{\rm OI63, HVC} $ vs. $ L_{\rm acc}$ slope is steeper than the $L_{\rm OI63, LVC} $ vs. $ L_{\rm acc}$ slope with a 2$\sigma$ confidence,  consistent with the findings in \cite{2013ApJ...772...60R}.
Note that using only our sources we find the same slopes as those reported in Equations~\ref{Equ2:LVC_acc} and ~\ref{Equ3:HVC_acc} within the, now larger, uncertainties (0.60$\pm$0.06 and 0.62$\pm$0.16).

For our sources we can further correlate the line luminosity of individual LVC components with $ L_{\rm acc}$, see  Figure~\ref{Fig:line_acc_HVC}.  The Pearson correlation coefficients for the LVC-NC and LVC-BC  are  0.71 and 0.92 with  two-sided p-values of 1.4$\times$10$^{-4}$ and 7.6$\times$10$^{-10}$, respectively. This means that their luminosities are correlated with the accretion luminosity with the LVC-BC displaying a higher correlation than the NC and the total LVC  (see Fig.~\ref{Fig:line_acc}(a)).
The outlier-resistant two-variable linear regression gives the following relations:
\begin{equation}
Log~L_{\rm OI63, LVC-BC}=(0.51\pm0.05)Log~L_{\rm acc} -(4.22\pm0.07)
\end{equation}
\begin{equation}
Log~L_{\rm OI63, LVC-NC}=(0.49\pm0.10)Log~L_{\rm acc} -(4.53\pm0.14)
\end{equation}
\noindent The $L_{\rm OI63, LVC-BC} $  and the $L_{\rm OI63, LVC-NC}$ vs. $ L_{\rm acc}$ slopes are the same within the uncertainties.

\subsection{Inner disk evolution traced by the [O~{\scriptsize I}] emission}
 The evolutionary stage of a star+disk system is often assessed from its SED. However, optical forbidden lines can aid in understanding the evolutionary stage of a system by probing the evolution of the gas content (see e.g., Figure~5 in \citealt{2017RSOS....470114E}). 

The frequency of different \OIa\ line components differ for full and TDs, see Table~\ref{Table:statistics} where we provide statistics\footnote{Only LVC-NC means that the \OIa\ profile does not show any LVC-BC component. Similarly, only LVC-BC means that there is no LVC-NC component in the [O~I] profile.} for our sample. 
Full disks more frequently than TDs show an HVC in their \OIa\ profiles\footnote{Note that our 72\% fraction of HVCs in full disks is much higher than the  $\sim$30\% quoted in \cite{2018A&A...609A..87N} perhaps due a bias toward strong accretors in our sample.}, and a BC accompanied by an NC. In contrast,   TDs more frequently than full disks present only a LVC-NC or BC. In addition, TDs have simpler line profiles: $\sim$81\% (13/16) of our TDs only show one LVC-NC or one LVC-BC without any other LVC or HVCs components, while this fraction is down to 24\% (7/29)  for the full disks\footnote{We also note that among the 7 full disks, showing only simple \OIa\ line profiles, two of them, DS~Tau and RY~Lup, could be also transition disks based on the sub-millimeter/millimeter data \citep{2014A&A...564A..95P,2016ApJ...828...46A}}, see also Figures \ref{Fig:SED_LR1} and~\ref{Fig:SED_LR2} in Appendix~\ref{Appen:SED}. {\rev With a larger sample of 65 T~Tauri stars, Banzatti~et~al.~(2018) also find that those surrounded by a TD tend to show only a single component in their \OIa\ line profile (see their Section 5.5 for details). The much lower fraction of TDs with an HVC and a BC+NC with respect to full disks might indicate an evolution in disk winds (see also  \citealt{2017RSOS....470114E}). The TDs in our sample have lower mass accretion rates than full disks, hence lower HVC and LVC luminosities (Figure 8), and likely a more depleted inner gas disk. As also suggested in Banzatti~et~al.~(2018), winds in TDs might be launched from larger radii, have a larger opening angle, and not re-collimate into jets.}

\setcounter{table}{5}
\begin{table}
%\scriptsize
\renewcommand{\tabcolsep}{0.08cm}
\caption{Statistics on the line profiles }\label{Table:statistics}
\begin{center}
\begin{tabular}{cccccc}
\hline
        &Only NC  & Only BC &NC+BC  &HVC \\%&Single component\\
        \hline
% & \multicolumn{4}{c}{The sources in this work}\\
%\hline
Full disks&28\% (8/29) &24\% (7/29) &24\% (7/29)&72\% (21/29)\\
TDs       &44\% (7/16) &44\% (7/16) &13\% (2/16) &13\% (2/16)\\
%Full disks&24\% (7/29) &14\% (4/29) &31\% (9/29)&72\% (21/29)\\% &  28\% (8/29) \\  
%TDs       &44\% (7/16) &44\% (7/16) &13\% (2/16) &13\% (2/16)% &  81\% (13/16)\\    \hline
% & \multicolumn{4}{c}{The sources in this work and in \cite{2016ApJ...831..169S}}\\
%\hline
%Full disks&24\% (9/38) &39\% (15/38)&29\% (11/38)&63\% (24/38)    %\\  
%TDs &40\% (8/20)       &45\% (8/20) &15\% (3/20) &10\% (2/20)    %\\  
\hline
\end{tabular}
\end{center}
\end{table}

\begin{figure}
\begin{center}
\includegraphics[width=\columnwidth]{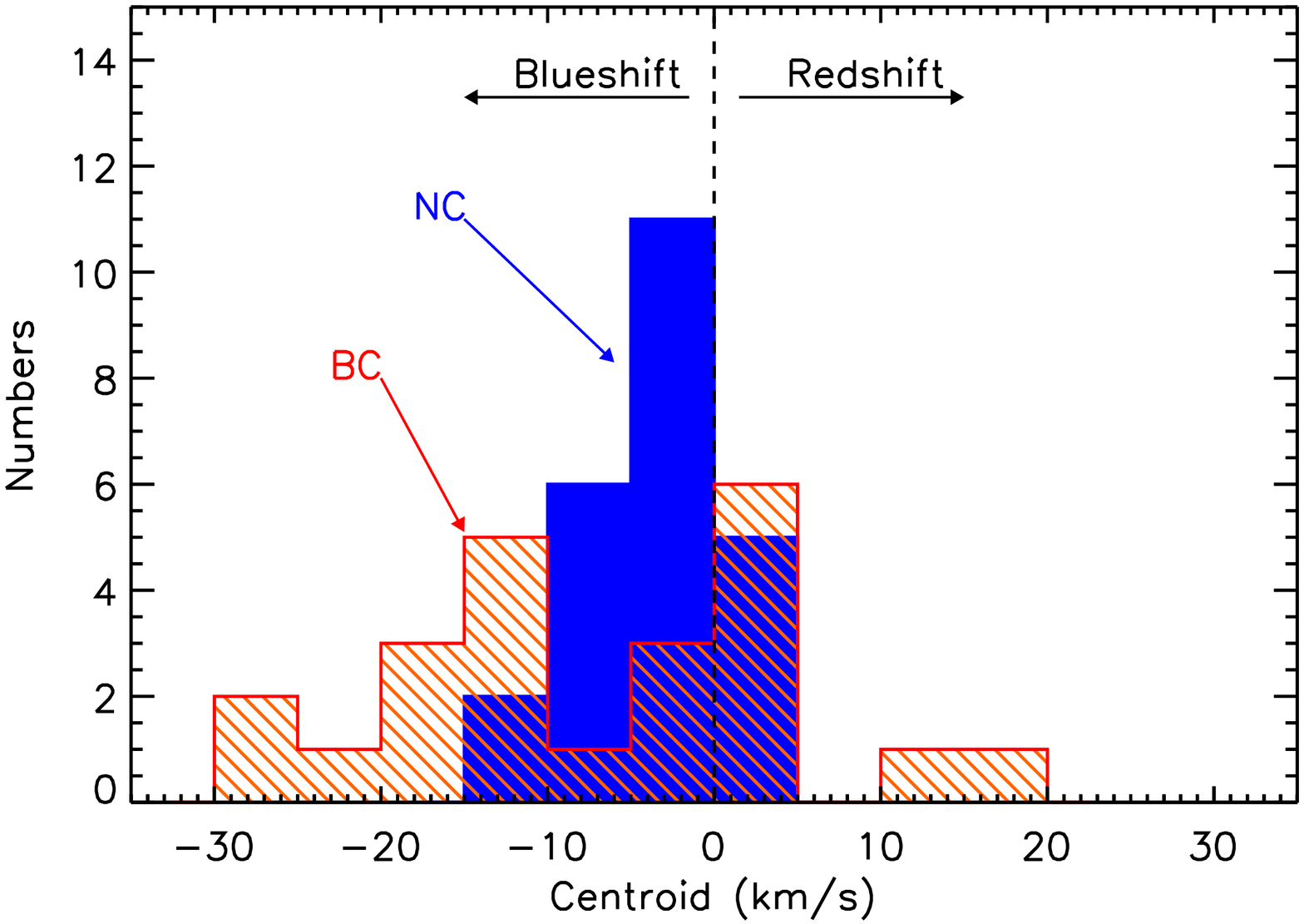}
 \includegraphics[width=1\columnwidth]{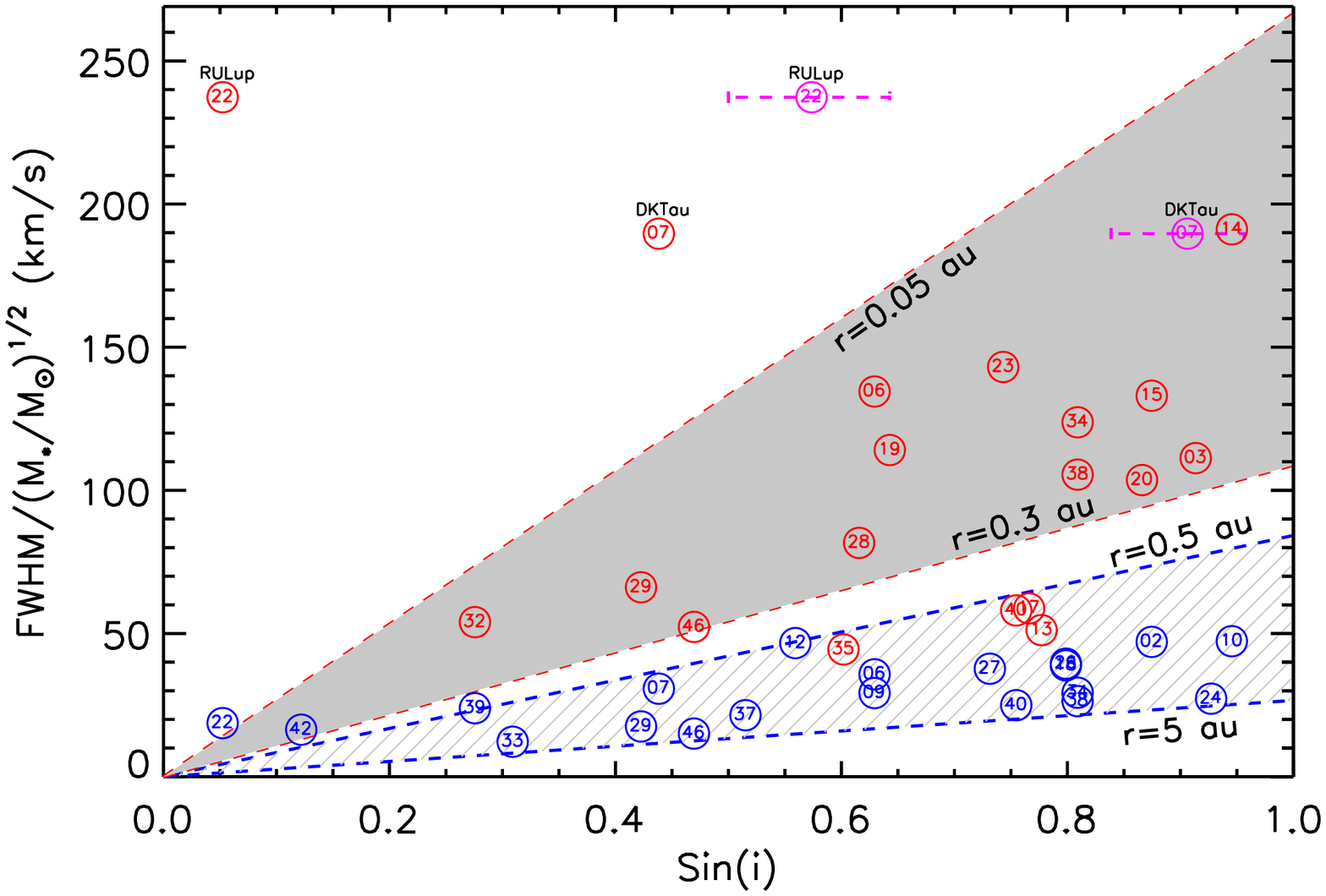}
 \caption{Upper panel: Distribution of centroids for the LVC-NC (blue-color filled histogram) and the LVC-BC (hatch-filled histogram). Note that many centroids are blueshifted with respect to the stellar velocity (see also Banzatti et al. 2018 for more details on the [O {\scriptsize I}] kinematics). Bottom panel: FWHM corrected for instrumental broadening and normalized by the stellar mass vs disk inclination,  LVC-BC in red and LVC-NC in blue. Dashed lines show Keplerian FWHM as a function of disk inclination at radii of 0.05, 0.3, 0.5, and 5\,au. The hatch-filled  area and grey color-filled area enclose most of the NC and BC, respectively. The magenta circles locate two sources (RU~Lup, ID~22, and DK~TauA, ID~7) for which we also use alternative disk inclinations.}\label{Fig:centroid_LVC}
 \end{center}
 \end{figure}

Furthermore,  the \OIa\ line profiles can be used to distinguish the evolutionary stage of systems with similar SEDs. An example is shown in Figure~\ref{Fig:SED_LR3} for  LkCa~15 and J1842, both classified as TDs. The central stars have similar spectral types (K5.5 vs. K3) and accretion rates ($1.8\times10^{-9}$ vs. $3.0\times10^{-9}$ $M_{\odot}~yr^{-1}$ ), and their disks present very similar SEDs.   
However, their forbidden line profiles are clearly different: LkCa~15 only presents a LVC-BC (with a FWHM, 44.6\,\kms, near the boundary to separate the LVC NCs and BCs) in the \OIa , while J1842 shows a LVC-NC, a LVC-BC, and a HVC in both the \SII\ and \OIa\  lines, and a LVC-NC and a LVC-BC in the \OIb\ line. Thus, while both sources would appear to be in the same evolutionary stage based on their SED, the forbidden line profiles demonstrate that the inner gaseous disk of LkCa~15  is in a more evolved stage than that of J1842.

\subsection{Mass outflow rates}\label{Sect:mloss_all}
Previous analysis of \OIa{} high-resolution spectra has demonstrated that a significant number of LVC-BC and NC peak centroids are blueshifted with respect to the stellar velocity (\citealt{2016ApJ...831..169S}; \citealt{2018arXiv180310287M}; Banzatti~et~al. 2018). With the largest blueshifts found in sources surrounded by lower inclination disks and an emitting region within $\sim$0.5\,au from the star, \cite{2016ApJ...831..169S} attributed the LVC-BC to the base of an MHD-driven wind. Although the focus of our paper is on forbidden line luminosities and line ratios, here we show that the kinematics of the LVC-BC and NC of our sample are consistent with previous findings. 

The upper panel of Fig.~\ref{Fig:centroid_LVC}  shows the distribution of the NC and BC centroids.
About 63\% (15/24) of NCs and 52\% (12/23) of BCs have blueshifts larger than 1.5\,\kms, while only 8\% (2/24) of NCs and 9\% (5/23) of BCs are redshifted. The larger proportion of blueshifts when compared to no shift or redshifts is consistent with a wind origin for both components. The lower panel of Fig.~\ref{Fig:centroid_LVC} shows the relation between disk inclination and the BC and NC FWHMs, corrected for instrumental broadening and normalized by stellar mass.

If we attribute the line width primarily to Keplerian broadening as in previous studies \citep{2016ApJ...831..169S,2018arXiv180310287M}, most BCs trace radii within 0.5\,au while  NCs probe gas further out but mostly within $\sim$5\,au.
A few sources appear as outliers in this plot and are worth discussing. First, the FWHM from DK~TauA and RU~Lup are much larger than those expected from pure Keplerian rotation using the disk inclinations given in Table~1. The disk inclination of DK~TauA is not well constrained: the 20$^{\circ}$ value we adopted here comes from the analysis of mm CO lines but the continuum emission points to a much higher inclination of 65$^{\circ}$ \citep{2017ApJ...844..158S}. In the case of RU~Lup the outer disk inclination of $3^{\circ}$ derived from ALMA mm imagery may not apply to the forbidden lines studied here. Indeed, spectroastrometry in the CO rovibrational band suggests a higher inclination of 35$^{\circ}$ for the inner disk \citep{2011ApJ...733...84P} and an even higher inclination is implied by MIDI visibilities (see \citealt{2018arXiv180502939V} and Banzatti et al. 2018 for further details).
If we use these alternative/higher disk inclinations, the BC FWHM of DK~TauA falls in the same region as the other BCs and that of RU~Lup becomes much closer to that region. 
Finally, the BCs from LkCa~15, Sz~73, RNO~90, and VV~CrA fall within the domain of LVC-NC, they could have been misclassified using our stringent cut in FWHM that does not take into account disk inclination.

Having established that our sample supports the scenario in which LVC-BC and NC trace a wind and their FWHMs are consistent with being broadened by Keplerian rotation, we use  the  \OIa\ luminosity and our constraints on the gas temperature, velocity, and emitting radii to compute wind mass loss rates. Fig.~\ref{Fig:twolineratio_com2}(a) shows that electron densities $n_e \geq 3 \times 10^6$\,cm$^{-3}$ are needed to reproduce the measured LVC line ratios. As these densities are larger than the \OIa\ critical density, which is $1.8 \times 10^6$\,cm$^{-3}$, the LVC emitting gas is most likely in LTE. In addition, very high densities are required to make the \OIa\ line optically thick (see Table~9 in \citealt{1989ApJ...342..306H}). Hence, we will use equations for optically thin LTE gas to relate  the LVC \OIa\ luminosity to a gas mass.
As a shock origin is more likely for the HVC emission and this lower density gas may not be in LTE, see \sect~\ref{Sect:shock_HVC} and Fig.~\ref{Fig:twolineratio_com2}~(b),  we will use equations derived for radiative shocks to estimate the mass loss rate from this high-velocity component.

\subsubsection{Mass loss rates from the LVC}\label{sect:Mloss_LVC}
As the  \OIa\ line is optically thin its luminosity ($L_{6300}$) can be written as:
\begin{equation}
L_{6300}=N_{\rm u}~A~h\nu \label{Equ6} 
\end{equation}
\noindent Where $N_{\rm u}$ is the total number of O atoms in the upper $^1D_{2}$ level, $A$ is the transition probability (6.503$\times10^{-3}$), and $h\nu$ is the associated energy they emit \citep{1997A&AS..125..149D}. In addition, for gas in LTE the total number of O atoms ($N$) is related to $N_{\rm u}$ as:
\begin{equation}
N_{\rm u}= N~\frac{g_{u}~e^{-h\nu/kT}}{Z(T)} \label{Equ7}
\end{equation}
\noindent Where the $g_{u}$ is the upper level statistical weight, $k$ is the Boltzmann constant, $T$ is the gas temperature, and $Z(T)$ is the partition function at temperature $T$. We calculate $Z(T)$ assuming a 5-level oxygen atom.
Considering that in disks the abundance of neutral oxygen is reduced because half of the cosmic O is in silicate grains \citep{2009ApJ...700.1299J} and the wind is launched from outside the dust sublimation radius (Fig.~\ref{Fig:centroid_LVC} lower panel),
we take $\alpha({\rm O})$=3.2$\times$10$^{-4}$ to calculate the total mass of gas $M$ from the total number of O~{\scriptsize I} atoms. Finally, the mass loss rate $\dot{M}_{\rm wind}$ can be written as:
\begin{equation}
\dot{M}_{\rm wind}=M\frac{V_{\rm wind}}{l_{\rm wind}}\label{Equ8}
\end{equation}
\noindent Where $V_{\rm wind}$ and $l_{\rm wind}$ are the velocity and wind height. We further write the unconstrained wind height as $l_{\rm wind}=f\times r_{\rm base}$, i.e. the wind vertical extent is $f$ times the emitting radius at the base of the wind. Combining Equations~\ref{Equ6} through to ~\ref{Equ8}, we have: 
\begin{align}
\label{Equ9}
\dot{M}_{\rm wind}=&\frac{V_{\rm wind}}{l_{\rm wind}} \frac{\eta m_{\rm H}}{\alpha({\rm O})}~e^{h\nu/kT}{Z(T)} \frac{L_{6300}}{g_{\rm u}~A~h\nu}\\\nonumber &={\rm C(T)}\Big(\frac{V_{\rm wind}}{10~{\rm km~s^{-1}}}\Big)\Big(\frac{l_{\rm wind}}{1~{\rm au}}\Big)^{-1}\Big(\frac{L_{6300}}{L_{\odot}}\Big)M_{\odot}～yr^{-1}
\end{align}
\noindent Where $\eta$=1.34 is the ratio of total gas mass to Hydrogen mass. From this equation it is clear that the estimated mass loss rate has a strong dependence on the gas temperature $T$ and scales linearly with $V_{\rm wind}$ and $l_{\rm wind}$. LVC gas temperatures range from 5,000 to 10,000~K (\sect~\ref{Sect:LVC_lineratio}) and within this range $C(T)$ varies by an order of magnitude: from 2.4$\times$10$^{-4}$ for $T$=5,000\,K down to 2.6$\times$10$^{-5}$ for $T$=10,000\,K.

\begin{figure*}
\begin{center}
\includegraphics[width=1\columnwidth]{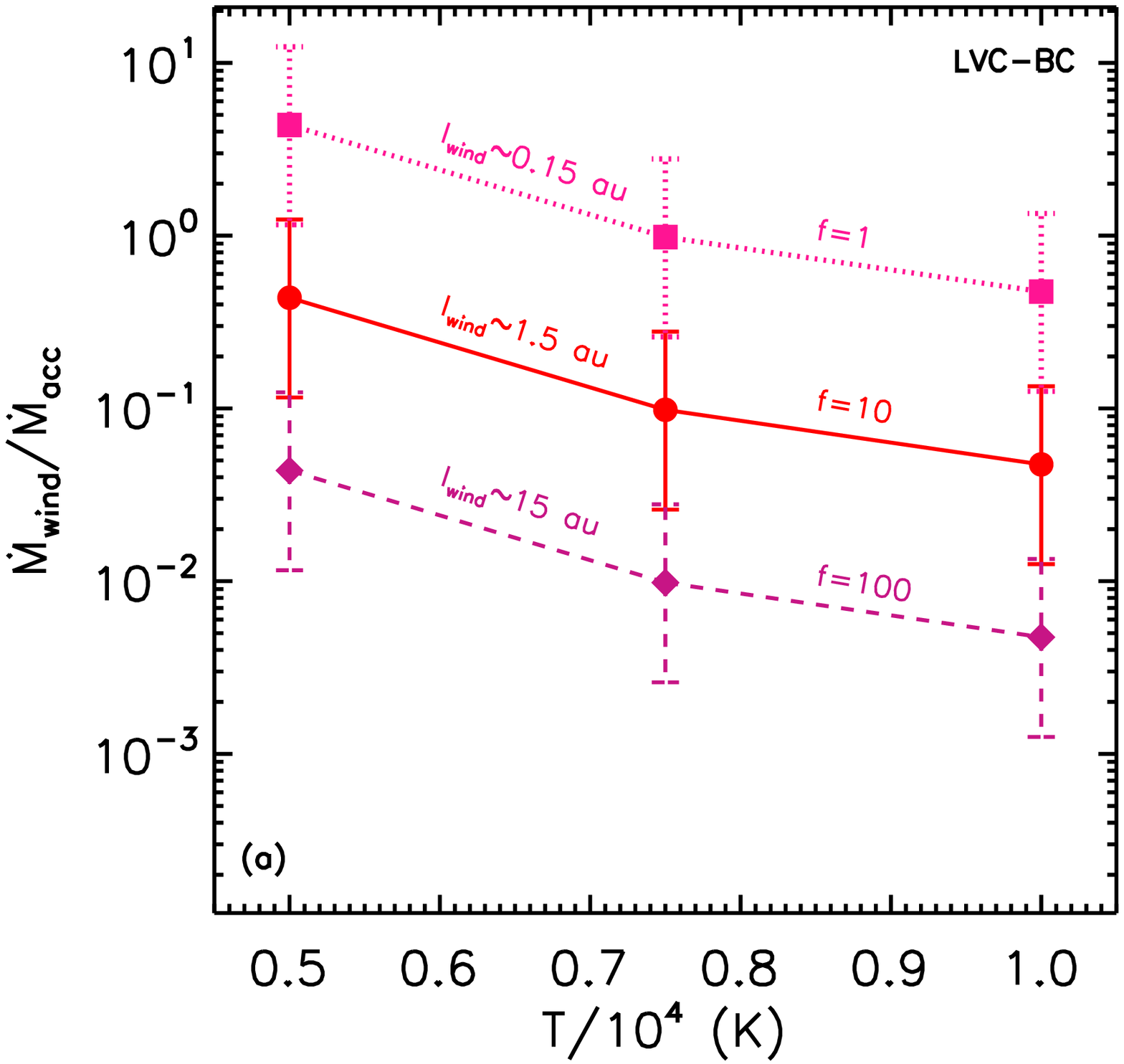}
\includegraphics[width=1\columnwidth]{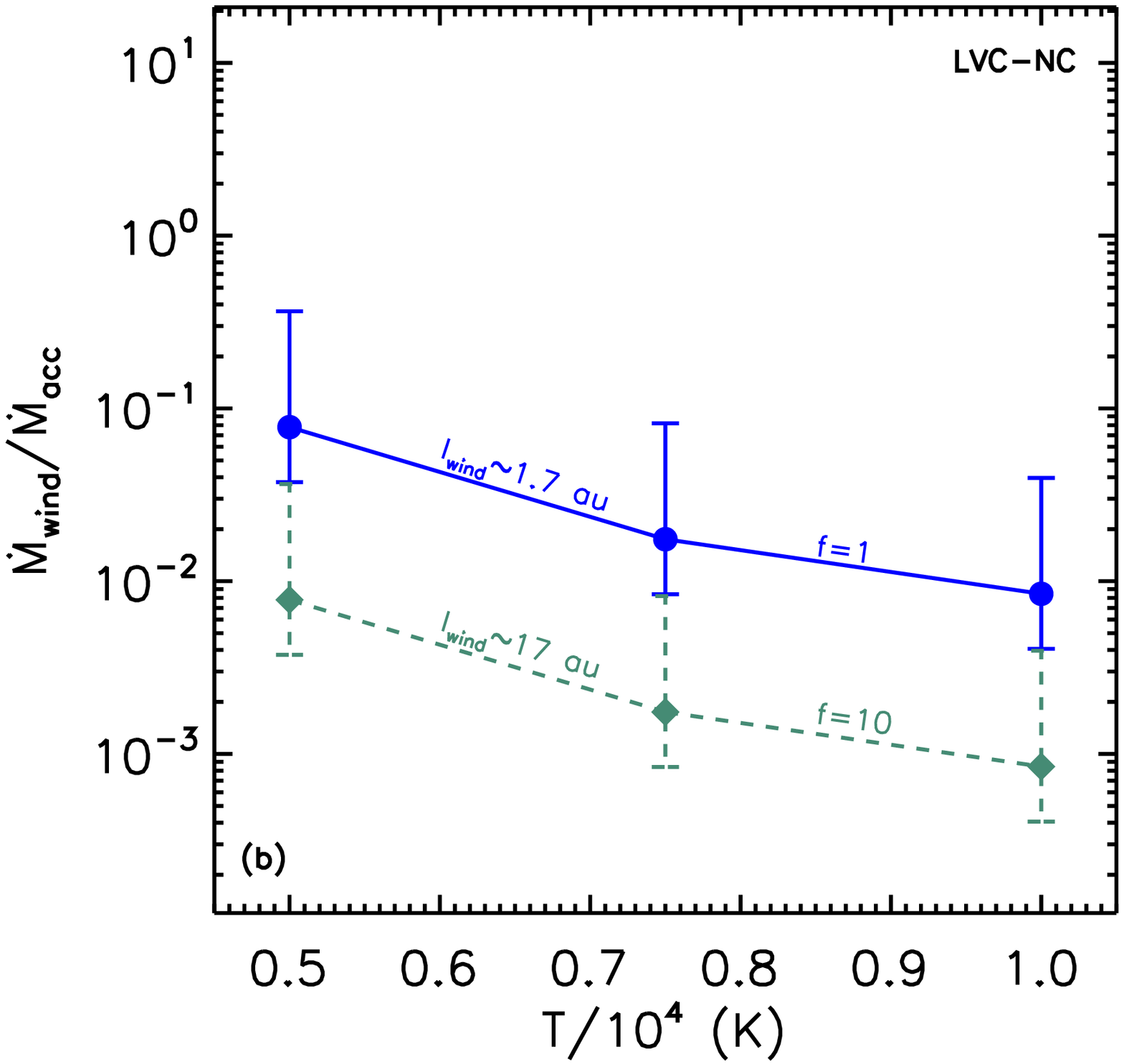}
 \caption{Median $\dot{M}_{\rm wind}$/$\dot{M}_{\rm acc}$ ratios for our sample as a function of gas temperature and for different wind heights ($l_{\rm wind}=f\times r_{\rm base}$), LVC-BC left panel and LVC-NC right panel. In each panel, the error bars show the upper and lower quartiles of the $\dot{M}_{\rm wind}$/$\dot{M}_{\rm acc}$ ratio distributions.}\label{Fig:mloss_LVC}
 \end{center}
 \end{figure*}
 
 \begin{figure*}
\begin{center}
\includegraphics[width=1\columnwidth]{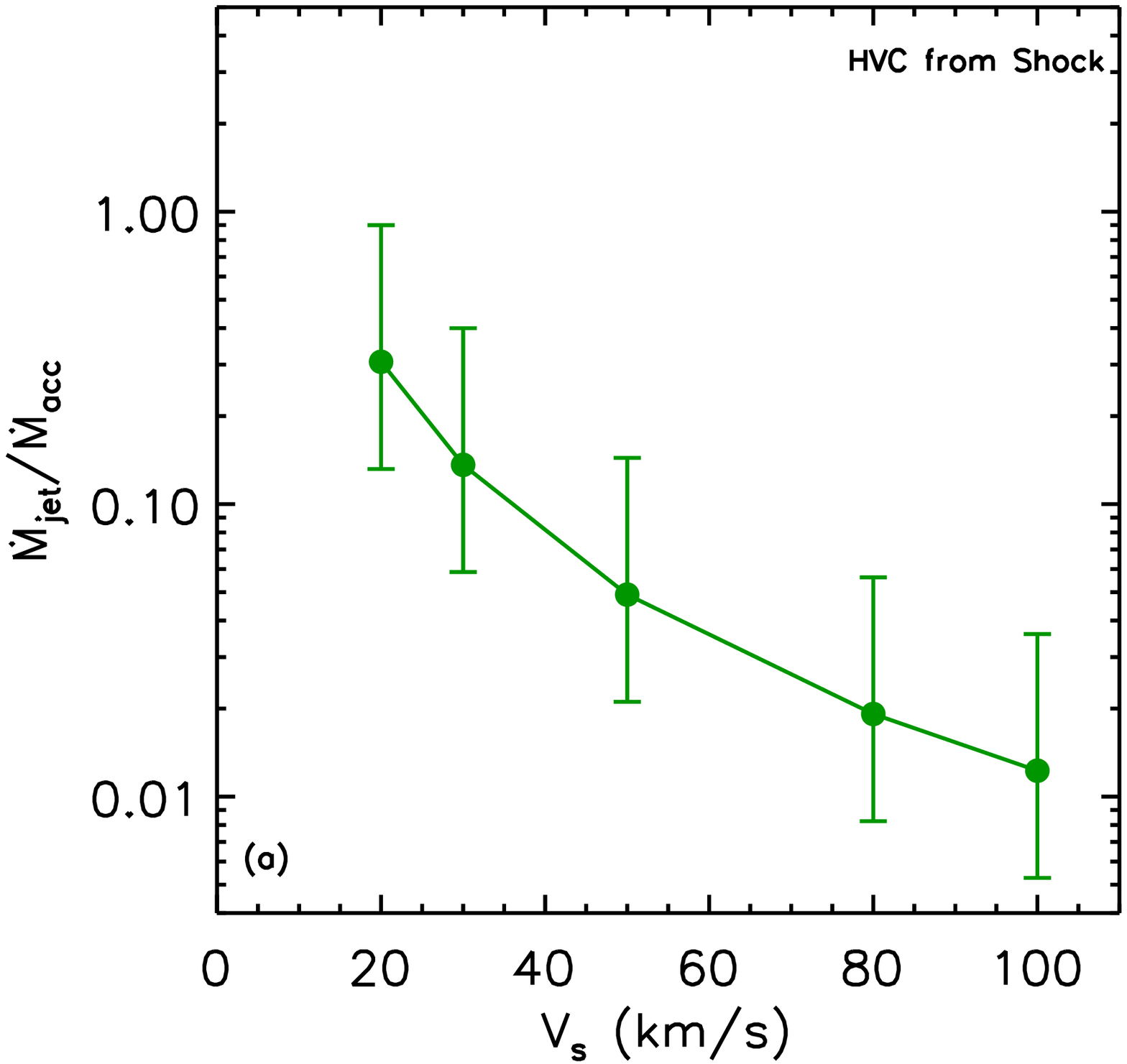}
\includegraphics[width=1\columnwidth]{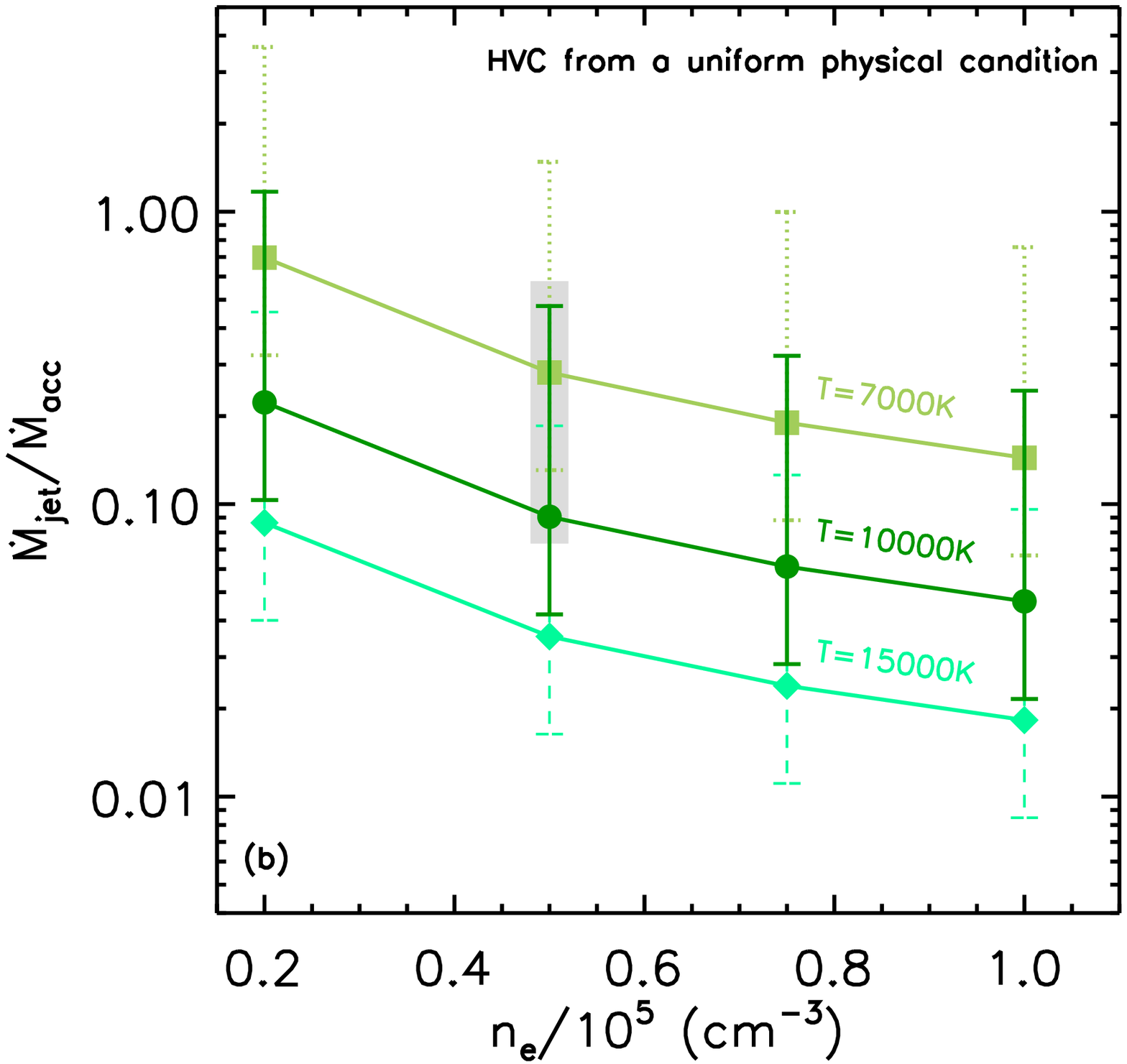}
 \caption{Left panel: Median $\dot{M}_{\rm jet}$/$\dot{M}_{\rm acc}$ ratio as a function of shock velocity assuming that HVC arises in post-shock gas. Right panel: Median $\dot{M}_{\rm jet}$/$\dot{M}_{\rm acc}$ as a function of electron density and for different gas temperatures assuming that HVC are produced in uniform physical conditions. Error bars show the upper and lower quartiles of the median $\dot{M}_{\rm jet}$/$\dot{M}_{\rm acc}$ ratios for our sample while the grey region for sources in \cite{2018A&A...609A..87N} which are reported in their paper only for  $T$=10,000~K, $n_{\rm e}$=$5\times10^{4}$\,cm$^{-3}$.}\label{Fig:mloss_HVC}
 \end{center}
 \end{figure*}

 For each source with an \OIa\ detection we calculate its own $\dot{M}_{\rm wind}$ by taking as $V_{\rm wind}$ the median of projected peak velocities and as $r_{\rm base}$  the Keplerian radius from half of the line FWHM, corrected for the instrumental broadening and projected (see Fig.~11 lower panel). For the BC these values correspond to  $-$15\,\kms\ and 0.15\,au while for the NC they are $-$6\,\kms\ and 1.7\,au\footnote{Unprojected values are for the BC $-$12\,\kms\ and 0.33\,au while for the NC they are $-$4\,\kms\ and 4.3\,au.}. We then divide $\dot{M}_{\rm wind}$ by each source  $\dot{M}_{\rm acc}$ and show in Figure~\ref{Fig:mloss_LVC} (panels a and b) how the median and upper and lower quartile ratios depend on $T$ and $l_{\rm wind}$. As expected from Equation~\ref{Equ9},  $\dot{M}_{\rm wind}$, and hence $\dot{M}_{\rm wind}/\dot{M}_{\rm acc}$ ratios, increase with decreasing gas temperature and wind height. Since the extent of the \OIa\ emitting region is unknown, we vary the factor $f$ within a range of plausible values for which $l_{\rm wind}$ is similar in the LVC-BC and NC.

 For the BC, $f=1$ produces unreasonably large $\dot{M}_{\rm wind}/\dot{M}_{\rm acc}$ ratios for most temperatures, except for 10,000\,K. However, such a high temperature is unlikely otherwise emission lines from ionized oxygen would have been also easily detected (see Sect.~5.7 in \citealt{2016ApJ...831..169S}).  Heating processes not driven by photons (e.g., ambipolar drift heating, see \citealt{1993ApJ...408..115S}) could result in higher gas temperatures without affecting the ionization state, but their efficiency has not been sufficiently explored in this context. We will further discuss the implications of Fig.~\ref{Fig:mloss_LVC} in \sect~\ref{Sect:discussion}. However, it is already worth noting that, for likely wind extents, the mass loss rates implied by the BC are much higher than those implied by the NC. This results from the fact that \OIa\ luminosities as well as the median peak centroid are larger for the LVC-BC than for the NC component (see Figs.~\ref{Fig:line_acc_HVC} and \ref{Fig:centroid_LVC}) and the wind heights of the NC are likely higher than for the BC.

 {\rev This last point is important, hence we summarize here additional arguments supporting a large vertical extent of the LVC-NC. First, it is highly unlikely that the wind height of the LVC-NC is small (f$<$1) compared to the Keplerian radius because with our high-resolution ($\Delta$v$\sim$6km/s) spectra we would see double peaked NC profiles. Next, as we will discuss, it is very likely that the extent of the NC is larger than that of the BC. The LVC-BC traces an MHD wind because of its small Keplerian radius (0.05--0.5\,au) and blueshifted centroids; thermal pressure is insufficient to launch the gas flow and MHD forces are required. The NC may similarly also trace an MHD wind. In both cases the wind is not expected to extend much beyond the Alfv\'{e}n surface since  the magnetic field cannot hold the gas in rigid rotation and the wind will be accelerated and collimated into jets beyond the Alfv\'{e}n surface \citep{1982MNRAS.199..883B}. In MHD simulations, the height of the Alfv\'{e}n surface above the disk surface is about 0.5--1.5 the wind launching radius \citep{2006MNRAS.365.1131P,2017ApJ...845...75B}. Since the NC is launched further out (1--5\,au), its vertical extent is therefore likely to be larger than that of the BC.  If the LVC-NC is partly supported by a photoevaporative (thermal) wind, it should be even more extended vertically given the gas temperature (5,000--10,000K) inferred from the line ratios (see Fig.~\ref{Fig:twolineratio_com2}). As an example, in the X-ray-driven photoevaporative  wind model \citep{2016MNRAS.460.3472E}, the winds can extend to about 35 au above the disk. While the wind height cannot be determined without spatially resolved observations, our finding that the mass loss rates implied by the BC are higher than those implied by the NC should be reliable.}

\subsubsection{Mass loss rates from the HVC}\label{sect:Mloss_HVC} 
As discussed in Section~\ref{Sect:shock_HVC} shock models reproduce most of the [O{\scriptsize I}] and [S{\scriptsize II}] HVC line ratios. In fast shocks, most of the mechanical energy of the preshock gas is converted into heat (e.g. \citealt{1979ApJ...227..131S}, \citealt{1979ApJS...41..555H}). The mass loss rate can then be estimated assuming that a certain fraction of the kinetic energy is radiated away in the \OIa\ line. 
Hence, we use the equations for radiative shocks as presented e.g. in \cite{2009ApJ...703.1203H}. If the hydrogen density is high ($n_{\rm H}\sim10^{6}$~cm$^{-3}$), the line luminosity can be computed as:
\begin{equation}
L_{6300}=\frac{y_{0}}{2}\dot{M}_{\rm loss}V_{\rm s}^{2}
\end{equation} 
\noindent With $y_{0}$ being to the fraction of cooling that goes into the \OIa\ line and $V_{\rm s}$ the shock velocity. The mass-loss rate is then:
\begin{equation}
\dot{M}_{\rm jet}=L_{6300} \frac{2}{y_{0}}V_{\rm s}^{-2}
\end{equation}
At a density $n_{\rm H}\sim10^{6}$~cm$^{-3}$, $y_{0}$ is around 0.1-0.3, hence we take here $y_{0}=0.2$. As for the LVC, we first compute individual 
$\dot{M}_{\rm jet}$ divide them by each source $\dot{M}_{\rm acc}$ and show the behavior of median, upper and lower quartile ratios for a range of shock velocities compatible with observations (e.g., \citealt{1987ApJ...316..323H}). The faster the shock velocity the lower the 
$\dot{M}_{\rm jet}$/$\dot{M}_{\rm acc}$ ratio.  For a typical shock velocity of 30\,\kms\ \citep{1994ApJ...436..125H}, the  median $\dot{M}_{\rm jet}$/$\dot{M}_{\rm acc}$ ratio is around 0.1.

The approach we have taken above differs from the one typically used in the literature where it is assumed that the jet has uniform properties within the slit width and the line luminosity is calculated from a collisional excitation model akin to that described in Section~\ref{Sect:LVC_lineratio} for the LVC (see also Fig.~\ref{Fig:twolineratio_com2} left panel). More specifically and following \cite{1995ApJ...452..736H} and \cite{2018A&A...609A..87N}, the mass loss rate from the jet can be written as:
\begin{equation}
\dot{M}_{\rm jet}={\rm C(T, n_{\rm e})}\Big(\frac{V_{\perp}}{100~{\rm km~s^{-1}}}\Big)\Big(\frac{l_{\perp}}{100~{\rm au}}\Big)^{-1}\Big(\frac{L_{6300}}{L_{\odot}}\Big)M_{\odot}～yr^{-1}
\end{equation}
\noindent Where $V_{\perp}$ is the component of the jet velocity in the plane of the sky (we assume 100\,\kms\ as \citealt{2018A&A...609A..87N}) and  $l_{\perp}$ is the projected size of the slit aperture (0\farcs861) on the plane of the sky. C(T, $n_{\rm e}$) depends on the gas temperature and electron density and is calculated assuming a 5-level oxygen atom as in Section~\ref{Sect:LVC_lineratio}. For $T$=10,000~K, $C(T, n_{\rm e})$ are 9.0$\times$10$^{-5}$, 6.1$\times$10$^{-5}$, and 4.6$\times$10$^{-5}$ for  $n_{\rm e}=5.0\times10^{4}$, 7.5$\times10^{4}$ and  $1.0\times10^{5}~{\rm cm^{-3}}$, respectively. 
Taking the same temperature (10,000~K) and electron density ($5.0\times10^{4}$~cm$^{-3}$)  as \cite{2018A&A...609A..87N}, we find a median  $\dot{M}_{\rm jet}$/$\dot{M}_{\rm acc}$ ratio of $\sim$0.1. 
The right panel of Fig.~\ref{Fig:mloss_HVC} also shows the $\dot{M}_{\rm jet}$/$\dot{M}_{\rm acc}$ for the \OIa\ detected HVCs in \cite{2018A&A...609A..87N} after implementing the new {\it Gaia} distances  \citep{2018arXiv180410121B} and assuming $\alpha({\rm O})=3.2\times10^{-4}$, T=10,000~K, and  $n_{\rm e}=5.0\times10^{4}$~cm$^{-3}$. The median  $\dot{M}_{\rm jet}$/$\dot{M}_{\rm acc}$ ratio is $\sim$0.18 for this sample \footnote{The difference between our 0.18 value and the 0.07 value in \cite{2018A&A...609A..87N} is due to: (1) we use a different oxygen abundance ($3.2\times10^{-4}$ vs. $4.6\times10^{-4}$); (2) our value gives the median of the distribution while \cite{2018A&A...609A..87N} report the most common Log~$\dot{M}_{\rm jet}$/$\dot{M}_{\rm acc}$ value; (3) we use the new {\it Gaia} DR2 distances}. Furthermore, the same panel shows the dependence of $\dot{M}_{\rm jet}$/$\dot{M}_{\rm acc}$ on $T$ and  $n_{\rm e}$: the ratio decreases with increasing gas temperature and electron  density. For values of $T$$\sim$7,000--15,000\,K and $n_{\rm e}$=$2-6\times10^{4}$~cm$^{-3}$ that are appropriate for the base of the T~Tauri jets \citep{2007ApJ...660..426H,2011A&A...532A..59A,2013ApJ...778...71G,2014A&A...565A.110M}, the median $\dot{M}_{\rm jet}$/$\dot{M}_{\rm acc}$ ratios vary from 0.8 to 0.04. 

\section{Discussion}\label{Sect:discussion}
It is already well established that the HVC and the LVC components of forbidden lines trace different physical environments, where the HVC is more spatially extended and formed in shock-excited collimated jets, and the LVC, with a higher density than the HVC, is confined to scales of less than 5 AU (e.g., \citealt{1995ApJ...452..736H}, \citealt{1997A&AS..126..437H}, \citealt{2016ApJ...831..169S}).  Our study expands upon the literature by combining the \SII\ transition with the well-studied \OIa\ and \OIb\ lines, demonstrating that the same kinematic components appear in all 3 lines, thus enabling the use of component line ratios to constrain the properties of the emitting gas (\sect~\ref{com_SII_OI6300} and Fig.~\ref{Fig:threelines_HVC}). By including detections and upper limits we found that the HVC SII40/OI63  ratios are statistically higher than the LVC (both BC and NC) and that there is a low probability that the BC  OI55/63 ratios are drawn from the same parent population as the NC and HVC (Table~\ref{Table:problineratio}). These differences and the mean line ratios of the three components corroborate previous suggestions, mostly based on kinematics (e.g., \citealt{2016ApJ...831..169S}, \citealt{2018arXiv180310287M}, Banzatti et al. 2018), that the LVC-BC traces a higher density/hotter region than the LVC-NC, probably the base of an MHD disk wind. Finally, we could show that thermally excited gas with temperatures 5,000$-$10,000\,K and electron densities $\sim 10^7-10^8$\,cm$^{-3}$ can explain most of the LVC-BC and NC line ratios while the HVC ratios are best reproduced by radiative shock models where forbidden lines arise in the hot ($\sim$7,000$-$10,000\,K) but less dense (total H density $\sim 10^6$\,cm$^{-3}$) post-shock cooling zone (\sect~\ref{Sect:LVC_lineratio} and ~\ref{Sect:shock_HVC}).

Armed with the physical properties of the emitting gas in each component, we computed mass loss rates for the LVC-BC, NC, and HVC (\sect~\ref{Sect:mloss_all}). To evaluate the efficiency of winds in removing disk mass we also computed mass accretion rates by converting the luminosity of several permitted lines covered in our spectra to the accretion luminosity (\sect~\ref{sect:Macc}). For the HVC, we relied on radiative shock models that convert a fraction of the mechanical energy of the pre-shock gas into the \OIa\ luminosity. We found relatively low $\dot{M}_{\rm jet}/\dot{M}_{\rm acc}$  with a median value $\sim$0.1 for typical shock velocities of 30\,\kms, but this ratio could be as low as 0.01 for a high shock velocity of 100\,\kms\ or as high as 0.3 for a low shock velocity of 20\,\kms. These values are similar to those previously reported in the literature, which were instead calculated  assuming a collisional excitation model, and hence sensitive to several not well constrained quantities such as gas temperature and electron density in addition to jet velocity (e.g., \citealt{1995ApJ...452..736H}, \citealt{2018A&A...609A..87N}, and \sect~\ref{sect:Mloss_HVC}).

A first estimate for mass loss rates for the LVC, assuming a spherical outflow geometry, comes from \cite{2014A&A...569A...5N} who find  $\dot{M}_{\rm wind}/\dot{M}_{\rm acc}$ between 0.1 and 1 for their sample of T~Tauri stars in Lupus and $\sigma$~Ori. These values are on the high end of the HVC $\dot{M}_{\rm jet}/\dot{M}_{\rm acc}$ ratios, suggesting that the LVC may have a higher mass loss rate than the jet. Our work is the first to provide $\dot{M}_{\rm wind}/\dot{M}_{\rm acc}$ separately for the BC and NC. One important finding is that the mass loss rate from the BC exceeds that from the NC by at least a factor of 5. A comparison of the mass loss rates between the BC and HVC is more uncertain. To this end we explored the sensitivity of wind mass loss rates in thermally excited gas to the parameters that are the least constrained and most affect the estimates, namely the gas temperature and $l_{\rm wind}$ (\sect~\ref{sect:Mloss_LVC}). 
For the BC, a wind extent equal to the radial extent inferred from Keplerian broadening of the \OIa\ lines ($f=1$ Fig.~\ref{Fig:mloss_LVC} left panel) provides unreasonably large $\dot{M}_{\rm wind}/\dot{M}_{\rm acc}$, i.e. larger than 1, for most sources. Most likely the vertical extent of the BC, which remains unconstrained from our data, is larger than the Keplerian radius inferred from the BC FWHM. If we assume a vertical extent comparable to the NC radius inferred from Keplerian broadening ($f=1$ Fig.~\ref{Fig:mloss_LVC} right panel and $f=10$  Fig.~\ref{Fig:mloss_LVC} left panel) the median $\dot{M}_{\rm wind}/\dot{M}_{\rm acc}$ from the BC ranges from $\sim$0.05 for gas as hot as 10,000\,K up to $\sim$0.4 for gas at 5,000\,K. As already discussed in \sect~\ref{sect:Mloss_LVC}, a temperature of 10,000\,K is unlikely for the LVC gas, hence higher mass loss rates are more realistic.  However, given the sensitivity of  $\dot{M}_{\rm wind}/\dot{M}_{\rm acc}$ on $l_{\rm wind}$ we cannot conclusively state that the LVC-BC is carrying away more mass than the HVC. 
What is more certain is that the NC $\dot{M}_{\rm wind}/\dot{M}_{\rm acc}$ is lower than the BC for reasonable $l_{\rm winds}$ by at least a factor of 5. Thus, most of the disk mass appears to be lost close to the central star in the region traced by the BC.

How do our results compare with theoretical models of winds? Both photoevaporative and MHD winds have been proposed to explain forbidden line emission. However, as already pointed out by \cite{2016ApJ...831..169S} and confirmed by \cite{2018arXiv180310287M} and Banzatti~et~al.~(2018), the only component that may be tracing a thermal photoevaporative wind is the LVC-NC, as the BC is formed well inside the gravitational potential well of the star. Our finding that most NC ratios can be reproduced by thermally excited gas and that the line luminosity correlates with accretion luminosity agree with the most recent predictions from X-ray-driven photoevaporative winds \citep{2016MNRAS.460.3472E}. In these models, stellar EUV photons heat the forbidden line emitting region and, due to their low penetration depth, constrain it to a thin vertically extended (up to $\sim$35\,au) zone above the inner disk (see e.g. their Fig.~4). In addition, \citet{2016MNRAS.460.3472E} claim that most of EUV luminosity comes from accretion and that would lead to a correlation between the [O{\scriptsize I}] luminosity and $L_{\rm acc}$. 
The vertical extent of the emitting region results in a wide range of wind velocities that can broaden the [O{\scriptsize I}] lines up to $\sim$30\,\kms\, for close to edge-on disks while keeping moderate blueshifts of up to $\sim$7\,\kms\, for disks inclined by $\sim 50^\circ$ with respect to the observer\footnote{Note that the $l_{\rm wind}$ in these models is so large, corresponding to $f=20$ with respect to the NC $r_{\rm base}$, that the wind mass loss is too low to remove any significant mass}.
While the predicted FWHM and peak centroids are consistent with most of the LVC-NC values we report here, more detailed kinematic comparisons disfavor a thermal outflow origin because several \OIa\ widths are still larger than predicted (\citealt{2018arXiv180310287M}, Banzatti et al. 2018), the largest
observed blueshifts occur for a lower disk inclination than predicted, and the centroids of the BC and NC are correlated (see Banzatti et al. 2018 for further details).  For these reasons we favor a similar origin for both the BC and the NC in an MHD wind.

MHD wind models come in two broad categories: (1) winds arising close to the co-rotation radius ($\leq$0.1\,au), such as X-winds (e.g., \citealt{2007prpl.conf..261S}), conical winds (e.g., \citealt{2009MNRAS.399.1802R}), magnetospheric ejections (e.g., \citealt{2009A&A...508.1117Z}) or accretion powered stellar winds (e.g., \citealt{2005ApJ...632L.135M}) and (2) winds arising over a broad range of radii in the disk, from 0.1 to 10\,au, sometimes called D-winds (e.g., \citealt{2007prpl.conf..277P}). A key difference between these two genres of models is that the former are important in removing angular momentum from the accreting star, enabling it to spin down, and the latter in removing angular momentum from the disk itself, enabling accretion to occur through the disk rather than relying on some kind of magnetic viscosity (e.g., MRI). Note that these types of winds are not mutually exclusive.
All of these scenarios can reproduce the terminal velocity of the fast ($\sim 100$\,\kms) outflowing gas traced by the HVC. However, only D-winds can account for the high angular resolution observations showing a decrease in outflow velocity with increasing jet radius, implying an "onion-like" layering of a faster flow engirdled by a slower flow (e.g., \citealt{2000ApJ...537L..49B,2002ApJ...576..222B}, \citealt{2002ApJ...570..724P,2003ApJ...590..340P}, \citealt{2004A&A...416..213T}, \citealt{2010ApJ...722.1360B}, \citealt{2011A&A...532A..59A}). Also D-winds provide a natural explanation for possible rotation signatures in jets, implying wind footpoints spanning a large range of disk radii, from  $\sim$0.1 to 5\,au (e.g., \citealt{2006A&A...453..785F}, \citealt{2009ASSP...13..247C}). 

Our results provide strong support for the presence of D-winds, although they do not help in determining whether D-winds are the dominant contributor to the jet traced by the HVC, or whether the jet is primarily fed by an MHD wind from near the co-rotation radius. {\rev However, the Keplerian radii for the BC infered from their FWHM are larger than the co-rotation radii ($\sim$0.05\,au), and the ones for NCs are well beyond the  co-rotation radii, implying a large range of wind radii (from $\sim$0.05 out to 5\,au), similar to those expected in D-winds and inferred from jet rotation signatures.} The observed correlation between the [O~{\scriptsize I}] line and accretion luminosity might indicate that the wind truly drives accretion: as the \OIa\ line is optically thin a higher luminosity implies more mass removed in the wind, and hence more angular momentum promoting mass accretion from the disk onto the central star.

Another prediction from some D-wind models is that most of the mass is lost within a few au from the central star (e.g., \citealt{1992ApJ...394..117P}), in agreement with our finding that the BC has larger $\dot{M}_{\rm wind}/\dot{M}_{\rm acc}$ than the NC. This last result also places interesting constraints on more recent disk simulations that include non-ideal MHD effects (e.g., \citealt{2014ApJ...791..137B}, \citealt{2015ApJ...801...84G}, \citealt{2017A&A...600A..75B} ). 
These simulations show that disk winds are launched from radii extending out to $\sim 10-20$\,au with vigorous mass loss rates out to these radii (see also \citealt{2017ApJ...845...75B}). 
However, even for gas at 5,000\,K we obtain an $\dot{M}_{\rm wind}/\dot{M}_{\rm acc}$ of 0.4 from the upper quartile of the $f=1$ LVC-NC curve (Fig.~\ref{Fig:mloss_LVC} right panel) suggesting that the mass loss rate is not as large as predicted out to the $\sim$5\,au radii we can trace with the NC. If the theoretically predicted winds exist, most of the gas must be cooler ($< 5,000$K) than what can be traced with the forbidden lines analyzed here. Other wind diagnostics that can probe cooler neutral hydrogen in the flow would be valuable in determining the presence of more extended disk winds.

\section{Summary}
We analyzed optical high-resolution spectra, covering the \SII, \OIb, and \OIa\ forbidden lines, from a sample of 48 T~Tauri stars,   31 of which are surrounded by full disks while 17 by TDs. We detected the \OIa\  from 45 sources, the \OIb\  from 26 sources, and the \SII\  from 22 sources.  Following \cite{2016ApJ...831..169S}, we decomposed the line profiles into HVC, LVC-NC, and LVC-BC. 
As in previous studies, we attribute the line width primarily to Keplerian broadening, with forbidden emission arising both in the inner (BC within 0.5\,au) and outer (NC within 1$-$5\,au) disk. Many components show  peak centroids blueshifted by more than 1.5\,\kms\  with respect to the stellar velocity. Thus, the LVC is most likely tracing unbound slow wind gas. After flux calibrating our spectra we derived line luminosity and line ratios for individual kinematic components to assess the properties of the emitting gas and with the goal of measuring mass loss rates. We also estimated mass accretion rates using 12 accretion-related permitted lines to then evaluate the mass loss over mass accretion rate.
Our main results can be summarized as follows:
\begin{enumerate}
\item About 72\% of full disks present a HVC, while this number is down to 13\% for TDs. Furthermore, TDs more frequently than full disks show only a LVC-NC (44\% for TDs vs. 28\% for the full disks), while full disks tend to show line profiles with multiple components. These findings point to a depletion of gas in the inner disk of TDs, in agreement with their lower average mass accretion rate.
  \item The HVC and LVC \OIa\ luminosity is confirmed to correlate with the 
accretion luminosity. When the \OIa\ LVC is decomposed into LVC-BC and NC, we see that the LVC-BC luminosity is more tightly correlated with accretion luminosity than the NC. 
 \item The profiles of individual kinematic components are similar in the \SII, \OIb, and \OIa\ lines, when detected. This indicates that they trace a similar region, hence their line ratios can be used to infer the physical properties of the emitting gas. 
 \item The HVC has statistically different SII40/OI63 ratios than the LVC-BC and NC. Its OI55/63 ratios are also statistically different from the LVC-BC, while they are not distinguishable from the LVC-NC. These differences and the mean line ratios corroborate previous suggestions that the BC traces hot/dense gas close to the base of an MHD wind.
 \item Most LVC-BC and NC ratios can be explained by thermally excited gas with electron densities of $\sim10^{7}-10^{8}$\,cm$^{-3}$ and temperatures 5,000$-$10,000~K. HVC line ratios are better explained by shock models with a pre-shock number density of nucleons of $\sim10^{6}-10^{7}$cm$^{-3}$. 
\item Building on the \OIa\ HVC luminosity,  converting a fraction of the shock mechanical energy into radiant energy, and adopting a typical shock velocity of 30\,\kms, we find a median $\dot{M}_{\rm jet}$/$\dot{M}_{\rm acc}$  ratio of 0.1. This value is comparable to previous results from the literature which were calculated with a very different approach. The agreement between these different approaches to convert HVC luminosity into a mass outflow rate suggests this ratio may be robust. An $\dot{M}_{\rm jet}$/$\dot{M}_{\rm acc}$ of  0.1 reflects a minimal impact on disk mass removal by outflowing material seen in the jet.  
\item  Building on our finding that the [OI] LVC emission is most likely thermal, gas is in LTE, and the \OIa\ line is optically thin, we estimate for the first time $\dot{M}_{\rm wind}$/$\dot{M}_{\rm acc}$  ratios separately for the BC and NC.  We show that $\dot{M}_{\rm wind}$ is a factor of 5 higher in the BC (inner disk) than the NC (outer disk). However absolute $\dot{M}_{\rm wind}$ values are sensitive to the gas temperature and wind height, making direct comparisons between outflow rates in the LVC and the HVC uncertain. However, for plausible wind heights, we find that the mass flowing out in the BC is likely at least as large as in the HVC and may be considerably higher. If so, then the inner disk wind traced by the BC may play an important role in the evolution of the disk mass.
\end{enumerate}
 Taken together our results favor D-wind models. In particular, we find that winds are launched from a range of disk radii beyond the gas co-rotation radius even in the LVC-BC alone. In addition, most of the mass is lost close to the star, within a few au. For plausible wind heights,  LVC-NC $\dot{M}_{\rm wind}$/$\dot{M}_{\rm acc}$ ratios are lower than $\sim 0.4$, and thus lower than the ratios predicted by recent non-ideal MHD simulations for radii $\sim 1-20$\,au (e.g. \citealt{2017ApJ...845...75B}). Additional wind diagnostics tracing cooler gas would be helpful to test  if indeed radially extended   winds could drive accretion.

\acknowledgements
We thank Kelle Cruz and Scott Dahm for doing the observations with Keck/HIRES. I.P., U.G., and S.E. acknowledge support from a Collaborative NSF Astronomy \& Astrophysics Research Grant (ID: 1715022, ID:1713780, and ID:1714229). This material is based upon work supported by the National Aeronautics and Space Administration under Agreement No. NNX15AD94G for the program "Earths in Other Solar Systems". The results reported herein benefitted from collaborations and/or information exchange within NASA Nexus for Exoplanet System Science (NExSS) research coordination network sponsored by NASA's Science Mission Directorate. The data presented here were obtained at the W. M. Keck Observatory, which is operated as a scientific partnership among the California Institute of Technology, the University of California and the National Aeronautics and Space Administration. The Observatory was made possible by the generous financial support of the W. M. Keck Foundation. The authors wish to recognize and acknowledge the very significant cultural role and reverence that the summit of Mauna Kea has always had within the indigenous Hawaiian community.  We are most fortunate to have the opportunity to conduct observations from this mountain. This research has made use of the Keck Observatory Archive (KOA), which is operated by the W. M. Keck Observatory and the NASA Exoplanet Science Institute (NExScI), under contract with the National Aeronautics and Space Administration.
\facility{Keck:I (HIRES)}

\setcounter{table}{1}
%\scriptsize
\begin{table*}[hp]
%\begin{rotatetable}
\scriptsize
\begin{center}
\renewcommand{\tabcolsep}{0.03cm}    
\caption{Parameters for line decomposition of \SII, \OIb, and \OIa\ line profiles}\label{Tab:para_LVC1}
\begin{tabular}{ccccccccccccccccccccccccccccccccccccc}
\hline
   &   &  \multicolumn{4}{c}{\SII}           &~~~~   & \multicolumn{4}{c}{\OIb}    &~~~~   & \multicolumn{4}{c}{\OIa} &\\
      \cline{3-6} \cline{8-11} \cline{13-16} 
   &   & $FWHM$ &$v_{\rm c}$ &$EW$   &Log~$L_{\rm 4069}$  & &$FWHM$   &$v_{\rm c}$ &$EW$ &Log~$L_{\rm 5577}$ &  &$FWHM$   &$v_{\rm c}$ &$EW$   &Log~$L_{\rm 6300}$ &preliminary &refined\\
ID&Name  &(\kms)&(\kms)    &(\AA) &($L_{\odot}$)            & &(\kms) &(\kms)     &(\AA) &($L_{\odot}$)& &(\kms) &(\kms)     &(\AA) &($L_{\odot}$) &Class &Class \\
\hline     
1 & DPTau&   62.8&$  -91.9$&   1.01&$  -5.29$& &   45.4&$  -90.6$&   0.08&$  -6.32$& &   64.4&$  -89.3$&   0.89&$  -5.24$&HVC-B&    \\
  &     &  115.6&$   31.2$&   5.92&$  -4.52$& &  116.7&$   13.6$&   0.47&$  -5.54$& &  111.8&$   23.5$&   6.03&$  -4.41$&LVC-BC&HVC-R\\
  &     &   27.3&$    2.4$&   1.23&$  -5.20$& &   24.4&$    0.4$&   0.60&$  -5.44$& &   22.5&$    0.8$&   3.99&$  -4.59$&LVC-NC&    \\
\hline
2 & CXTau&\nodata&\nodata&\nodata&\nodata& &\nodata&\nodata&\nodata&$<  -6.52$& &   27.9&$   -2.6$&   0.05&$  -6.40$&LVC-NC&    \\
\hline
3 & FPTau&\nodata&\nodata&\nodata&$<  -5.51$& &   84.6&$    1.4$&   0.14&$  -6.14$& &   66.0&$    1.1$&   0.38&$  -5.68$&LVC-BC&    \\
\hline
4 & FNTau&   13.8&$  -44.1$&   0.09&$  -6.18$& &\nodata&\nodata&\nodata&$<  -6.47$& &   10.6&$  -43.0$&   0.03&$  -6.25$&HVC-B&    \\
  &     &   15.3&$ -117.8$&   0.14&$  -5.99$& &\nodata&\nodata&\nodata&$<  -6.36$& &   13.7&$ -121.6$&   0.06&$  -6.00$&HVC-B&    \\
  &     &   40.5&$  -79.2$&   0.95&$  -5.17$& &\nodata&\nodata&\nodata&$<  -5.91$& &   38.3&$  -77.4$&   0.47&$  -5.11$&HVC-B&    \\
  &     &   30.4&$  -11.1$&   0.27&$  -5.72$& &   22.8&$  -12.3$&   0.02&$  -6.45$& &   26.9&$  -12.2$&   0.34&$  -5.26$&LVC-NC&    \\
\hline
5 & V409Tau&  113.3&$  -85.4$&   0.92&$  -5.58$& &\nodata&\nodata&\nodata&$<  -5.81$& &  107.6&$  -64.8$&   0.22&$  -5.50$&HVC-B&    \\
  &     &   70.8&$   42.5$&   0.29&$  -6.08$& &\nodata&\nodata&\nodata&$<  -5.95$& &   79.3&$   51.5$&   0.10&$  -5.83$&HVC-R&    \\
\hline
6 & BPTau&   94.7&$   11.3$&   0.36&$  -5.22$& &   88.3&$    4.0$&   0.17&$  -5.24$& &   98.9&$   12.4$&   0.26&$  -5.04$&LVC-BC&    \\
  &     &   37.5&$  -18.2$&   0.11&$  -5.73$& &   27.5&$   -6.1$&   0.03&$  -6.01$& &   26.3&$   -2.6$&   0.10&$  -5.44$&LVC-NC&    \\
\hline
7 & DKTau&   45.1&$ -125.6$&   0.16&$  -5.30$& &\nodata&\nodata&\nodata&$<  -5.69$& &   41.0&$ -135.7$&   0.09&$  -5.27$&HVC-B&    \\
  &     &  124.2&$  -18.1$&   0.29&$  -5.05$& &   96.8&$   -1.4$&   0.16&$  -5.02$& &  153.7&$  -27.4$&   0.42&$  -4.59$&LVC-BC&    \\
  &     &\nodata&\nodata&\nodata&$<  -5.79$& &   14.2&$    1.0$&   0.02&$  -5.88$& &   24.9&$   -5.1$&   0.13&$  -5.09$&LVC-NC&    \\
\hline
8 & HNTau&  116.2&$  -78.5$&   2.80&$  -4.25$& &  183.1&$  -57.9$&   0.23&$  -5.19$& &  132.4&$  -67.7$&   2.99&$  -4.07$&HVC-B&    \\
  &     &   60.2&$   -9.6$&   1.44&$  -4.54$& &   45.3&$   -0.0$&   0.10&$  -5.56$& &   50.4&$   -6.5$&   1.33&$  -4.42$&LVC-BC&HVC-B\\
\hline
9 & UXTau&\nodata&\nodata&\nodata&$<  -5.20$& &\nodata&\nodata&\nodata&$<  -5.18$& &   34.6&$   -1.2$&   0.11&$  -4.70$&LVC-NC&    \\
\hline
10 & GKTau&  160.6&$  -22.4$&   0.36&$  -4.87$& &  116.6&$  -35.5$&   0.04&$  -5.57$& &  179.1&$  -30.7$&   0.38&$  -4.58$&HVC-B&    \\
  &     &   36.0&$   -1.0$&   0.13&$  -5.31$& &   51.5&$    3.1$&   0.05&$  -5.48$& &   38.6&$   -3.3$&   0.17&$  -4.94$&LVC-NC&    \\
\hline
11 & GITau&   86.3&$  -51.9$&   0.20&$  -4.79$& &   38.9&$  -52.6$&   0.03&$  -5.73$& &   85.9&$  -67.5$&   0.19&$  -4.85$&HVC-B&    \\
  &     &   33.6&$   46.3$&   0.15&$  -4.91$& &   54.1&$   40.6$&   0.05&$  -5.44$& &   31.7&$   46.7$&   0.17&$  -4.90$&HVC-R&    \\
  &     &   39.3&$    6.2$&   0.08&$  -5.20$& &   59.0&$  -11.5$&   0.06&$  -5.38$& &   44.7&$    1.3$&   0.16&$  -4.92$&LVC-BC&    \\
\hline
12 & DMTau&\nodata&\nodata&\nodata&$<  -6.17$& &   24.8&$   -2.0$&   0.05&$  -6.51$& &   26.1&$    0.5$&   0.35&$  -5.69$&LVC-NC&    \\
\hline
13 & LKCa15&\nodata&\nodata&\nodata&$<  -5.47$& &\nodata&\nodata&\nodata&$<  -5.54$& &   44.6&$   -0.6$&   0.14&$  -5.06$&LVC-BC&    \\
\hline
14 & DSTau&\nodata&\nodata&\nodata&$<  -4.86$& &\nodata&\nodata&\nodata&$<  -4.93$& &  151.1&$   16.3$&   0.19&$  -5.05$&LVC-BC&    \\
\hline
15 & Sz65A&\nodata&\nodata&\nodata&$<  -4.92$& &\nodata&\nodata&\nodata&$<  -4.87$& &  109.9&$    3.6$&   0.25&$  -4.74$&LVC-BC&    \\
\hline
16 & Sz68A&\nodata&\nodata&\nodata&$<  -6.55$& &\nodata&\nodata&\nodata&$<  -5.49$& &\nodata&\nodata&\nodata&$<  -5.50$&\nodata&\nodata\\
\hline
17 & Sz73&   82.5&$  -88.4$&   0.70&$  -4.89$& &\nodata&\nodata&\nodata&$<  -5.39$& &   71.1&$  -94.4$&   1.67&$  -4.43$&HVC-B&    \\
  &     &\nodata&\nodata&\nodata&$<  -5.09$& &\nodata&\nodata&\nodata&$<  -5.27$& &   94.2&$   64.2$&   0.25&$  -5.26$&HVC-R&    \\
  &     &\nodata&\nodata&\nodata&$<  -5.35$& &   41.1&$  -11.9$&   0.11&$  -5.66$& &   50.8&$  -19.2$&   0.69&$  -4.82$&LVC-BC&    \\
\hline
18 & HMLup&   68.3&$ -119.7$&   0.50&$  -5.78$& &\nodata&\nodata&\nodata&$<  -5.70$& &   90.4&$ -116.9$&   0.59&$  -5.38$&HVC-B&    \\
  &     &\nodata&\nodata&\nodata&$<  -6.41$& &\nodata&\nodata&\nodata&$<  -6.31$& &   22.1&$   -3.8$&   0.22&$  -5.81$&LVC-NC&    \\
\hline
19 & GWLup&\nodata&\nodata&\nodata&$<  -5.99$& &   68.8&$   -7.5$&   0.09&$  -5.93$& &   69.4&$   -9.7$&   0.29&$  -5.41$&LVC-BC&    \\
\hline
20 & GQLup&\nodata&\nodata&\nodata&$<  -4.56$& &   46.1&$   10.7$&   0.04&$  -5.04$& &   91.4&$    4.0$&   0.28&$  -4.18$&LVC-BC&    \\
\hline
21 & Sz76&\nodata&\nodata&\nodata&$<  -5.80$& &   69.9&$   31.7$&   0.22&$  -5.86$& &   50.1&$   38.8$&   0.39&$  -5.66$&HVC-R&    \\
  &     &   77.9&$  -16.0$&   0.48&$  -5.64$& &   53.1&$  -26.4$&   0.24&$  -5.81$& &   65.5&$  -18.3$&   1.21&$  -5.16$&LVC-BC&    \\
\hline
22 & RULup&  130.0&$ -137.8$&   1.68&$  -3.68$& &\nodata&\nodata&\nodata&$<  -4.67$& &  109.9&$ -149.3$&   0.61&$  -4.01$&HVC-B&    \\
  &     &   68.5&$  -38.9$&   0.26&$  -4.49$& &\nodata&\nodata&\nodata&$<  -5.10$& &   40.5&$  -32.3$&   0.07&$  -4.92$&HVC-B&    \\
  &     &  150.3&$  -13.1$&   0.43&$  -4.28$& &  120.5&$  -11.7$&   0.21&$  -4.47$& &  176.4&$  -10.1$&   0.41&$  -4.18$&LVC-BC&    \\
  &     &\nodata&\nodata&\nodata&$<  -5.21$& &\nodata&\nodata&\nodata&$<  -5.57$& &   13.9&$  -10.9$&   0.06&$  -5.02$&LVC-NC&    \\
\hline
23 & IMLup&\nodata&\nodata&\nodata&$<  -4.98$& &\nodata&\nodata&\nodata&$<  -4.83$& &  117.1&$  -10.0$&   0.16&$  -4.79$&LVC-BC&    \\
\hline
24 & RYLup&\nodata&\nodata&\nodata&$<  -5.18$& &\nodata&\nodata&\nodata&$<  -5.13$& &   30.7&$    2.5$&   0.09&$  -4.64$&LVC-NC&    \\
\hline
25 & Sz102&  206.1&$  -32.5$&  20.69&$  -4.54$& &\nodata&\nodata&\nodata&$<  -5.54$& &  209.3&$  -38.1$&  22.13&$  -4.63$&HVC-B&    \\
  &     &   58.3&$  -25.8$&  17.62&$  -4.61$& &   30.7&$  -16.0$&   0.89&$  -5.92$& &   40.0&$  -24.5$&  17.99&$  -4.72$&LVC-NC&HVC-B\\
  &     &   85.3&$   20.6$&  53.70&$  -4.13$& &  108.9&$    5.9$&   4.45&$  -5.22$& &   93.6&$   11.8$& 127.33&$  -3.87$&LVC-BC&HVC-R\\
\hline
\end{tabular}  
\end{center}
\normalsize
%\end{rotatetable}
\end{table*}  
\normalsize 

\setcounter{table}{1}
\begin{table*}[hp]
%\begin{rotatetable}
\begin{center}
\scriptsize
\renewcommand{\tabcolsep}{0.03cm}    
\caption{Parameters for line decomposition of \SII, \OIb, and \OIa\ line profiles}\label{Tab:para_LVC2}
\begin{tabular}{ccccccccccccccccccccccccccccccccccccc}
\hline
   &   &  \multicolumn{4}{c}{\SII}           &~~~~   & \multicolumn{4}{c}{\OIb}    &~~~~   & \multicolumn{4}{c}{\OIa} &\\
      \cline{3-6} \cline{8-11} \cline{13-16} 
   &   & $FWHM$ &$v_{\rm c}$ &$EW$   &Log~$L_{\rm 4069}$  & &$FWHM$   &$v_{\rm c}$ &$EW$ &Log~$L_{\rm 5577}$ &  &$FWHM$   &$v_{\rm c}$ &$EW$   &Log~$L_{\rm 6300}$ &preliminary &refined\\
ID&Name  &(\kms)&(\kms)    &(\AA) &($L_{\odot}$)            & &(\kms) &(\kms)     &(\AA) &($L_{\odot}$)& &(\kms) &(\kms)     &(\AA) &($L_{\odot}$) &Class &Class \\
\hline
26 & Sz111&\nodata&\nodata&\nodata&$<  -6.16$& &\nodata&\nodata&\nodata&$<  -5.91$& &   28.0&$   -0.4$&   0.41&$  -5.24$&LVC-NC&    \\
\hline
27 & Sz98&  167.4&$  -50.9$&   2.05&$  -4.46$& &  108.9&$  -12.4$&   0.21&$  -5.13$& &  152.5&$  -37.7$&   1.07&$  -4.40$&HVC-B&    \\
  &     &   30.4&$   -8.1$&   0.21&$  -5.46$& &\nodata&\nodata&\nodata&$<  -5.75$& &   28.7&$   -6.7$&   0.12&$  -5.35$&LVC-NC&    \\
\hline
28 & EXLup&\nodata&\nodata&\nodata&\nodata& &\nodata&\nodata&\nodata&$<  -4.98$& &   25.1&$ -119.3$&   0.08&$  -4.53$&HVC-B&    \\
  &     &   84.3&$ -181.2$&   0.17&$  -3.86$& &\nodata&\nodata&\nodata&$<  -4.48$& &   79.6&$ -179.2$&   0.26&$  -4.03$&HVC-B$^{a}$&    \\
  &     &  121.9&$  -88.0$&   0.81&$  -3.19$& &\nodata&\nodata&\nodata&$<  -4.69$& &   48.9&$  -79.5$&   0.14&$  -4.29$&HVC-B&    \\
  &     &\nodata&\nodata&\nodata&\nodata& &\nodata&\nodata&\nodata&$<  -4.88$& &   31.5&$  102.8$&   0.06&$  -4.64$&HVC-R&    \\
  &     &\nodata&\nodata&\nodata&\nodata& &\nodata&\nodata&\nodata&$<  -4.88$& &   31.1&$   56.9$&   0.11&$  -4.41$&HVC-R&    \\
  &     &   37.8&$  -15.0$&   0.10&$  -4.11$& &   65.1&$  -15.8$&   0.07&$  -4.53$& &   59.6&$  -17.6$&   0.45&$  -3.79$&LVC-BC&    \\
\hline
29 & As205A&\nodata&\nodata&\nodata&\nodata& &\nodata&\nodata&\nodata&$<  -5.09$& &   56.3&$ -221.9$&   0.11&$  -4.69$&HVC-B&    \\
  &     &  154.2&$ -112.1$&   0.29&$  -4.29$& &\nodata&\nodata&\nodata&$<  -4.49$& &  225.5&$  -85.4$&   0.28&$  -4.27$&HVC-B$^{a}$&    \\
  &     &   60.4&$  -19.0$&   0.12&$  -4.67$& &   50.8&$   -0.7$&   0.03&$  -5.18$& &   54.5&$  -13.5$&   0.16&$  -4.50$&LVC-BC&    \\
  &     &   18.9&$    1.5$&   0.03&$  -5.36$& &   12.7&$    1.1$&   0.02&$  -5.40$& &   14.5&$   -0.7$&   0.11&$  -4.66$&LVC-NC&    \\
\hline
30 & DoAr21&\nodata&\nodata&\nodata&\nodata& &\nodata&\nodata&\nodata&$<  -4.31$& &\nodata&\nodata&\nodata&$<  -4.51$&\nodata& \\
\hline
31 & DoAr24E&\nodata&\nodata&\nodata&\nodata& &\nodata&\nodata&\nodata&$<  -5.69$& &\nodata&\nodata&\nodata&$<  -5.62$&\nodata& \\
\hline
32 & DoAr44&\nodata&\nodata&\nodata&$<  -5.04$& &   35.0&$    3.8$&   0.02&$  -5.68$& &   59.6&$   -1.1$&   0.15&$  -4.82$&LVC-BC&    \\
\hline
33 & SR21A&\nodata&\nodata&\nodata&$<  -3.29$& &\nodata&\nodata&\nodata&$<  -4.45$& &   16.3&$   -6.9$&   0.01&$  -4.81$&LVC-NC&    \\
\hline
34 & V853Oph&   30.0&$  -33.8$&   1.26&$  -5.27$& &\nodata&\nodata&\nodata&$<  -6.10$& &   28.5&$  -31.3$&   0.93&$  -5.00$&HVC-B&    \\
  &     &   68.4&$  -16.0$&   1.19&$  -5.30$& &   37.5&$  -12.4$&   0.24&$  -5.62$& &   70.2&$  -14.4$&   1.31&$  -4.85$&LVC-BC&    \\
  &     &\nodata&\nodata&\nodata&$<  -6.52$& &\nodata&\nodata&\nodata&$<  -6.33$& &   16.6&$   -1.6$&   0.31&$  -5.48$&LVC-NC&    \\
\hline
35 & RNO90&\nodata&\nodata&\nodata&$<  -4.14$& &   56.1&$   -3.0$&   0.06&$  -4.73$& &   57.5&$   -3.4$&   0.23&$  -4.17$&LVC-BC&    \\
\hline
36 & V2508Oph&\nodata&\nodata&\nodata&$<  -4.49$& &\nodata&\nodata&\nodata&$<  -4.67$& &  116.2&$  -38.4$&   0.11&$  -4.85$&HVC-B&    \\
\hline
37 & V1121Oph&\nodata&\nodata&\nodata&$<  -5.64$& &\nodata&\nodata&\nodata&$<  -5.54$& &   21.1&$   -7.2$&   0.03&$  -5.48$&LVC-NC&    \\
\hline
38 & J1842&   36.5&$ -122.1$&   0.06&$  -5.60$& &\nodata&\nodata&\nodata&$<  -5.43$& &   45.6&$ -126.3$&   0.10&$  -5.18$&HVC-B&    \\
  &     &   83.9&$  -16.4$&   0.28&$  -4.94$& &   93.0&$   -3.1$&   0.08&$  -5.29$& &  109.2&$  -11.5$&   0.35&$  -4.64$&LVC-BC&    \\
  &     &   25.9&$    1.8$&   0.08&$  -5.50$& &   23.4&$   -1.0$&   0.03&$  -5.71$& &   27.3&$   -0.5$&   0.23&$  -4.82$&LVC-NC&    \\
\hline
39 & J1852&\nodata&\nodata&\nodata&$<  -5.86$& &\nodata&\nodata&\nodata&$<  -5.88$& &   23.5&$   -2.4$&   0.09&$  -5.35$&LVC-NC&    \\
\hline
40 & VVCrA&\nodata&\nodata&\nodata&$<  -3.60$& &   85.2&$ -416.1$&   0.06&$  -5.06$& &  133.2&$ -361.6$&   0.78&$  -3.88$&HVC-B&    \\
  &     &  135.5&$ -209.3$&   1.88&$  -3.52$& &   94.7&$ -234.8$&   0.30&$  -4.37$& &  119.7&$ -205.4$&   1.58&$  -3.57$&HVC-B$^{a}$&    \\
  &     &  147.8&$  -88.8$&   3.89&$  -3.21$& &  101.5&$  -78.5$&   0.14&$  -4.71$& &  114.2&$  -87.0$&   1.73&$  -3.53$&HVC-B&    \\
  &     &   35.7&$  -22.5$&   0.48&$  -4.11$& &   51.4&$  -19.8$&   0.18&$  -4.59$& &   42.2&$  -22.1$&   0.69&$  -3.93$&LVC-BC&    \\
  &     &\nodata&\nodata&\nodata&$<  -4.34$& &\nodata&\nodata&\nodata&$<  -4.99$& &   18.1&$   -6.7$&   0.29&$  -4.30$&LVC-NC&    \\
\hline
41 & SCrAA+B&   44.1&$ -129.3$&   0.24&$  -4.67$& &\nodata&\nodata&\nodata&$<  -5.02$& &   45.6&$ -127.7$&   0.51&$  -3.91$&HVC-B&    \\
  &     &  132.7&$  -72.3$&   0.60&$  -4.27$& &  121.1&$  -40.4$&   0.09&$  -4.65$& &  128.0&$  -49.5$&   0.66&$  -3.80$&HVC-B&    \\
  &     &   58.5&$   59.1$&   0.07&$  -5.19$& &\nodata&\nodata&\nodata&$<  -4.85$& &   68.2&$   68.0$&   0.08&$  -4.70$&HVC-R&    \\
\hline
42 & TWHya&   15.9&$   -0.3$&   0.08&$  -6.29$& &    9.6&$    0.1$&   0.08&$  -5.89$& &   12.8&$    0.0$&   0.57&$  -5.03$&LVC-NC&    \\
\hline
43 & TWA3A&\nodata&\nodata&\nodata&$<  -7.03$& &   56.0&$   -1.1$&   0.11&$  -6.98$& &   61.9&$    1.3$&   0.59&$  -6.09$&LVC-BC&    \\
\hline
44 & V1057Cyg&\nodata&\nodata&\nodata&$<  -1.61$& &\nodata&\nodata&\nodata&$<  -2.51$& &   50.0&$ -154.6$&   0.23&$  -1.88$&HVC-B&    \\
  &     &\nodata&\nodata&\nodata&$<  -1.55$& &\nodata&\nodata&\nodata&$<  -2.45$& &   58.0&$  -98.6$&   0.24&$  -1.86$&HVC-B&    \\
  &     &\nodata&\nodata&\nodata&$<  -1.27$& &\nodata&\nodata&\nodata&$<  -2.17$& &  110.9&$  -60.0$&   0.25&$  -1.85$&HVC-B&    \\
\hline
45 & V1515Cyg&\nodata&\nodata&\nodata&$<  -3.05$& &\nodata&\nodata&\nodata&$<  -3.28$& &   32.1&$ -120.8$&   0.03&$  -3.45$&HVC-B&    \\
  &     &\nodata&\nodata&\nodata&$<  -2.72$& &\nodata&\nodata&\nodata&$<  -2.94$& &   69.9&$  -63.2$&   0.18&$  -2.73$&HVC-B&    \\
\hline
46 & HD143006&\nodata&\nodata&\nodata&$<  -4.37$& &\nodata&\nodata&\nodata&$<  -4.73$& &   66.3&$    2.8$&   0.09&$  -4.45$&LVC-BC&    \\
  &     &\nodata&\nodata&\nodata&$<  -4.91$& &\nodata&\nodata&\nodata&$<  -5.27$& &   19.1&$    2.1$&   0.02&$  -5.02$&LVC-NC&    \\
\hline
47 & DICep&\nodata&\nodata&\nodata&$<  -4.75$& &\nodata&\nodata&\nodata&$<  -4.69$& &   27.3&$   -7.5$&   0.11&$  -4.18$&LVC-NC&    \\
\hline
48 & As353A&\nodata&\nodata&\nodata&\nodata& &\nodata&\nodata&\nodata&$<  -4.77$& &   48.4&$ -280.0$&   0.21&$  -4.25$&HVC-B&    \\
  &     &  318.0&$ -128.6$&   1.96&$  -3.67$& &\nodata&\nodata&\nodata&$<  -3.96$& &  322.8&$ -126.6$&   0.94&$  -3.61$&HVC-B$^{a}$&    \\
  &     &   95.1&$  -43.2$&   0.41&$  -4.35$& &\nodata&\nodata&\nodata&$<  -4.66$& &   62.7&$  -27.9$&   0.13&$  -4.47$&LVC-BC&Suspicious\\
  &     &\nodata&\nodata&\nodata&\nodata& &\nodata&\nodata&\nodata&$<  -4.99$& &   29.1&$   -4.2$&   0.09&$  -4.62$&LVC-NC&    \\
\hline
\end{tabular}  
\end{center}
$^{a}$ The fitted parameters for the components from \SII\ are unreliable.
%\end{rotatetable}
\end{table*}  
\normalsize

\setcounter{table}{4}
\begin{table*}
\scriptsize
\renewcommand{\tabcolsep}{0.03cm}
\caption{The accretion luminosity and accretion rates of the sources in this work}\label{Table:acc_line}
\begin{center}
\begin{tabular}{lcccccccccccccccccccccccccccccc}
\hline
 &                   &\multicolumn{12}{ c}{Log~$L_{\rm acc}$}  &                                                       &\\
 \cline{3-14}
        &           &(H$\zeta$)              & (H$\delta$)       &(H$\gamma$)       &(H$\beta$)        &(H$\alpha$)    & (He {\scriptsize I}~$\lambda$4026)  & (He {\scriptsize I}~$\lambda$4471) &(He {\scriptsize I}~$\lambda$5876)  &(He {\scriptsize I}~$\lambda$6678) &(He {\scriptsize II}~$\lambda$4686)  &(Ca {\scriptsize II}~$\lambda$3934)&(Ca {\scriptsize II}~$\lambda$3968)  &Log~$\overline{L_{\rm acc}}$ &Log~$\dot{M}_{acc}$   \\
ID & Name                       &($L_{\odot}$)&($L_{\odot}$)&($L_{\odot}$)&($L_{\odot}$)&($L_{\odot}$)&($L_{\odot}$)&($L_{\odot}$)  &($L_{\odot}$)&($L_{\odot}$) &($L_{\odot}$)&($L_{\odot}$)   &($L_{\odot}$) &($L_{\odot}$) & ($M_{\odot}~yr^{-1}$)\\
\hline
\hline
1 &DPTau  &\nodata  &$-1.74$  &$-1.73$  &$-1.69$  &$-2.17$  &$-1.74$  &$-1.55$  &$-1.91$  &$-1.95$  &$-1.87$  &$-1.35$  &$-1.46$  &$-1.69$  &$ -9.07$  \\ 
2 &CXTau  &\nodata  &\nodata  &$-2.44$  &$-2.32$  &$-2.62$  &\nodata  &$-2.66$  &$-3.12$  &n  &n  &\nodata  &\nodata  &$-2.56$  &$ -9.37$  \\ 
3 &FPTau  &\nodata  &$-2.32$  &$-2.32$  &$-2.15$  &$-2.16$  &\nodata  &$-2.24$  &$-2.53$  &$-2.23$  &$-2.31$  &\nodata  &\nodata  &$-2.27$  &$ -9.17$  \\ 
4 &FNTau  &\nodata  &$-1.50$  &$-1.47$  &$-1.48$  &$-1.87$  &$-1.58$  &$-1.44$  &$-1.33$  &$-1.12$  &$-1.92$  &\nodata  &\nodata  &$-1.53$  &$ -7.97$  \\ 
5 &V409Tau  &\nodata  &n  &n  &n  &a  &n  &n  &n  &n  &n  &$-1.53$  &$-1.79$  &$-1.64$  &$ -8.44$  \\ 
6 &BPTau  &$-0.99$  &$-1.16$  &$-1.19$  &$-1.16$  &$-1.24$  &$-1.25$  &$-1.24$  &$-1.22$  &$-1.18$  &$-1.14$  &$-1.13$  &$-1.02$  &$-1.17$  &$ -8.14$  \\ 
7 &DKTauA  &$-0.67$  &$-0.89$  &$-0.87$  &$-0.76$  &$-1.18$  &$-0.79$  &$-0.98$  &$-0.74$  &$-0.72$  &$-0.74$  &$-0.76$  &$-0.64$  &$-0.79$  &$ -7.86$  \\ 
8 &HNTauA  &\nodata  &$-1.10$  &$-1.26$  &$-1.00$  &$-1.92$  &$-0.54$  &$-0.78$  &$-1.46$  &$-0.85$  &$-1.10$  &$-0.89$  &$-0.67$  &$-0.93$  &$ -8.37$  \\ 
9 &UXTauA  &\nodata  &n  &n  &n  &$-1.56$  &n  &n  &n  &n  &n  &$-1.16$  &$-1.46$  &$-1.51$  &$ -8.81$  \\ 
10 &GKTauA  &\nodata  &$-1.31$  &$-1.56$  &$-1.34$  &$-1.64$  &$-1.41$  &$-1.47$  &$-1.16$  &n  &$-1.30$  &$-0.76$  &$-0.88$  &$-1.38$  &$ -8.35$  \\ 
11 &GITau  &\nodata  &$-0.55$  &$-0.59$  &$-0.89$  &$-1.46$  &$-0.71$  &$-0.68$  &$-0.78$  &$-0.75$  &$-0.64$  &\nodata  &\nodata  &$-0.69$  &$ -7.61$  \\ 
12 &DMTau  &\nodata  &$-1.92$  &$-1.94$  &$-1.89$  &$-1.92$  &$-2.43$  &$-1.91$  &$-1.91$  &$-2.33$  &$-1.93$  &\nodata  &\nodata  &$-1.92$  &$ -8.79$  \\ 
13 &LKCa15  &\nodata  &$-1.70$  &$-1.88$  &$-1.84$  &$-1.65$  &$-1.66$  &$-1.78$  &$-1.51$  &n  &$-1.22$  &\nodata  &\nodata  &$-1.70$  &$ -8.75$  \\ 
14 &DSTau  &\nodata  &$-1.30$  &$-1.20$  &$-1.45$  &$-1.67$  &$-1.28$  &$-1.39$  &$-1.12$  &$-0.93$  &$-1.13$  &$-1.76$  &$-1.51$  &$-1.28$  &$ -8.42$  \\ 
15 &Sz65  &$-1.92$  &$-1.87$  &$-2.21$  &a  &\nodata  &n  &n  &$-2.76$  &\nodata  &n  &$-1.66$  &$-1.88$  &$-1.94$  &$ -8.90$  \\ 
16 &Sz68A  &n  &n  &n  &n  &\nodata  &n  &n  &n  &\nodata  &n  &$-1.16$  &$-1.20$  &$-1.18$  &$ -8.13$  \\ 
17 &Sz73  &$-1.14$  &$-1.21$  &$-1.29$  &$-1.33$  &\nodata  &$-1.22$  &$-1.30$  &$-1.41$  &\nodata  &$-1.27$  &$-1.16$  &$-1.05$  &$-1.22$  &$ -8.54$  \\ 
18 &HMLup  &$-1.54$  &$-1.62$  &$-1.64$  &$-1.43$  &\nodata  &$-1.69$  &$-1.66$  &$-1.52$  &\nodata  &$-1.98$  &$-1.23$  &$-1.29$  &$-1.61$  &$ -8.45$  \\ 
19 &GWLup  &$-1.61$  &$-1.74$  &$-1.80$  &$-1.87$  &\nodata  &$-1.84$  &$-1.99$  &$-1.91$  &\nodata  &$-1.94$  &$-1.89$  &$-1.96$  &$-1.87$  &$ -8.71$  \\ 
20 &GQLup  &$-0.32$  &$-0.49$  &$-0.66$  &$-0.40$  &\nodata  &$-0.53$  &$-0.61$  &$-0.26$  &\nodata  &$-0.51$  &$-0.10$  &$-0.11$  &$-0.36$  &$ -7.39$  \\ 
21 &Sz76  &$-1.92$  &$-2.12$  &$-2.16$  &$-2.52$  &\nodata  &$-2.82$  &$-2.82$  &$-3.46$  &\nodata  &n  &$-1.85$  &$-2.22$  &$-2.23$  &$ -8.96$  \\ 
22 &RULup  &$-0.34$  &$-0.16$  &$ 0.03$  &$-0.01$  &\nodata  &b  &b  &$ 0.11$  &\nodata  &$-0.32$  &$ 0.18$  &$ 0.13$  &$-0.01$  &$ -6.76$  \\ 
23 &IMLup  &$-1.68$  &$-1.92$  &a  &a  &\nodata  &n  &n  &$-2.41$  &\nodata  &n  &$-1.49$  &$-1.72$  &$-1.75$  &$ -8.68$  \\ 
24 &RYLup  &n  &n  &n  &n  &\nodata  &n  &n  &n  &\nodata  &n  &$-1.39$  &$-1.40$  &$-1.40$  &$ -8.59$  \\ 
25 &Sz102  &$-2.00$  &$-1.92$  &$-1.91$  &$-1.71$  &\nodata  &$-1.81$  &$-1.55$  &$-2.26$  &\nodata  &$-1.91$  &$-1.80$  &$-1.57$  &$-1.89$  &   \\ 
26 &Sz111  &$-1.49$  &$-1.65$  &$-1.70$  &$-1.86$  &\nodata  &$-1.67$  &$-1.82$  &$-1.87$  &\nodata  &$-1.69$  &$-1.99$  &$-1.89$  &$-1.74$  &$ -8.79$  \\ 
27 &Sz98  &$-1.39$  &$-1.66$  &$-1.66$  &$-1.53$  &\nodata  &$-1.56$  &$-1.80$  &$-1.65$  &\nodata  &$-1.50$  &$-1.42$  &$-1.36$  &$-1.53$  &$ -8.58$  \\ 
28 &EXLup  &$ 0.39$  &$ 0.53$  &$ 0.79$  &$ 0.62$  &\nodata  &b  &$ 0.89$  &$ 0.39$  &\nodata  &$ 0.19$  &$ 0.99$  &$ 0.99$  &$ 0.72$  &$ -6.12$  \\ 
29 &As205A  &$-0.44$  &$-0.26$  &$-0.20$  &$-0.10$  &\nodata  &$-0.05$  &$-0.39$  &$ 0.02$  &\nodata  &$-0.30$  &$ 0.23$  &$ 0.20$  &$-0.07$  &$ -6.58$  \\ 
30 &DoAr21  &\nodata  &\nodata  &\nodata  &n  &\nodata  &\nodata  &n  &n  &\nodata  &n  &\nodata  &\nodata  &   &  \\ 
31 &DoAr24E  &\nodata  &\nodata  &\nodata  &n  &\nodata  &\nodata  &n  &n  &\nodata  &n  &\nodata  &\nodata  &   &  \\ 
32 &DoAr44  &$-0.64$  &$-0.87$  &$-0.91$  &$-0.66$  &\nodata  &$-1.04$  &$-1.30$  &$-0.70$  &\nodata  &$-1.20$  &$-0.42$  &$-0.41$  &$-0.73$  &$ -8.05$  \\ 
33 &SR21A  &n  &n  &n  &n  &\nodata  &n  &n  &n  &\nodata  &n  &n  &n  &   &  \\ 
34 &V853Oph  &$-1.21$  &$-1.43$  &$-1.50$  &$-1.48$  &\nodata  &$-1.41$  &$-1.39$  &$-1.38$  &\nodata  &$-1.43$  &$-1.65$  &$-1.54$  &$-1.46$  &$ -8.08$  \\ 
35 &RNO90  &$-0.01$  &$ 0.13$  &$-0.01$  &$ 0.02$  &\nodata  &$ 0.32$  &$ 0.12$  &$ 0.05$  &\nodata  &n  &\nodata  &\nodata  &$ 0.06$  &$ -7.26$  \\ 
36 &V2508Oph  &$-0.56$  &$-0.63$  &$-0.70$  &$-0.61$  &\nodata  &$-0.76$  &$-0.89$  &$-0.74$  &\nodata  &$-0.86$  &$-0.56$  &$-0.50$  &$-0.66$  &$ -7.35$  \\ 
37 &V1121Oph  &$-1.01$  &$-1.39$  &$-1.47$  &$-1.06$  &\nodata  &$-1.35$  &$-1.56$  &$-1.10$  &\nodata  &$-1.00$  &$-0.90$  &$-0.95$  &$-1.12$  &$ -8.30$  \\ 
38 &J1842  &$-1.09$  &$-1.34$  &$-1.42$  &$-1.11$  &\nodata  &$-1.33$  &$-1.69$  &$-1.08$  &\nodata  &$-1.18$  &$-0.91$  &$-1.11$  &$-1.18$  &$ -8.52$  \\ 
39 &J1852  &$-1.62$  &$-1.83$  &$-1.94$  &$-1.61$  &\nodata  &$-1.82$  &n  &$-1.69$  &\nodata  &$-1.49$  &$-1.66$  &$-1.75$  &$-1.69$  &$ -9.03$  \\ 
40 &VVCrA  &$-0.22$  &$ 0.24$  &$ 0.25$  &$ 0.12$  &\nodata  &b?  &$-0.12$  &$-0.65$  &\nodata  &n  &$ 1.07$  &$ 0.70$  &$ 0.21$  &$ -6.43$  \\ 
41 &SCrA~A+B  &a  &a  &a  &a  &\nodata  &b  &$-0.75$  &$ 0.14$  &\nodata  &$-0.59$  &a  &a  &$-0.66$  &$ -7.43$  \\ 
42 &TWHya  &$-1.43$  &$-1.48$  &$-1.42$  &$-1.25$  &\nodata  &$-1.68$  &$-1.64$  &$-1.49$  &\nodata  &$-1.64$  &$-1.83$  &$-1.79$  &$-1.53$  &$ -8.67$  \\ 
43 &TWA3A  &$-3.51$  &$-3.70$  &$-3.54$  &$-3.58$  &\nodata  &$-3.87$  &$-3.84$  &$-3.47$  &\nodata  &n  &$-3.18$  &$-3.18$  &$-3.48$  &$-10.15$  \\ 
44 &V1057Cyg  &n  &n  &n  &n  &\nodata  &n  &n  &n  &\nodata  &n  &a  &n  &  &$\sim-$4.3  \\ 
45 &V1515Cyg  &n  &n  &n  &n  &\nodata  &n  &n  &n  &\nodata  &n  &$ 1.09$  &a  &  &$\sim-4.5$  \\ 
46 &HD143006  &n  &n  &n  &$-0.68$  &\nodata  &n  &n  &n  &\nodata  &n  &$-0.65$  &b  &$-0.66$  &$ -7.99$  \\ 
47 &DICep  &$-0.05$  &$-0.04$  &$ 0.19$  &$ 0.53$  &\nodata  &n  &$ 0.04$  &$ 0.27$  &\nodata  &$-0.24$  &$ 0.54$  &$ 0.45$  &$ 0.26$  &$ -6.97$  \\ 
48 &As353A  &n  &a  &a  &a  &\nodata  &n  &$ 0.08$  &$-0.22$  &\nodata  &n  &$-0.35$  &a  &$-0.13$  &$ -6.81$  \\ 
\hline
\end{tabular}
\end{center}
Note: In the table, '...' marks the lines which are not covered or fully covered by the extracted spectra, 'a' for the ones which appear as emission, but are strongly affected by the absorption due to the winds, 'n' for those which do not clearly appear as emission lines, and 'b' for the ones which are seriously contaminated by other nearby emission lines. $^{c}$ For the two FU~Ori objects, their accretion rates are obtained from \cite{2006ApJ...648.1099G} by fitting the SEDs,  assuming a central stellar mass of 1\,\Msun. 
\normalsize
\end{table*}

\clearpage
\newpage

\bibliographystyle{yahapj}
\bibliography{references}

\clearpage

\appendix

\section{Disk Classification and \OIa\ line profiles} \label{Appen:SED}

\begin{figure*}
\begin{center}
\includegraphics[width=1\columnwidth]{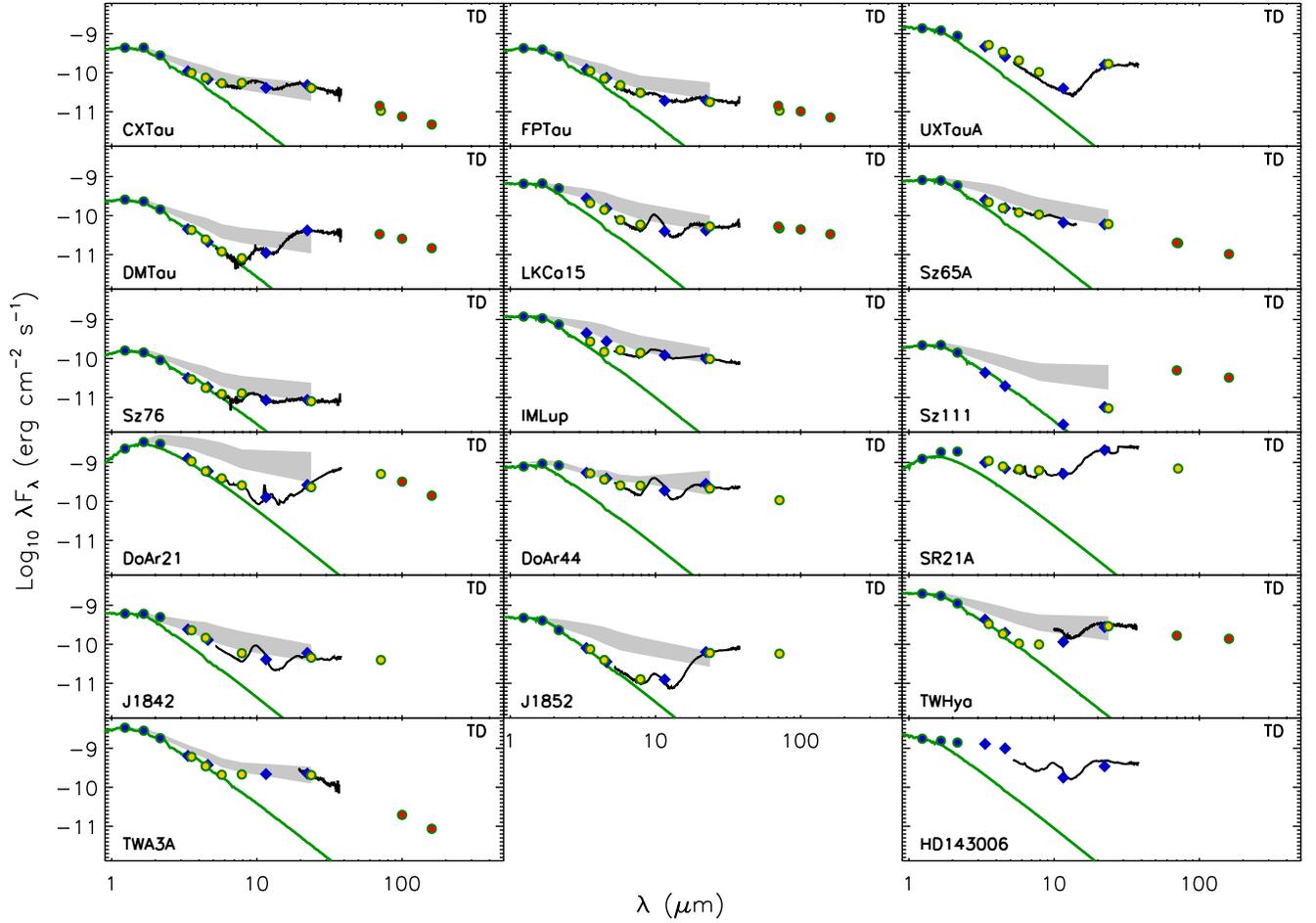}
\caption{SEDs for the TDs analyzed in this paper. Filled circles are photometric data from 2MASS (blue), Spitzer (yellow), and Herschel (red). Open circles are used for WISE photometry. The black solid line shows the Spitzer IRS spectrum while the green solid line the stellar photosphere.  The grey region shows the upper and lower quartiles of the Lynds~1641 classical T~Tauri median SED. The median SED has been reddened  with the extinction of each source, and then normalized to the J-band flux.}\label{Fig:SED1}
\end{center}
\end{figure*}

\begin{figure*}
\begin{center}
\includegraphics[width=1\columnwidth]{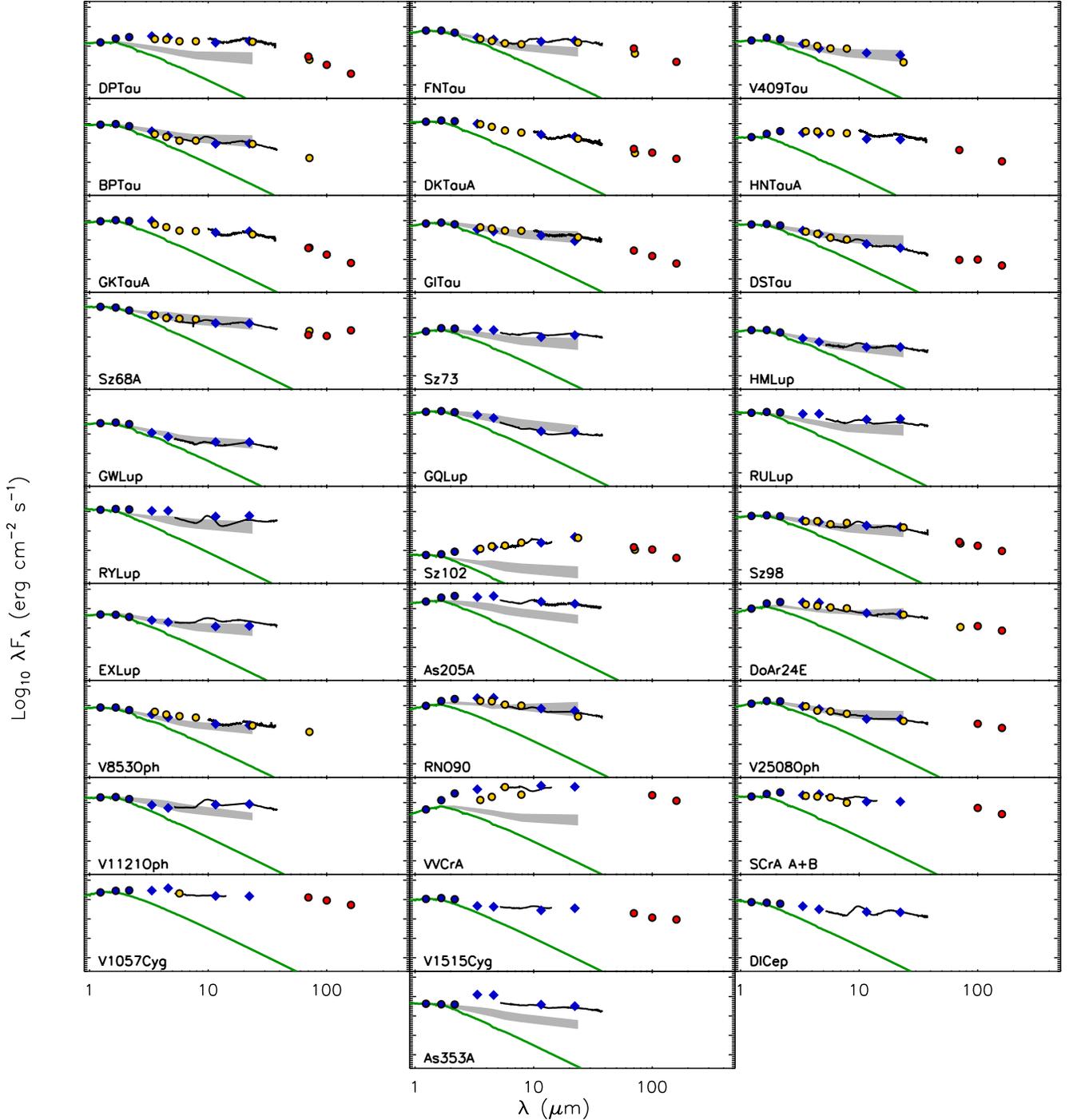}
\caption{Same as Fig.~\ref{Fig:SED1}, but for full disks.}\label{Fig:SED2}
\end{center}
\end{figure*}

Fig.~\ref{Fig:SED1} and ~\ref{Fig:SED2} show the SEDs of our sources. SEDs are constructed using the the Two-Micron All Sky Survey \citep[2MASS, ][]{2006AJ....131.1163S}, the Spitzer and Herschel photometric data, the Spitzer IRS spectra, and  the Wide-field Infrared Survey Explorer \citep[WISE,][]{2010AJ....140.1868W}. For the sources in Taurus, their Spitzer data are collected from \cite{2010ApJS..186..111L,2010ApJS..186..259R}, and the Herschel data are from \cite{2013ApJ...776...21H}. For the sources in Lupus and $\rho$~Oph, the Spitzer data are from \cite{2014yCat.2332....0E}, and the Herschel data are from \cite{2015A&A...578A..23B} and \cite{2015A&A...581A..30R}, respectively. For the sources in Corona Australis, the Spitzer data are collected from \cite{2008ApJ...677..630H} and \cite{2015ApJS..220...11D}, and the Herschel data are from \cite{2013A&A...551A..34S}. For the two sources in the TW~Hya association, their Spitzer data are collected from \cite{2010ApJS..186..111L}, and their Herschel data from \cite{2013A&A...555A..67R}. 

For each source in our sample, we construct a median SED from  CTTs  with similar spectral type in Lynds~1641 \citep{2009A&A...504..461F,2013ApJS..207....5F}. The age of Lynds~1641 is around 1-2\,Myr \citep{2013ApJS..207....5F}, which is similar to  most of the sources in this work, besides the TW~Hya association. In each panel of Fig.~\ref{Fig:SED1} and ~\ref{Fig:SED2}, we  show in grey the upper and lower quartiles of the median SEDs.  We classify a disk as TD if its SED is  below the lower quartile of the Lynds~1641 median SED at more than two bands, and otherwise as a full disk. 
As Lynds~1641 lacks G and F-type stars \citep{2013ApJS..207....5F} we cannot use our approach to classify SR~21A, HD~143006, V1057~Cyg, V1515~Cyg, and DI~Cep. In this case we adopt the TD classification for SR~21A and HD~143006, found in the literature \citep{2015MNRAS.450.3559N,2016A&A...592A.126V}, and consider the other three ones as full disks. 

\begin{figure*}
\begin{center}
\includegraphics[width=\columnwidth]{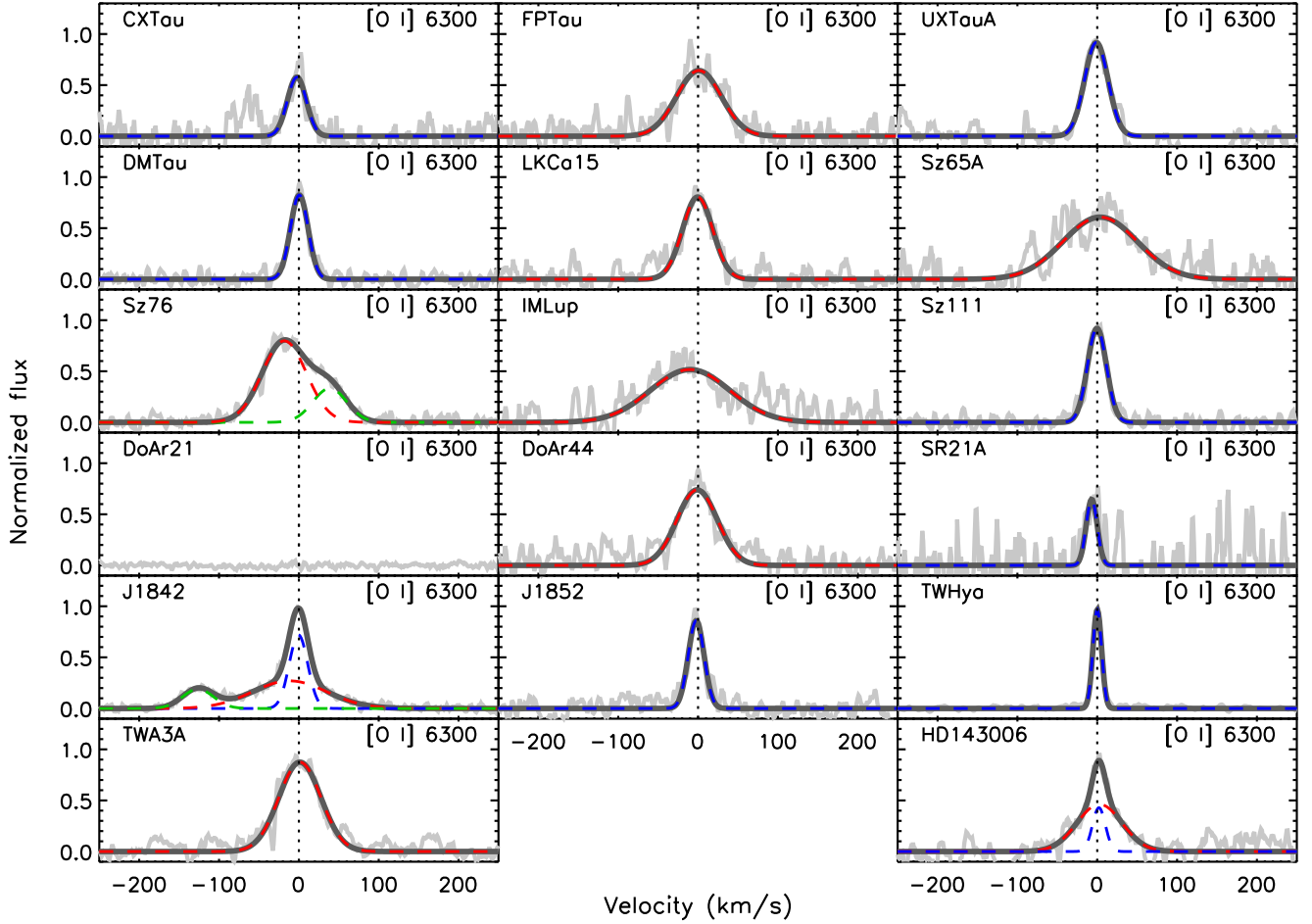}
\caption{\OIa\  line profiles for TDs. The SEDs of the corresponding sources are shown in Fig.~\ref{Fig:SED1}.}\label{Fig:SED_LR1}
\end{center}
\end{figure*}

\begin{figure*}
\begin{center}
\includegraphics[width=\columnwidth]{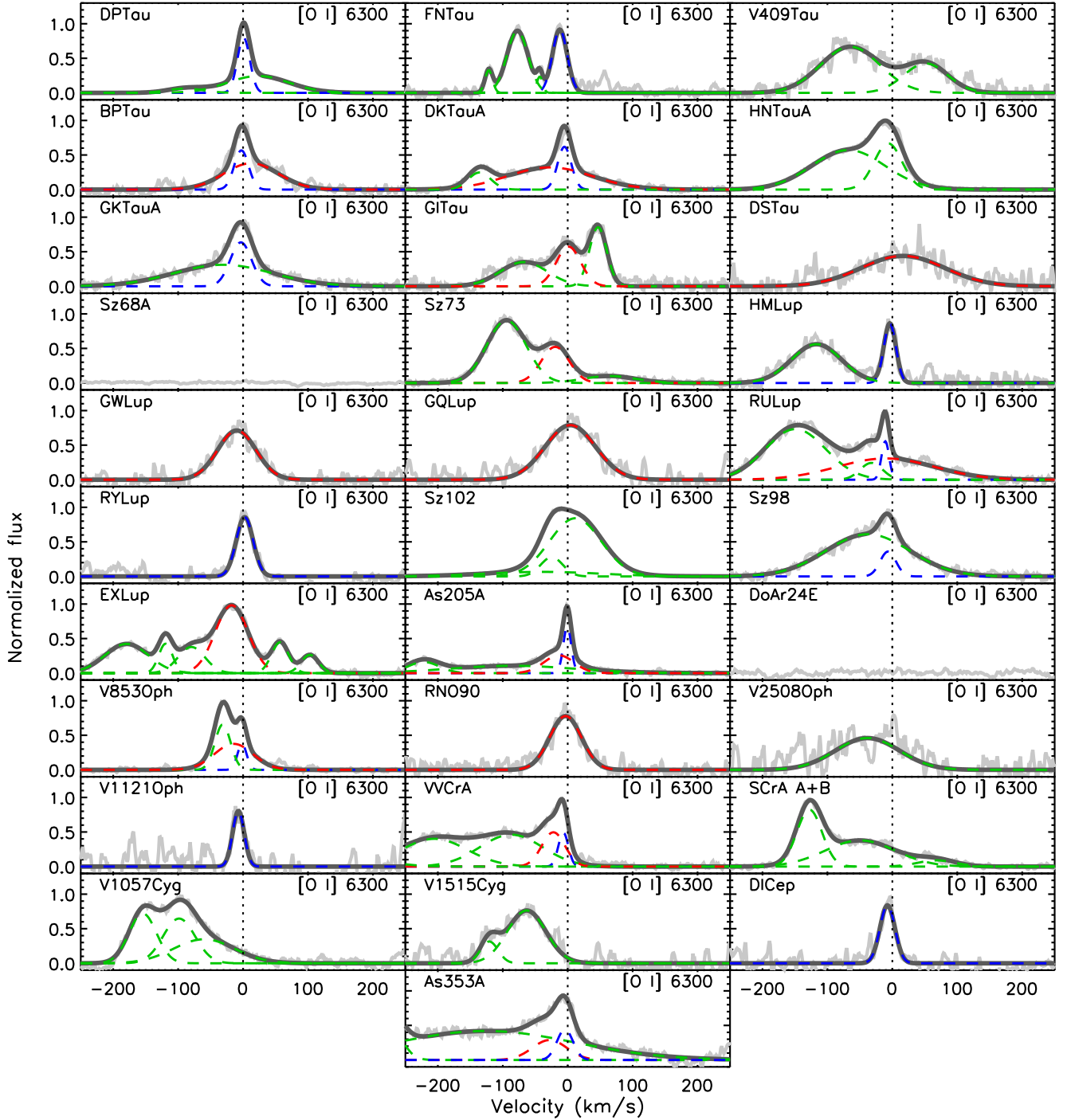}
\caption{\OIa\  line profiles for full disks. The SEDs of the corresponding sources are shown in Fig.~\ref{Fig:SED2}.}\label{Fig:SED_LR2}
\end{center}
\end{figure*}

Fig.~\ref{Fig:SED_LR1} and ~\ref{Fig:SED_LR2} show the \OIa\ line profiles for the TDs and full disks. It is clear that the TDs have simpler forbidden line profiles than the full disks.

\section{Flux calibration}\label{Sect:flux_calibration}

\begin{figure*}
\begin{center}
\includegraphics[width=0.65\columnwidth]{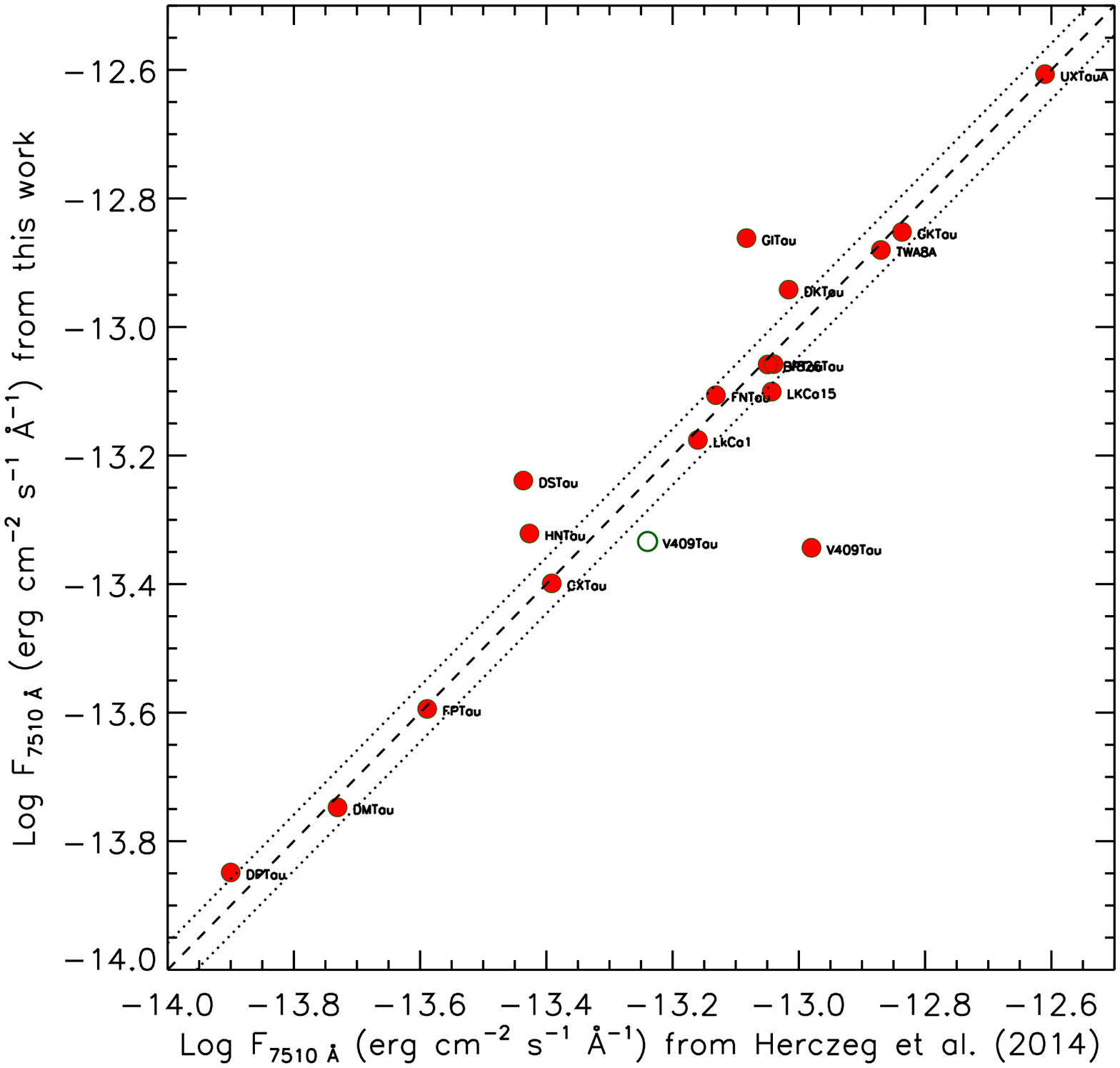}
\caption{Comparison of 7510\,\AA\ fluxes from our work and from  \citet{2014ApJ...786...97H}. The open circle also locates our 7510\,\AA\ flux for V409~Tau with respect to the SDSS photometry. The dash line shows the 1:1 line while dotted lines mark a $\pm$0.04~dex difference.}\label{Fig:flux_com}
\end{center}
\end{figure*}

We use 3 spectrophotometric standards each night to flux calibrate the spectra of our scientific targets. First, we obtain the BT-settl model atmosphere \citep{2011ASPC..448...91A} corresponding to each telluric standard spectral type in the night of January 23, 2008, and the white dwarf model atmosphere \citep{2010MmSAI..81..921K} for the white dwarf telluric standards in the night of May 23, 2008. Next, we fit its broad-band photometry using the aforementioned model with two free parameters, extinction and stellar angular radius as  in \citet{2009A&A...504..461F,2013ApJS..207....5F}.  Then, we shift and rotationally broaden the best-fit model atmosphere and degrade it to the Keck spectral resolution. In each order, the ratio between the observed and model spectrum is fitted with a 4-order polynomial function, from which we obtain the conversion from counts to absolute flux. The spectrophotometric standard that is observed closest in time to each of our target is chosen its flux calibration.
As our slit width is narrow (0\farcs861), we also correct for the missing flux due to the difference in point source functions between the target and the spectrophotometric standard. Atmospheric absorption due to the different airmass between the spectrophotometric standards and the scientific objects  is also corrected using optical extinction coefficients  proper for  Mauna Kea \footnote{The coefficients used are listed in the website: www.gemini.edu/sciops/telescopes-and-sites/observing-condition-constraints/extinction.}. The flux calibration did not account for atmospheric dispersion in the observations of DK~Tau where the slit position angle was offset from the parallactic angle.  For SCrA~A+B, the MAKEE pipeline cannot resolve the SCrA~A and B components, and thus the flux calibration over-correct the flux for the slit loss by  $\sim$20\%.

We use two methods to estimate the uncertainty on the flux-calibration. The first one is to flux calibrate one spectrophotometric standard using another close-in-time spectrophotometric, and compare the calibrated flux with the expected one. We found differences of $\sim$14\% for the night on January 23, 2008 and only $\sim$3\% for the night on May 23, 2008. To evaluate the uncertainty on line ratios close to the forbidden lines analyzed here, we also calculate the $F_{\rm cont, 4068}/F_{\rm cont, 6300}$ and the $F_{\rm cont, 5577}/F_{\rm cont, 6300}$, basically continuum fluxes near the \SII\, \OIb , and \OIa\ transitions, 
from the flux-calibrated spectra of one spectrophotometric standard using another one close in time. When comparing them with the expected values, we find differences 
of $\sim$5\% and 3\%  on January 23, 2008 and $\sim$5\% and 1\% on May 23, 2008
for the $F_{\rm cont, 4068}/F_{\rm cont, 6300}$ and $F_{\rm cont, 5577}/F_{\rm cont, 6300}$ ratios, respectively.

The data acquired on January 23, 2008 cover wavelengths near 7510\,\AA. Hence, we can further estimate the uncertainty on our flux calibration by comparing our calibrated  7510\,\AA\ fluxes with those reported in \cite{2014ApJ...786...97H}. Unfortunately, the  7510\,\AA\ wavelength is located in the gap between two HIRES orders. So, we did not measure the 7510\,\AA\ flux directly, but instead interpolate the flux using wavelengths near 7510\,\AA. Figure~\ref{Fig:flux_com} shows our flux values compared to those from \cite{2014ApJ...786...97H}. The mean difference between the two works  is about 0.07\,dex. Four sources, DS~Tau, HN~TauA, GI~Tau, and V409~Tau, show larger differences (0.10-0.37\,dex), which may be due to their brightness variability (see \citealt{2015AJ....150...32R} for V409~Tau and \citealt{2018ApJ...852...56G} for GI~Tau). V409~Tau, which sports the largest difference (0.37\,dex), has  been also observed with SDSS \citep{2012ApJS..203...21A} and is 11.934\,mag in $SDSS-i$ band. We convert the $i$-band photometry to flux and correct for the extinction $A_{\rm V}$=1 from \cite{2014ApJ...786...97H}, which gives a flux of 5.76$\times$10$^{-14}$\,erg~s$^{-1}$~cm$^{-2}$~\AA$^{-1}$ at 7625\,\AA\ (the center of $SDSS-i$ band). The difference between our 7510\,\AA\ flux and the  SDSS 7625\,\AA\ flux is $\sim$0.09\,dex, much lower than that with \cite{2014ApJ...786...97H}, which supports the hypothesis of variability driving a larger difference. If we exclude the four outliers in Figure~\ref{Fig:flux_com}, the mean difference between our 7510\,\AA\ fluxes and those from \cite{2014ApJ...786...97H} is reduced to $\sim$0.03\,dex. In summary, two different methods suggest that the uncertainty on our absolute flux calibration is likely better than 10\%.

\section{Comments on sources without  \OIa\ detection} \label{Appen:detail}

Our sample includes three sources (Sz~68A, DoAr~21, and DoAr~24E) with no \OIa\ detection.  Neither DoAr~21 nor  DoAr~24E show  emission lines in our spectra  that can be used to derive accretion luminosities and thus are inferred to have weak or absent disk accretion.  In the following, we provide additional information on these sources.

{\bf Sz~68A} is a a triple system with a close binary (separation 0\farcs126) and a third faint companion (M6) with a separation of 2\farcs808 \citep{2006A&A...459..909C}. Our spectroscopic observation targeted the close binary. We detect  the Ca~\,{\scriptsize II}\ lines at 3934 and 3968\,\AA\ but only weak H$\alpha$ emission ($EW\sim-$2.8\AA). This star has been classified as a weak-line T~Tauri star, a young non-accreting star, by \cite{2007ApJ...667..308C} and \cite{2017A&A...600A..20A}. The \OIa\ emission is not detected in our HIRES spectra and it is also absent in the X-Shooter spectra by \cite{2018A&A...609A..87N}.  \cite{2016ApJ...828...46A} reported cold CO gas emission around the binary and no gas or dust cavity at a resolution of 50$\times$40\,au. The SED of Sz~68A is compatible with that of a full disk. 

{\bf DoAr~21} is a close binary with a separation of 1.2--1.8\,au \citep{2008ApJ...675L..29L}.  The HIRES spectrum of DoAr~21, which covers from 4420 to 6310\,\AA, does not show any accretion-related emission lines. \cite{2013ApJ...769...21S}  detected Pf$\beta$  emission from DoAr~21 and  estimated a mass accretion rate of 1.7$\times10^{-8}$\,M\accunit. DoAr~21 is detected at far-infrared wavelengths \citep{2015A&A...581A..30R} and has reduced mid-infrared emission as TDs (see Fig.~\ref{Fig:SED1}  from our paper). However, based on high spatial resolution mid-infrared imagery, \cite{2009ApJ...703..252J} find very weak, almost absent, excess emission from this source at 9--18\,\micron{} and argue that low resolution data could be contaminated by an extended infrared source. Thus, they suggested that either the disk is in the final stage of clearing, or there is no disk around DoAr~21. 
The recent ALMA imaging of DoAr~21 at 870\,\mum\ shows no detection at a spatial resolution 0\farcs2 and a sensitivity 0.02~mJy \citep{2017ApJ...851...83C}. Therefore, even if there is a disk around DoAr~21, its disk mass should be very low.

{\bf DoAr~24E} has a faint companion at a separation of 2\farcs03\ \citep{2005A&A...437..611R}. Our HIRES spectrum covering  from 4420 to 6310\,\AA\ does not show any accretion-related emission lines. \cite{2006A&A...452..245N} detected  Pa$\beta$ emission from its near-infrared spectrum and estimated a mass accretion rate of 6.2$\times10^{-9}$\,M\accunit. The source SED is compatible with that of full disks. The continuum emission from the disk has been detected at far-infrared, sub-millimeter, and millimeter wavelengths \citep{2015A&A...581A..30R,2013ApJ...773..168M}. No [O\,{\scriptsize I}] and o-H$_{2}$O emission is detected around 63\,\mum\ \citep{2016A&A...594A..59R}.

\section{Corrected line profiles and kinematic decomposition}\label{Appendix:lineprofile}
The rationale for decomposing corrected line profiles with Gaussians is discussed in \sect~\ref{Sect:line_decomposition} while the best fit parameters are summarized in Table~\ref{Tab:para_LVC1}.
Fig.~\ref{Fig:line2} provides the best fit Gaussian profiles color coded by kinematic component and superimposed on the corrected \SII, \OIb, and \OIa\ profiles  for all the  sources in our sample.

\begin{figure*}
\begin{center}
\includegraphics[width=\columnwidth]{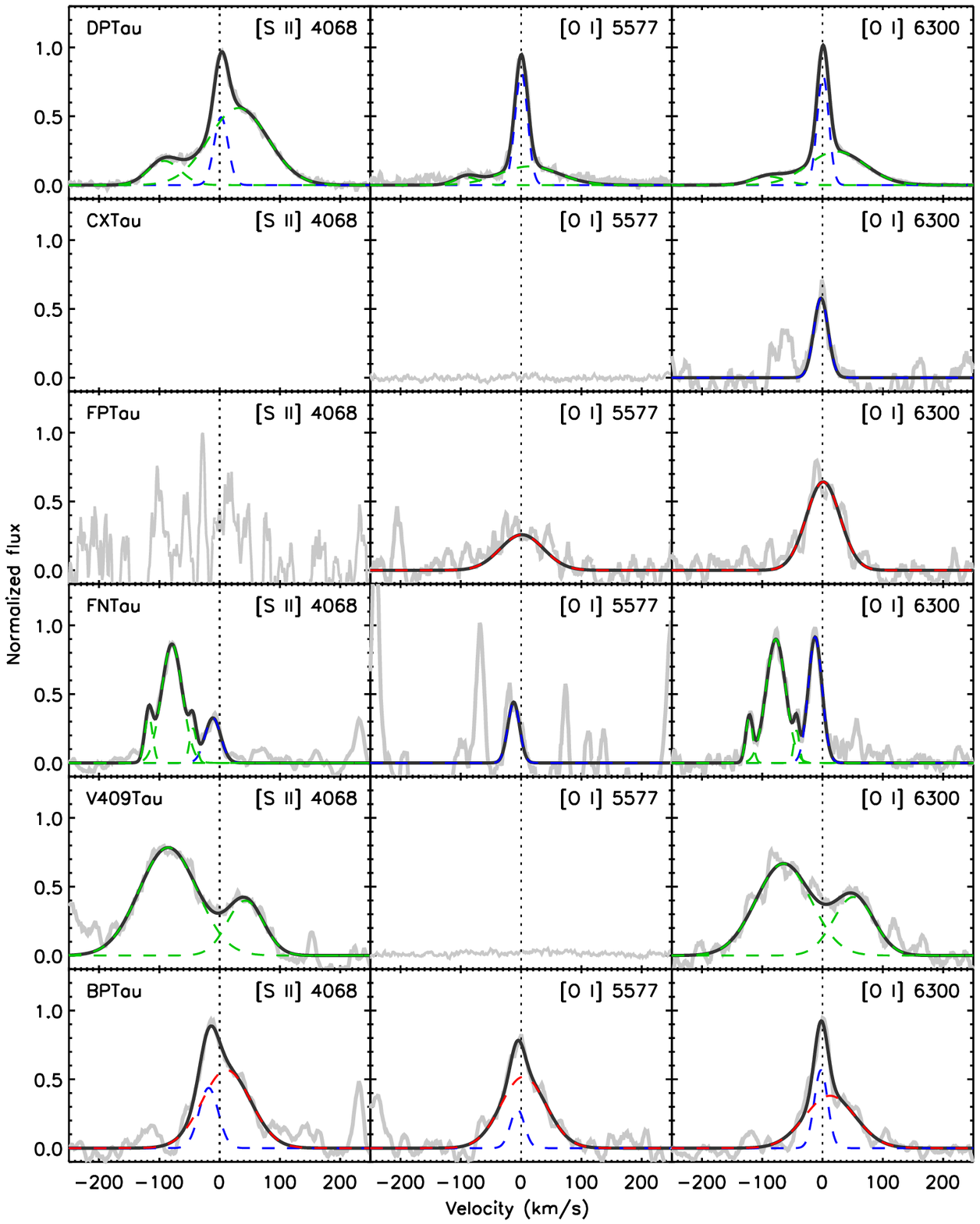}
\caption{\SII\,  \OIa\, and \OIb\ line profiles for our sample. In each panel, the green dashed line is for the HVC, the red dashed line for the LVC-BC, the blue dashed line for the LVC-NC, and the dark solid line is the sum of all components. For EX~Lup, Fe lines contaminating the profiles of the forbidden lines are marked with vertical dashed light blue lines.}\label{Fig:line2}
\end{center}
\end{figure*}
 
 \setcounter{figure}{19}
\begin{figure*}
\begin{center}
\includegraphics[width=\columnwidth]{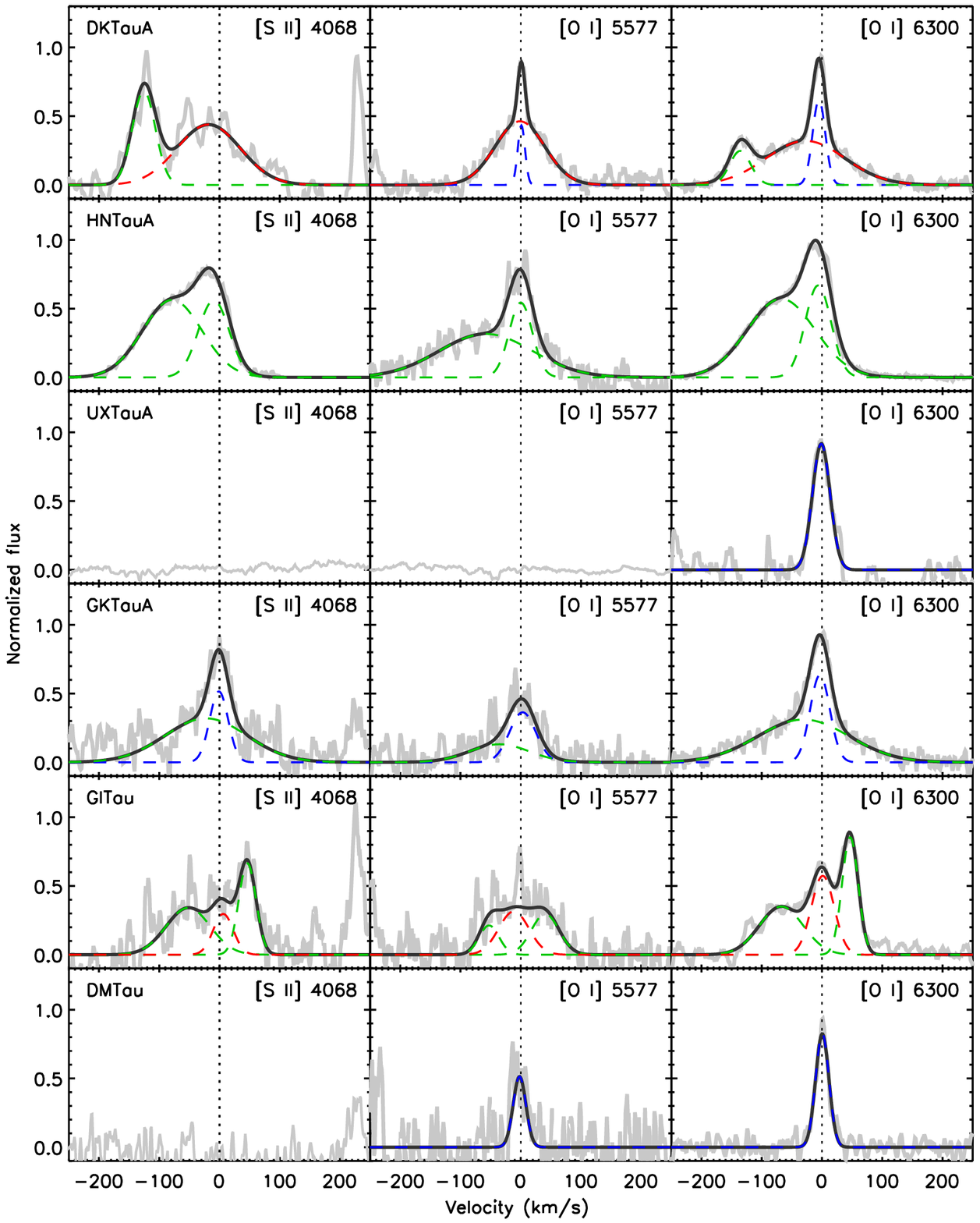}
\caption{continued.}\label{Fig:line2}
\end{center}
\end{figure*}
 
\setcounter{figure}{19}
\begin{figure*}
\begin{center}
\includegraphics[width=\columnwidth]{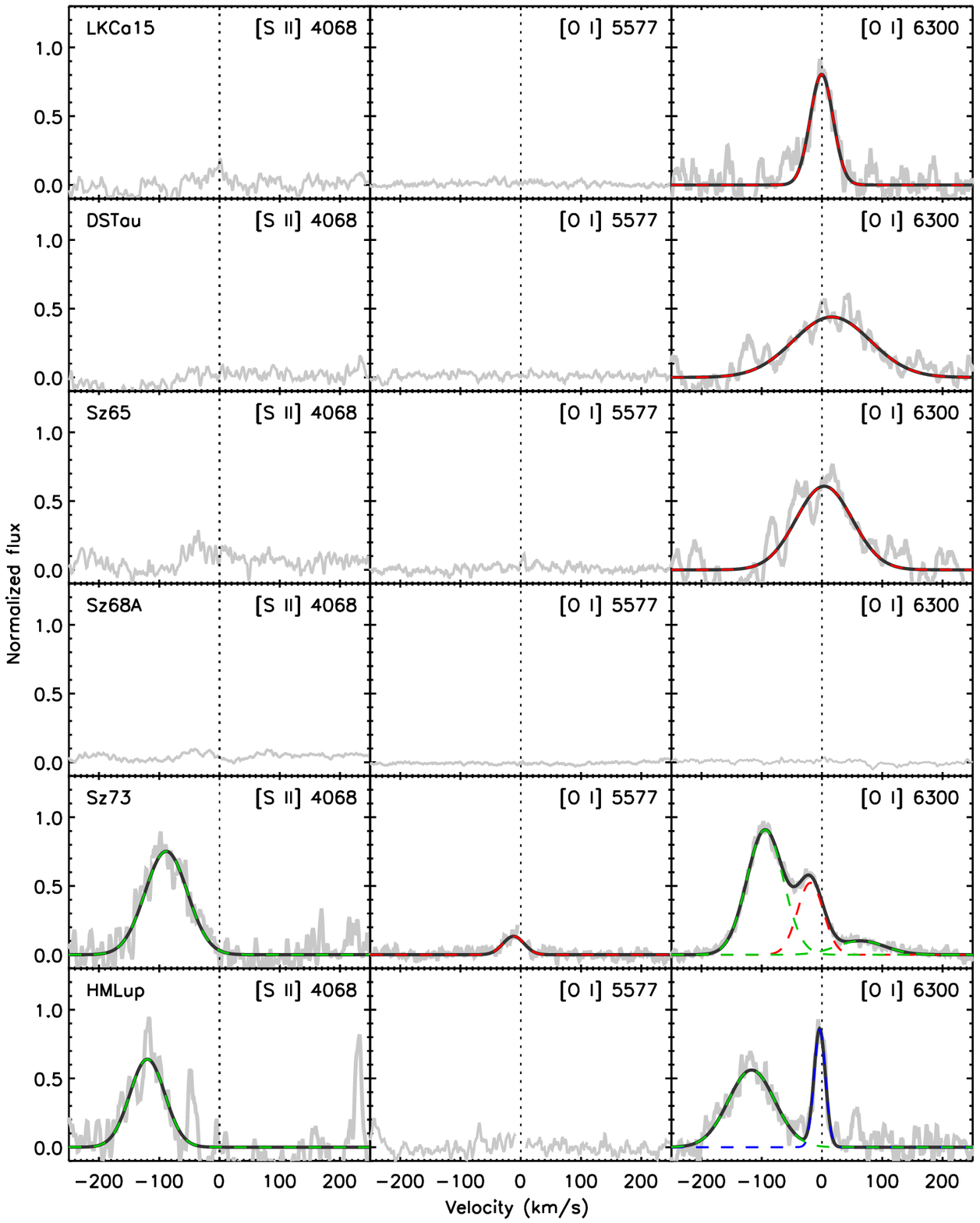}
\caption{continued.}\label{Fig:line3}
\end{center}
\end{figure*}
 
\setcounter{figure}{19}
\begin{figure*}
\begin{center}
\includegraphics[width=\columnwidth]{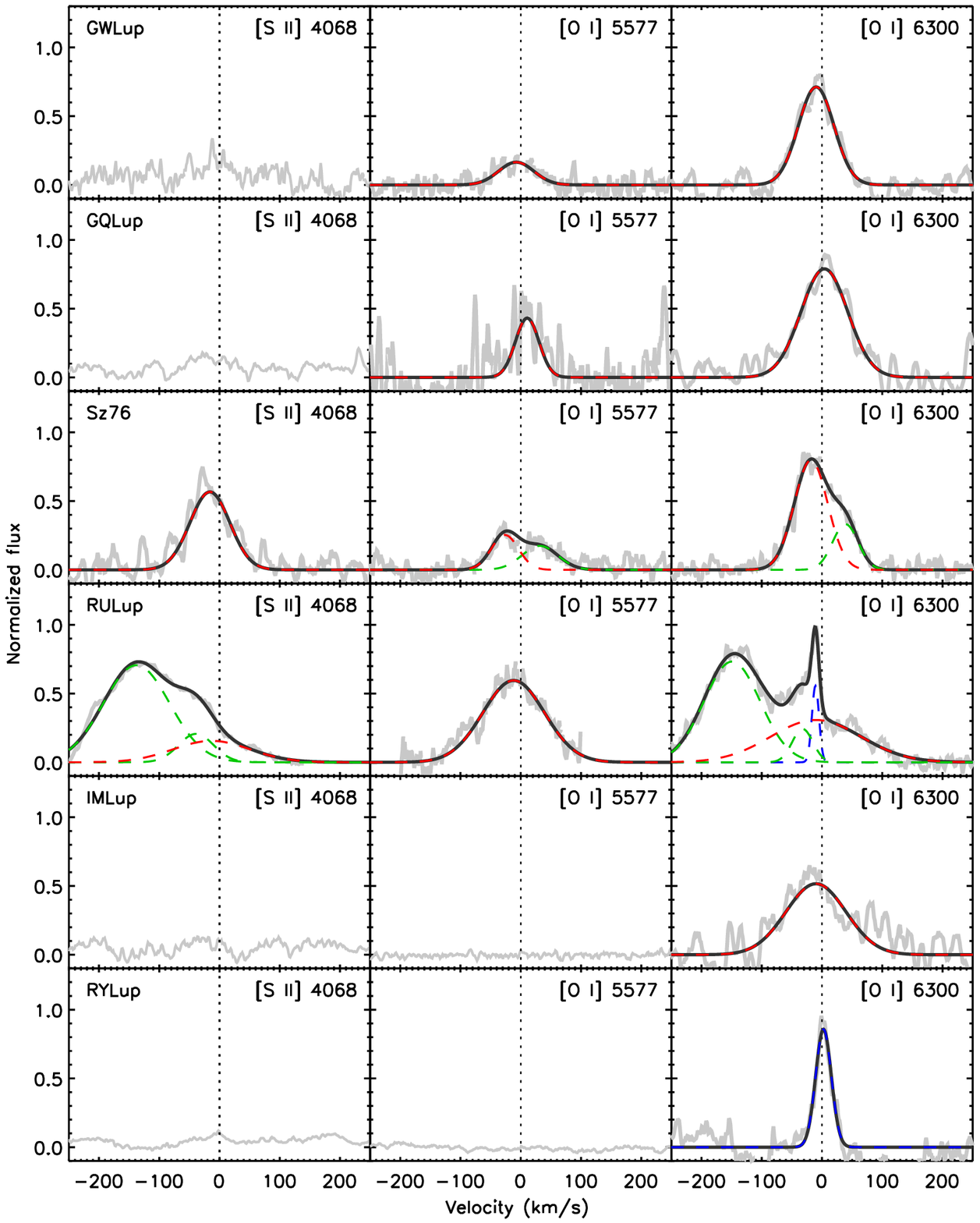}
\caption{continued.}\label{Fig:line4}
\end{center}
\end{figure*}

\setcounter{figure}{19}
\begin{figure*}
\begin{center}
\includegraphics[width=\columnwidth]{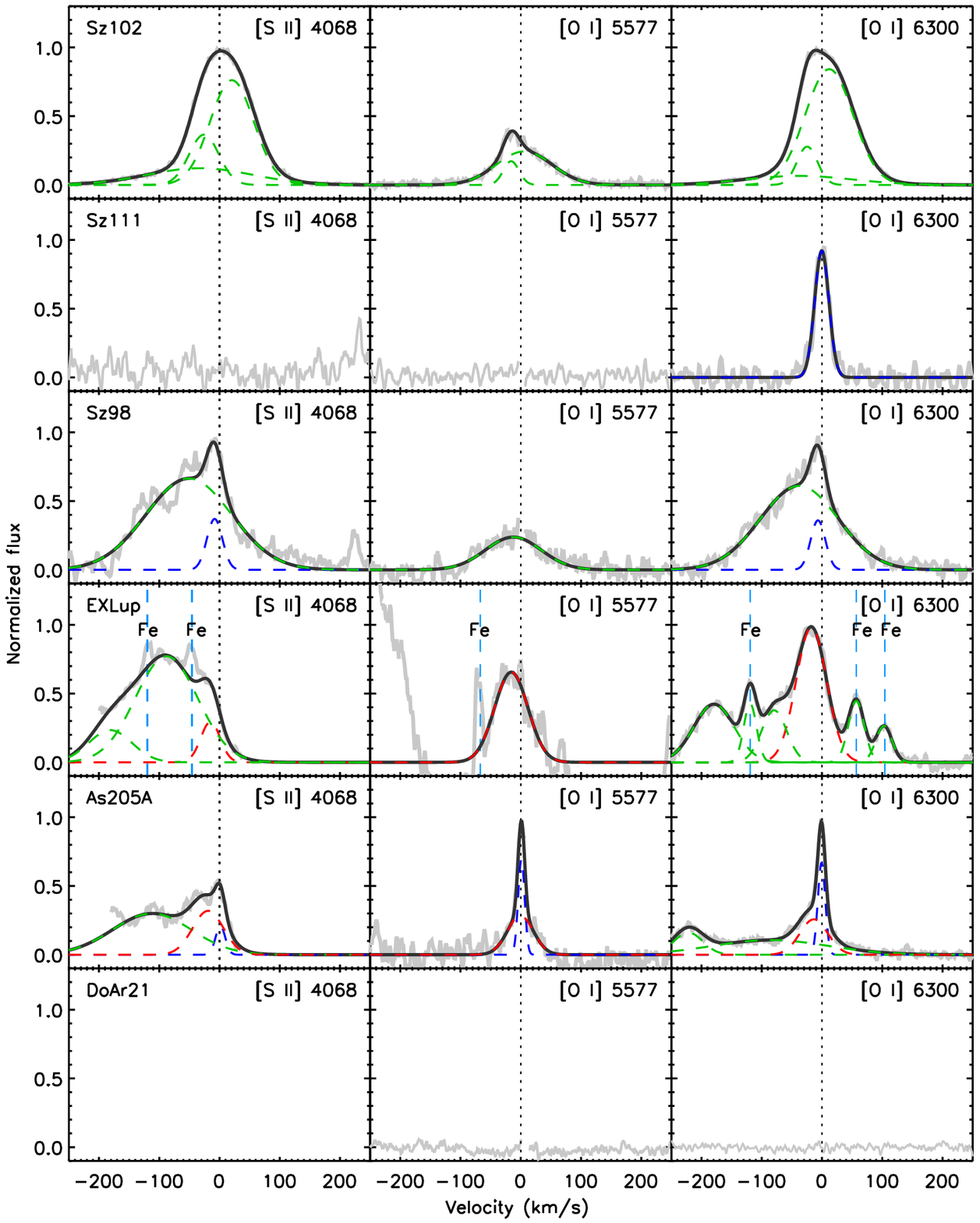}
\caption{continued.}\label{Fig:line5}
\end{center}
\end{figure*}

\setcounter{figure}{19}
\begin{figure*}
\begin{center}
\includegraphics[width=\columnwidth]{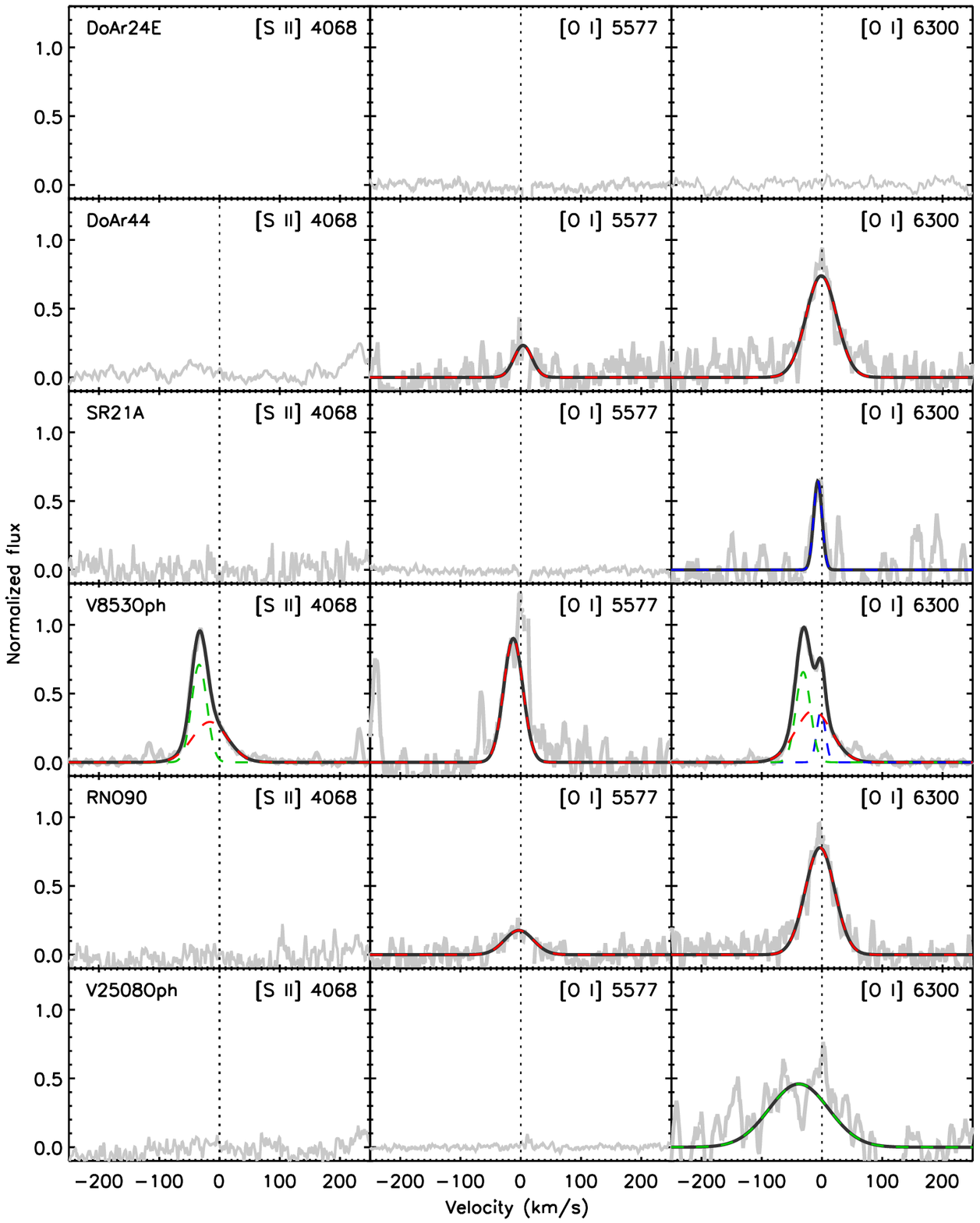}
\caption{continued.}\label{Fig:line5}
\end{center}
\end{figure*}

\setcounter{figure}{19}
\begin{figure*}
\begin{center}
\includegraphics[width=\columnwidth]{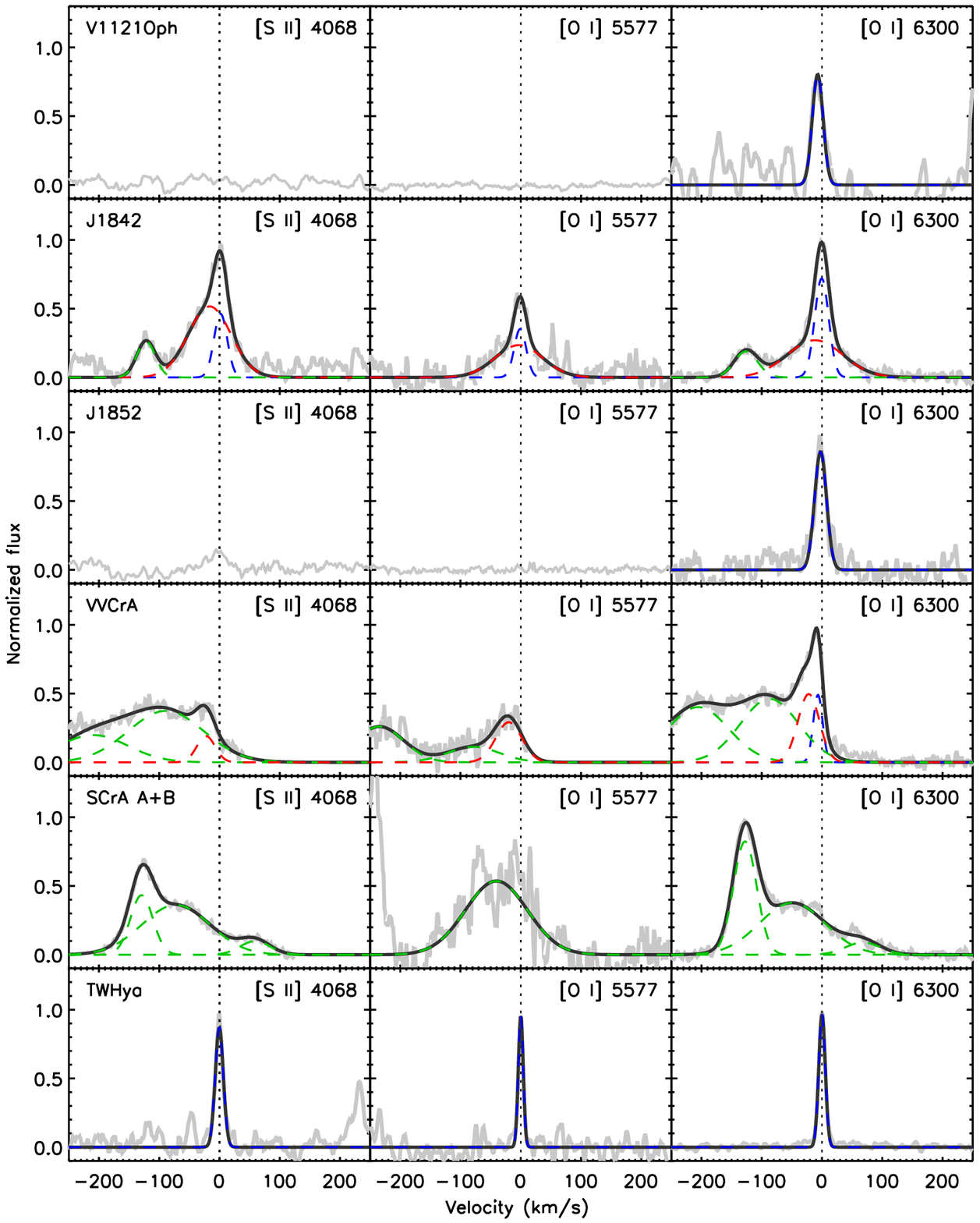}
\caption{continued.}\label{Fig:line5}
\end{center}
\end{figure*}

\setcounter{figure}{19}
\begin{figure*}
\begin{center}
\includegraphics[width=\columnwidth]{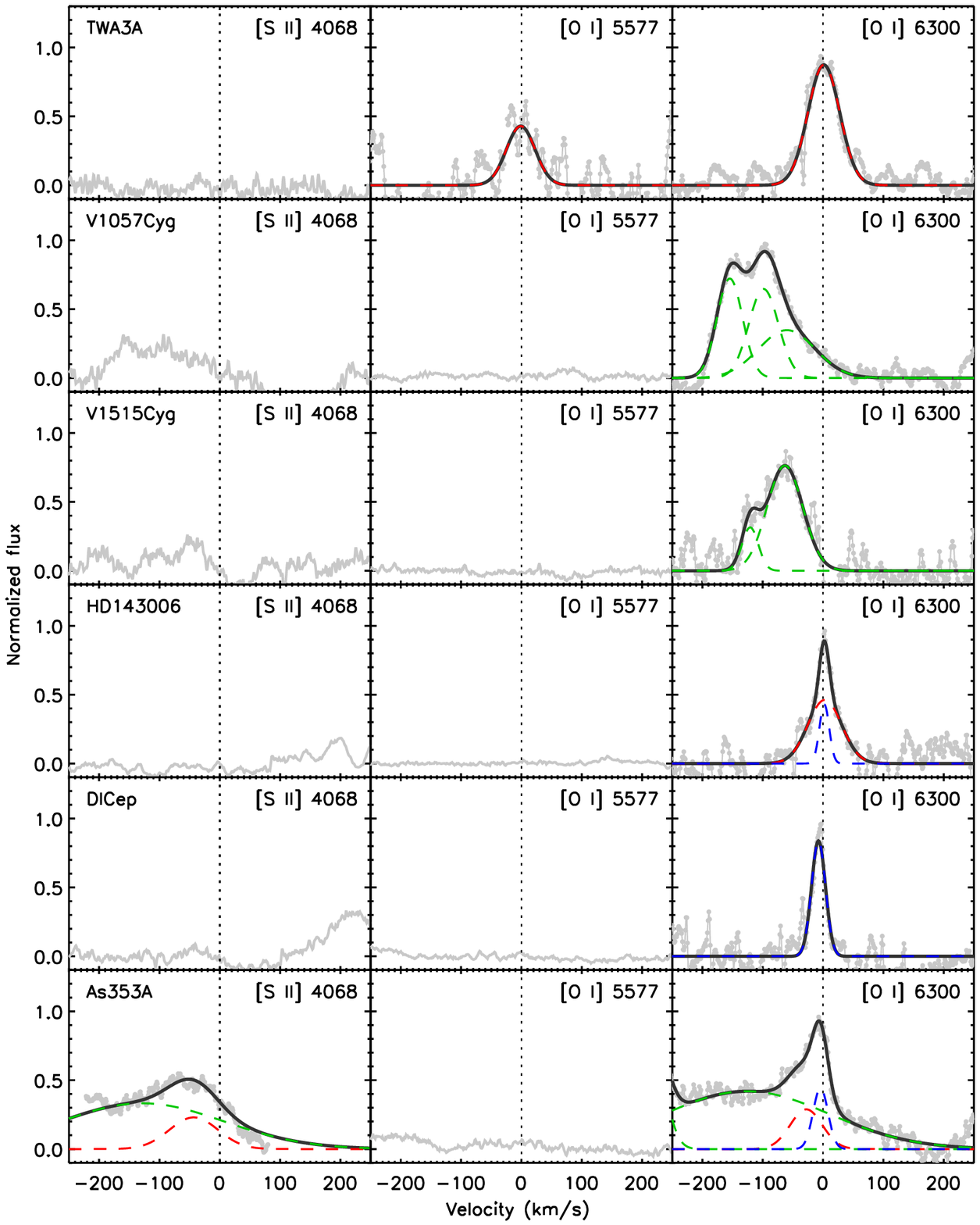}
\caption{continued.}\label{Fig:line5}
\end{center}
\end{figure*}

\section{Line decontamination}\label{Appen:line_decontamination}
\begin{figure*}
\begin{center}
\includegraphics[width=\columnwidth]{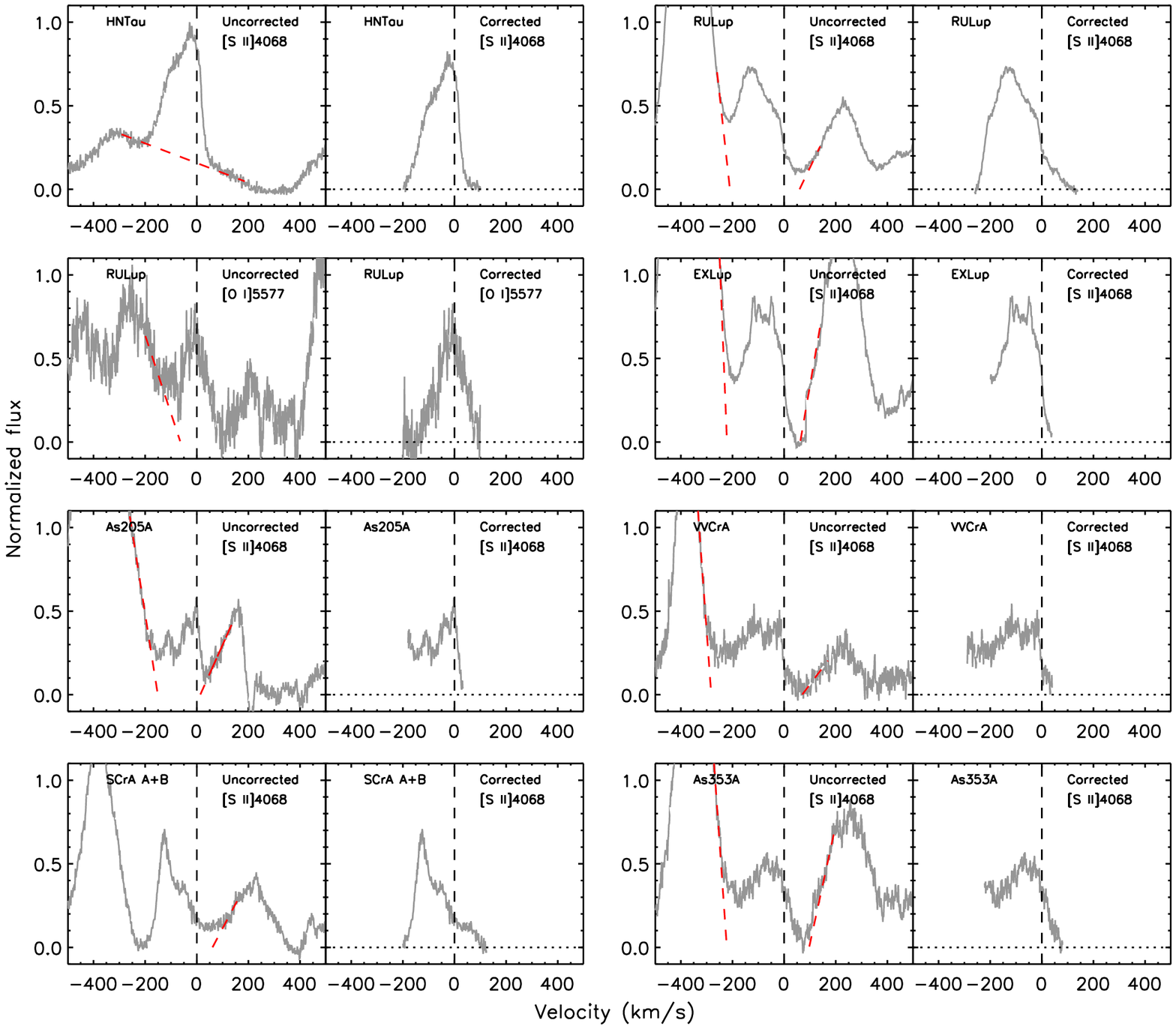}
\caption{Profiles that are contaminated by Fe~{\scriptsize I} lines (left panels). The right panels show our corrected profiles.}\label{Fig:Decont}
\end{center}
\end{figure*}

Seven sources present \SII\ line profiles that are highly contaminated by  Fe~{\scriptsize I} emission  at  4063.6 and 4071.7\,\AA. In addition, the \OIb\ line from RU~Lup is also contaminated by Fe~{\scriptsize I} emission at 5572.8\,\AA. Fig.~\ref{Fig:Decont} shows the contaminated  line profiles. We fit the  wings  of the Fe~{\scriptsize I} emission with linear functions and tentatively decontaminate the profiles by subtracting the fitted function from the observed spectra. The corrected \SII\ and \OIb\ line profiles  are also shown in Fig.~\ref{Fig:Decont}. Due to the strong contamination from the Fe~{\scriptsize I} line at 4063.6\,\AA, any \SII\  flux more blueshifted than $\sim$200\,\kms\ cannot be recovered. A comparison of \SII\ and \OIa\ profiles suggest that we achieve a good correction for less blueshifted emission  (see Fig.~\ref{Fig:threelines_HVC} in the main text).

\section{A comparison of the individual components in \SII\ and \OIa}\label{Appen:com_SII_OI6300}
\begin{figure}
\begin{center}
\includegraphics[width=0.45\columnwidth]{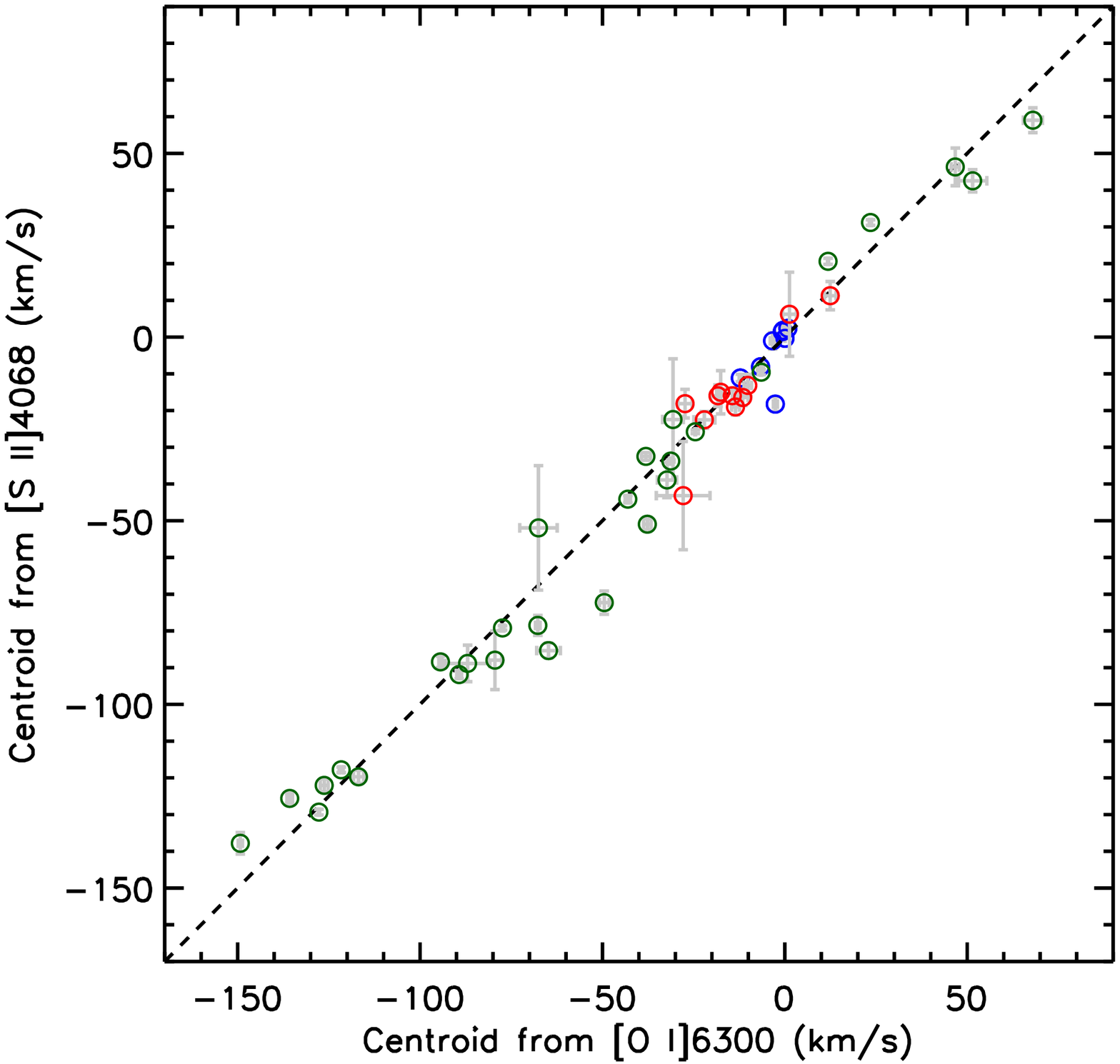}
\includegraphics[width=0.45\columnwidth]{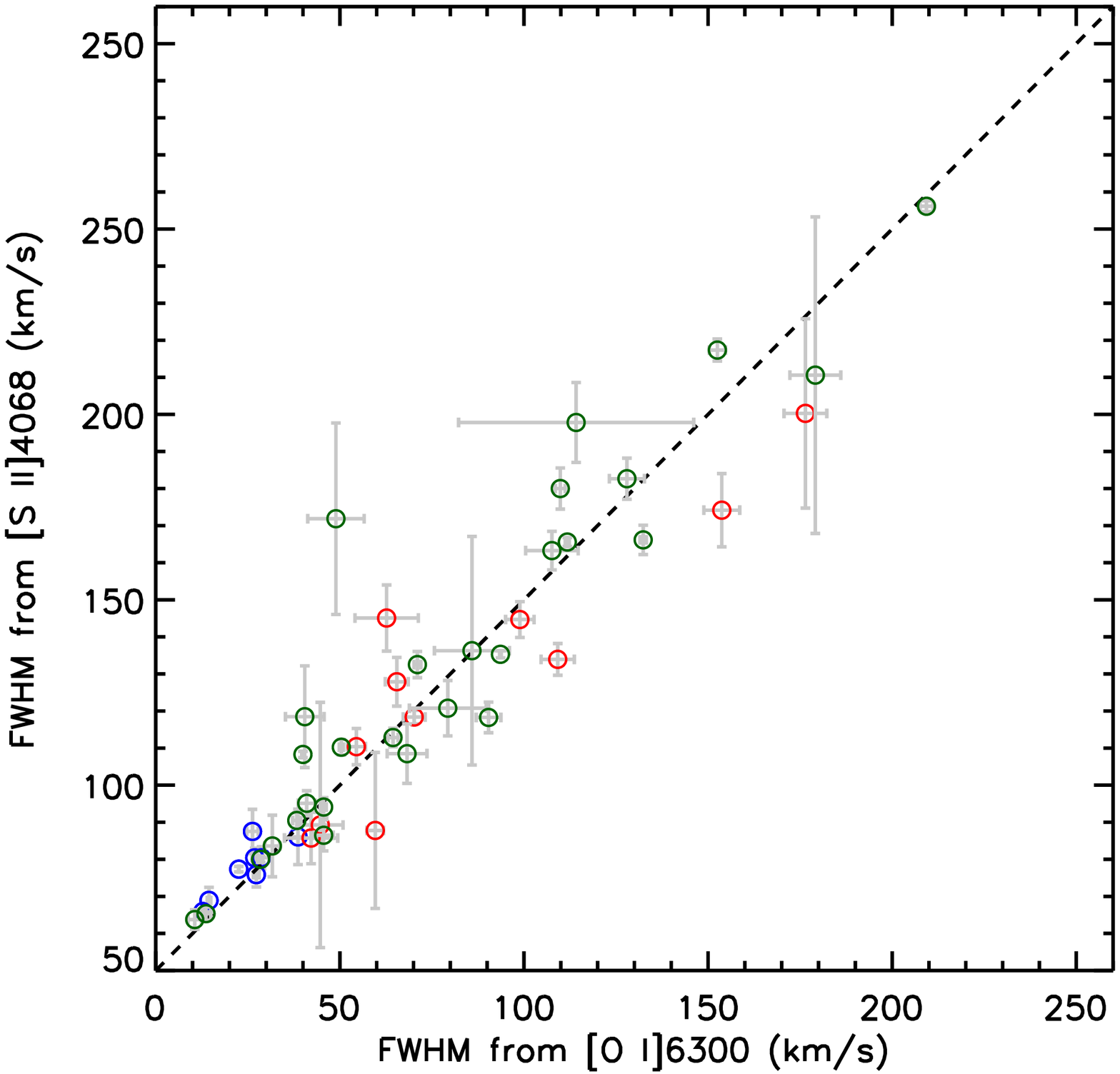}
\caption{Centroids (left) and FWHMs (right) of individual  \SII\ and \OIa\ components (HVC in green, LVC-BC in red, and LVC-NC in blue). A dashed line shows the 1:1 relation.  }\label{Fig:com_SII_OI6300}
\end{center}
\end{figure}

As discussed in the main text (\sect~\ref{com_SII_OI6300}), individual kinematic components present similar \SII , \OIb , and \OIa\ profiles indicating they trace the same physical region. Here, we provide further evidence to this finding by showing a 
comparison of centroids and FWHMs of individual \SII\ and \OIb\ components vs the 
\OIa\ components (see Figure~\ref{Fig:com_SII_OI6300}). Note the similarity in centroids and FWHMs for individual kinematic components.  

\begin{figure}
\begin{center}
\includegraphics[width=0.45\columnwidth]{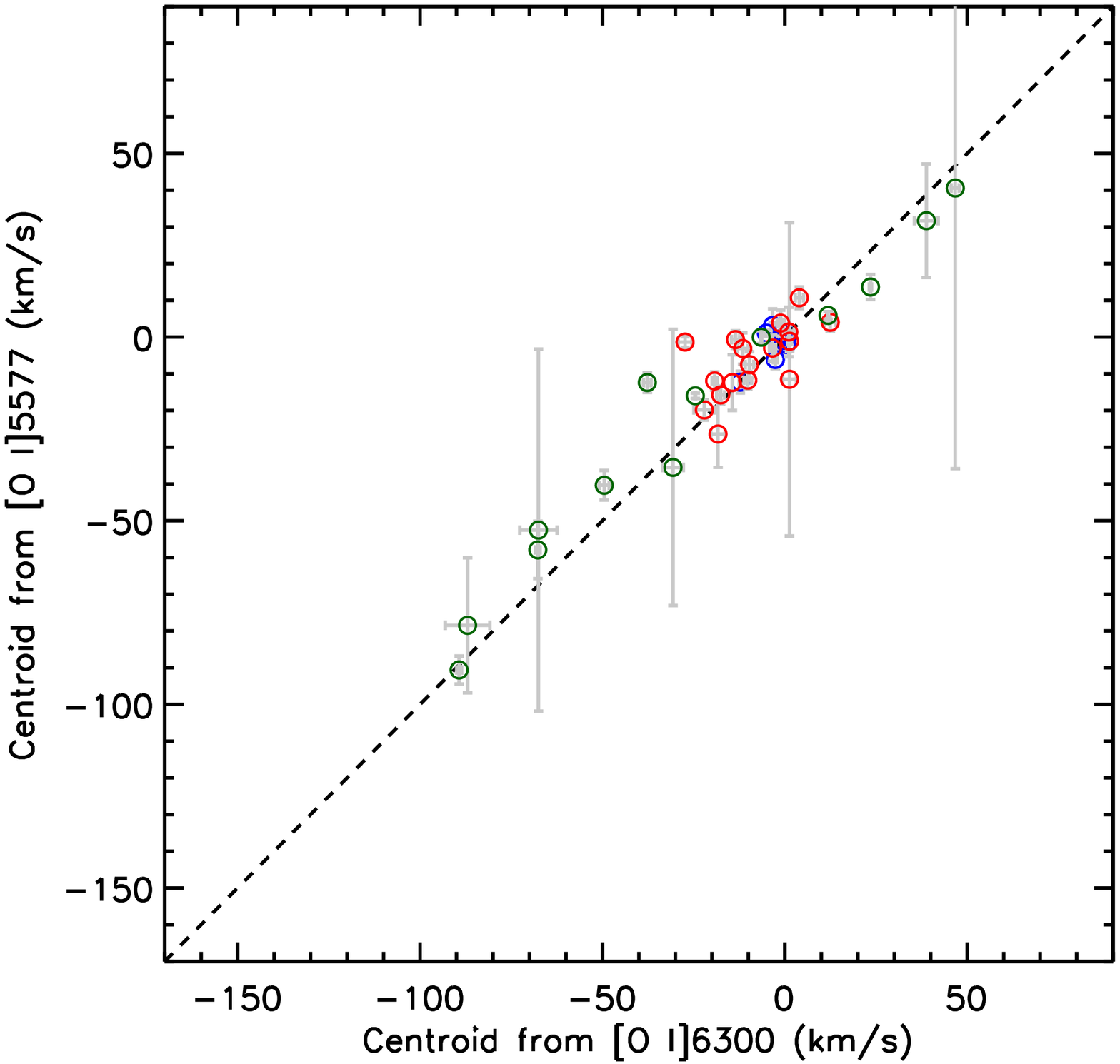}
\includegraphics[width=0.45\columnwidth]{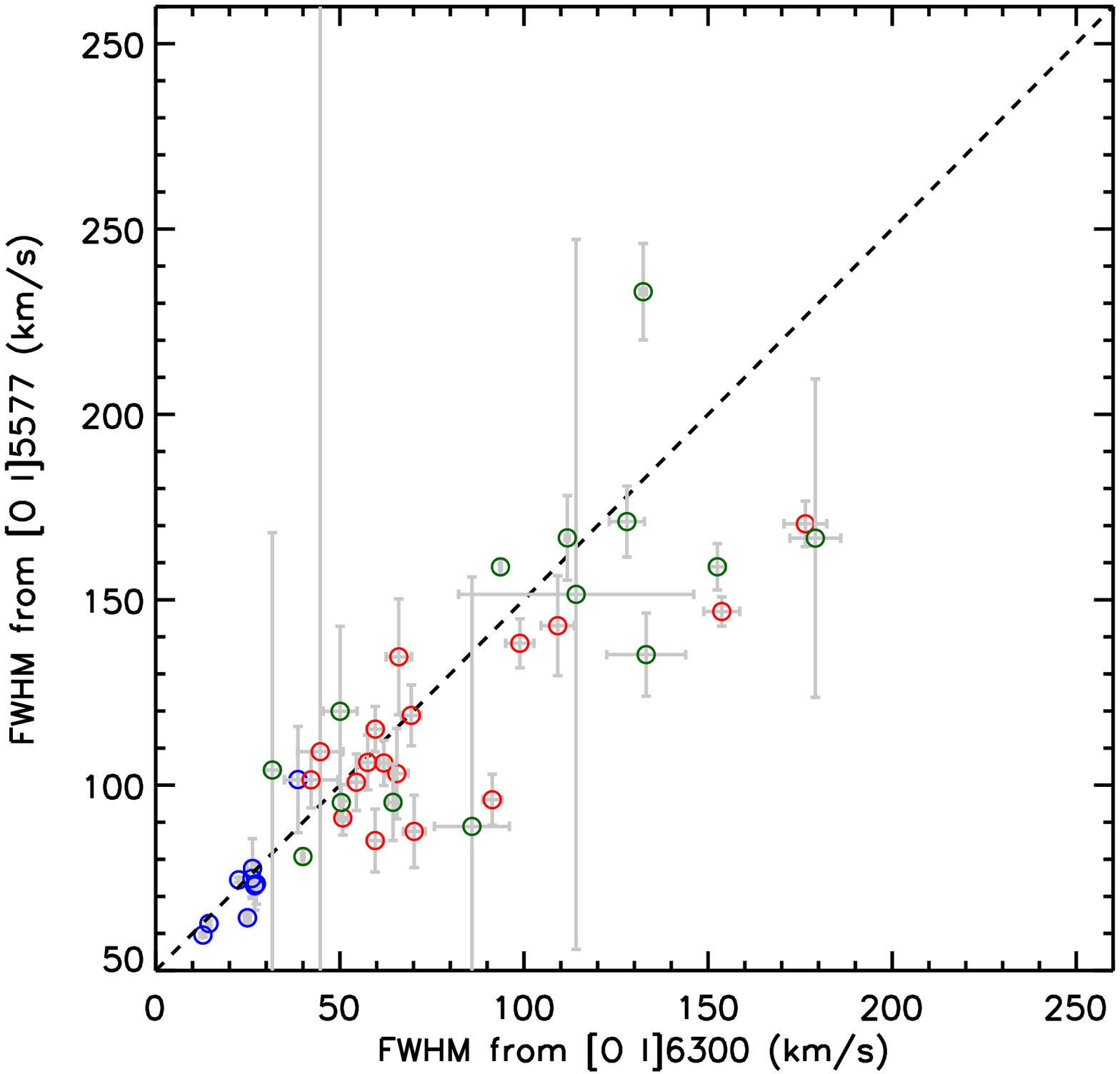}
\caption{Same as Fig.~\ref{Fig:com_SII_OI6300} but for the \OIb\ and \OIa\ lines. }\label{Fig:com_OI2}
\end{center}
\end{figure}

\section{Updated relation between line and accretion luminosity} \label{Appen:line_acc}
In view of the Gaia DR2 release \citep{2018arXiv180410121B}, we  revisit the relations between the luminosity of permitted lines ($L_{\rm line}$) and the accretion luminosity ($L_{\rm acc}$). We collect $L_{\rm line}$ and $L_{\rm acc}$ from \cite{2017A&A...600A..20A} and scale them to the new Gaia DR2 distance of each source\footnote{Note that \cite{2017A&A...600A..20A} take 200~pc as the distance of Lupus~III. However, the Gaia DR2 distance of most sources in Lupus~III is $\sim$160~pc}. 
We carry out a linear regression to fit: Log~$L_{\rm acc}$=a$\times$Log~$L_{\rm line}$+b. We have excluded weak (or dubious) accretors and sub-luminous objects from the fit. The best fit $a$ and $b$ coefficients and their uncertainties are listed in Table~\ref{Table:line_lacc}. Within 1$\sigma$, most of them are the same as those reported in \cite{2017A&A...600A..20A}.

\setcounter{table}{6}
\begin{table*}
\caption{Revised Log~$L_{\rm acc}$--Log~$L_{\rm line}$ linear fit}\label{Table:line_lacc}
\centering
\begin{tabular}{lcccccccccccc}
\hline
 Lines   &   $\lambda$ (\AA)     & a($\pm$err)   &b($\pm$err)   \\
\hline
H${\alpha}$  &6562.80  &  1.15($\pm$  0.05)  &  1.81($\pm$  0.20)\\
H${\beta}$  &4861.33  &  1.15($\pm$  0.03)  &  2.64($\pm$  0.16)\\
H${\gamma}$  &4340.46  &  1.12($\pm$  0.03)  &  2.75($\pm$  0.17)\\
H${\delta}$  &4101.73  &  1.09($\pm$  0.04)  &  2.74($\pm$  0.19)\\
%H${\epsilon}$  &3970.07  &  1.08($\pm$  0.04)  &  2.76($\pm$  0.19)\\
H${\zeta}$  &3889.05  &  1.06($\pm$  0.03)  &  2.76($\pm$  0.18)\\
\hline
Pa$\beta$  &12818.07  &  1.08($\pm$  0.07)  &  2.81($\pm$  0.36)\\
Pa$\gamma$  &10938.09  &  1.21($\pm$  0.06)  &  3.43($\pm$  0.29)\\
Pa$\delta$  &10049.37  &  1.27($\pm$  0.09)  &  3.93($\pm$  0.44)\\
\hline
Br$\gamma$  &21661.2  &  1.20($\pm$  0.11)  &  4.07($\pm$  0.57)\\
\hline
He~{\scriptsize I}  &4026.19  &  1.05($\pm$  0.04)  &  3.67($\pm$  0.22)\\
He~{\scriptsize I}  &4471.48  &  1.05($\pm$  0.04)  &  3.48($\pm$  0.23)\\
He~{\scriptsize I}  &4713.15  &  0.86($\pm$  0.08)  &  3.02($\pm$  0.50)\\
He~{\scriptsize I}  &5015.68  &  1.01($\pm$  0.04)  &  3.61($\pm$  0.24)\\
He~{\scriptsize I}  &5875.62  &  1.17($\pm$  0.04)  &  3.78($\pm$  0.22)\\
He~{\scriptsize I}  &6678.15  &  1.24($\pm$  0.06)  &  4.71($\pm$  0.34)\\
He~{\scriptsize I}  &7065.19  &  1.20($\pm$  0.05)  &  4.57($\pm$  0.29)\\
He~{\scriptsize II}  &4685.80  &  1.05($\pm$  0.05)  &  3.90($\pm$  0.33)\\
\hline
Ca~{\scriptsize II} (K)  &3933.66  &  1.05($\pm$  0.04)  &  2.60($\pm$  0.18)\\
Ca~{\scriptsize II} (H)  &3968.47  &  1.09($\pm$  0.03)  &  2.78($\pm$  0.16)\\
\hline
\end{tabular}
\normalsize
\end{table*}

\end{document}